%% file: main.tex
\documentclass[a4paper,11pt]{article}
\pdfoutput=1 

\usepackage{jheppub} 

\usepackage{cleveref}%
\usepackage{dsfont}%
\usepackage{mathrsfs}%
\usepackage{mathtools}%
\usepackage{physics}%
\usepackage{subcaption}%
\usepackage{subfiles}%
\usepackage{tensor}%
\usepackage{tikz}%
\usetikzlibrary{fadings}%
\usetikzlibrary{tikzmark}%
\usepackage{xfrac}%
\usepackage{xspace}%

\graphicspath{{figures/}}%
\newcommand{\eg}{\textit{e.g.\@\xspace}}%
\newcommand{\ie}{\textit{i.e.\@\xspace}}%
\DeclareMathOperator{\spn}{span}
\DeclareMathOperator{\swap}{\ensuremath{
    \begin{tikzpicture}[baseline=(char.base)]
      \draw[rounded corners=2] (0,0) -- ++ (0,0.125) -| (0.2,0.25);
      \draw[rounded corners=2, white, line width=3] (0.2,0) -- ++ (0,0.125) -| (0,0.25);
      \draw[rounded corners=2] (0.2,0) -- ++ (0,0.125) -| (0,0.25);
      \node (char) at (0,0) {};
    \end{tikzpicture}
  }}

\newcommand{\figscale}{0.5}

\definecolor{solarizedBase03}{HTML}{002B36}%
\definecolor{solarizedBase02}{HTML}{073642}%
\definecolor{solarizedBase01}{HTML}{586E75}%
\definecolor{solarizedBase00}{HTML}{657B83}%
\definecolor{solarizedBase0}{HTML}{839496}%
\definecolor{solarizedBase1}{HTML}{93A1A1}%
\definecolor{solarizedBase2}{HTML}{EEE8D5}%
\definecolor{solarizedBase3}{HTML}{FDF6E3}%
\definecolor{solarizedYellow}{HTML}{B58900}%
\definecolor{solarizedOrange}{HTML}{CB4B16}%
\definecolor{solarizedRed}{HTML}{DC322F}%
\definecolor{solarizedMagenta}{HTML}{D33682}%
\definecolor{solarizedViolet}{HTML}{6C71C4}%
\definecolor{solarizedBlue}{HTML}{268BD2}%
\definecolor{solarizedCyan}{HTML}{2AA198}%
\definecolor{solarizedGreen}{HTML}{859900}%

\title{\boldmath Observers seeing gravitational Hilbert spaces: abstract sources
  for an abstract path integral}

\author{Hong Zhe (Vincent) Chen}
\affiliation{Department of Physics, University of California,\\
  Santa Barbara, CA 93106, USA}

\emailAdd{hzchen@ucsb.edu}

\abstract{The gravitational path integral suggests a striking result: the
  Hilbert space of closed universes in each superselection sector, a so-called
  $\alpha$-sector, is one-dimensional. We develop an abstract formalism
  encapsulating recent proposals that modify the gravitational path integral in
  the presence of observers and allow larger Hilbert spaces to be associated
  with closed universes. Our formalism regards the gravitational path integral
  as a map from abstract objects called sources to complex numbers, and
  introduces additional objects called partial sources, which form sources when
  glued together. We apply this formalism to treat, on equal footing, universes
  with spatial boundaries, closed universes with prescribed observer worldlines,
  and closed universes containing observers entangled with external systems. In
  these contexts, the relevant gravitational Hilbert spaces contain states
  prepared by partial sources and can consequently have nontrivial
  $\alpha$-sectors supporting noncommuting operators. Within our general
  framework, the positivity of the gravitational inner product implies a bound
  on the Hilbert space trace of certain positive operators over each
  $\alpha$-sector. The trace of such operators, in turn, quantifies the
  effective size of this Hilbert space.}

\begin{document}
\maketitle
\flushbottom

\subfile{sections/section01_introduction}

\subfile{sections/section02_gpiBabyUniverses}

\subfile{sections/section03_partialSourceSectors}

\subfile{sections/section04_operatorsDimBounds}

\subfile{sections/section05_discussion}


\acknowledgments


I am very grateful to Donald Marolf for his guidance on this project. I would
also like to thank Goncalo Araujo-Regado, Elliott Gesteau, Maciej Kolanowski,
Bilyana Tomova, and Mykhaylo Usatyuk for interesting discussions. I am supported
by a Fundamental Physics Fellowship through the University of California, Santa
Barbara.

\paragraph{Note added.} In the v1 version of this paper, a puzzle was raised
concerning the bound \labelcref{eq:avgboundcircobs} on the size of the JT closed
universe Hilbert space with one observer, under the prescription of
ref.~\cite{Abdalla:2025gzn}. Partly inspired by remarks in
ref.~\cite{Blommaert:2025bgd}, we now present in this v2 version what is
believed to be a more accurate discussion of \cref{eq:avgboundcircobs} that
resolves the previously raised puzzle.









\bibliographystyle{JHEP.bst}
\bibliography{references.bib}

\end{document}

%% file: sections/section01_introduction.tex
\section{Introduction}
\label{sec:intro}

Recent years have seen significant advancements that clarified some aspects of
the black hole information problem. A central development was the discovery of
quantum extremal surfaces \cite{Penington:2019npb,Almheiri:2019psf} associated
with replica wormholes \cite{Penington:2019kki,Almheiri:2019qdq}. In particular,
a sum over topologies in entropy calculations using the gravitational path
integral was found to be crucial for seeing the expected decrease in the number of
states of an evaporating black hole, evoking
\cite{Marolf:2020xie,Marolf:2020rpm} old discussions
\cite{Coleman:1988cy,Giddings:1988cx,Giddings:1988wv} of so-called baby
universes and \(\alpha\)-states.

The same rules for the gravitational path integral, however, lead to the
striking conclusion that the Hilbert space of closed universes is
one-dimensional \cite{Marolf:2020xie,Usatyuk:2024mzs,Usatyuk:2024isz}. This
puzzling result is not unrelated to the successes in the black hole context
described above \cite{Marolf:2020rpm}. Indeed, the interior of an evaporated
black hole is similar to a closed universe. While the semiclassical descriptions
of quantum fields in the interiors of old black holes or in closed universes
admit many states, the gravitational path integral suggests that many or all of
these states in each case are in fact linearly dependent in quantum gravity
\cite{Marolf:2020xie,Marolf:2020rpm,Akers:2022qdl,Harlow:2025pvj,Akers:2025ahe}.

In an effort to describe the rich experiences of observers in closed universes,
refs.~\cite{Harlow:2025pvj,Abdalla:2025gzn} have suggested special rules for the
inclusion of observers in the gravitational path integral. While these rules
recover nontrivial Hilbert spaces for closed universes with observers, they are
somewhat ad hoc and are not derived from first principles. Indeed, the rules
proposed by refs.~\cite{Harlow:2025pvj,Abdalla:2025gzn} are distinct and it is
unclear whether one proposed set of rules is better than the other, or if either
really describe the experiences of an observer. We will instead take an agnostic
approach, aiming to identify the general structures shared by the proposals of
refs.~\cite{Harlow:2025pvj,Abdalla:2025gzn} which lead to nontrivial Hilbert
spaces in the presence of observers.

In this paper, we develop a general framework in which the gravitational path
integral \(\zeta:\mathcal{J}\to\mathbb{C}\) is a function from abstract objects
\(J\in\mathcal{J}\), called \emph{sources}, to complex numbers. Examples of
sources \(J\) include boundary manifolds specifying boundary conditions for bulk
fields. Other examples, relevant to the prescription of
ref.~\cite{Abdalla:2025gzn}, might be boundary manifolds connected by prescribed
bulk geodesics. However, to isolate essential features, we will remain as
abstract as possible while developing our general framework. We will therefore
be interested in identifying general structures and properties which define an
abstract set \(\mathcal{J}\) of sources and a function \(\zeta\) that we refer
to as the gravitational path integral.\footnote{This abstract approach to
  studying the gravitational path integral is similar to that taken by
  ref.~\cite{Colafranceschi:2023moh}. In this paper, we extend the abstraction
  to the inputs \(J\in\mathcal{J}\) of the path integral.}

In particular, we will argue that \(\mathcal{J}\) should have the structure of a
\(*\)-algebra\footnote{The \(*\)-involution can be understood as the complex
  conjugation and orientation- or time-reversal of sources. We also allow
  complex superpositions of sources. See \cref{sec:gpibaby} for details.},
equipped with a commutative multiplication \(\sqcup\) which one can view as
taking the disjoint union of sources. The gravitational path integral
\(\zeta:\mathcal{J}\to\mathbb{C}\) is therefore implicitly invariant under
permutations of constituent sources in disjoint unions \(\sqcup\). This property
of the path integral is the underlying structure responsible for the fact that
the Hilbert space of closed universes is one-dimensional.

More precisely, by inserting sources \(J\in\mathcal{J}\), say in the Euclidean
past, we can prepare a Hilbert space \(\mathcal{H}_\emptyset\) of states
associated with bulk slices that do not intersect sources \(J\). The slices have
no spatial boundary and their topology and local fields vary freely across
different configurations in the path integral. We will refer to such freely
fluctuating closed universes as baby universes. As we will later review, the
commutativity of the \(*\)-algebra \(\mathcal{J}\) implies that the baby
universe Hilbert space \(\mathcal{H}_\emptyset\) decomposes into one-dimensional
superselection sectors \(\mathcal{H}_\emptyset^\alpha\) --- so-called
\(\alpha\)-sectors. Because they are superselection sectors, each
\(\mathcal{H}_\emptyset^\alpha\) can be viewed as the baby universe Hilbert
space in a self-contained theory of quantum gravity. Meanwhile, the path
integral \(\zeta\) and the Hilbert space \(\mathcal{H}_\emptyset\) describe an
ensemble average over these theories labelled by \(\alpha\).\footnote{When the
  bulk spacetime dimension is greater than three, some authors, \eg{} in
  ref.~\cite{McNamara:2020uza}, advocate for the existence of a \emph{unique}
  \(\alpha\)-sector and thus the lack of an ensemble average. We will proceed in
  this paper without assuming that the path integral \(\zeta\) necessarily
  corresponds to a single \(\alpha\)-sector. \label{foot:uniquealpha}}

As we know concretely from AdS/CFT, if one considers not closed universes but
instead universes with spatial boundaries, the gravitational Hilbert space need
not be one-dimensional and can support noncommuting operators. In this paper, we
will demonstrate that there is a close analogy between universes with spatial
boundaries and the prescriptions of refs.~\cite{Harlow:2025pvj,Abdalla:2025gzn}
for including observers in closed universes. The common theme is that Hilbert
spaces with nontrivial \(\alpha\)-sectors supporting noncommuting operators
arise from cutting sources for the gravitational path integral.

To formalize this notion, we introduce abstract objects called \emph{cuts} and
\emph{partial sources}. For each cut \(C\), there is a set \(\mathcal{J}_C\) of
partial sources. A defining property of partial sources
\(J_C,J_C'\in\mathcal{J}_C\) is that they can be used to construct a complete
source \((J_C'|J_C)\in\mathcal{J}\) when appropriately glued together across the
cut \(C\).\footnote{As we will describe in more detail in
  \cref{sec:generalpartialsources}, \((\bullet|\bullet)\) is a
  \(\mathcal{J}\)-valued Hermitian form, involving a complex conjugation and
  orientation- or time-reversal of its first argument.} An example of a partial
source is part \(J_C=\mathscr{N}_\Sigma\) of a boundary manifold, itself with
boundary \(C=\Sigma\). To capture the prescriptions of
refs.~\cite{Abdalla:2025gzn,Harlow:2015lma} for including observers in closed
universes, one must consider other examples, obtained from cutting bulk
geodesics representing observer worldlines and cutting inner products of
external non-gravitational systems.

Before restricting to specific examples, we will see even from our abstract
construction that the set \(\mathcal{J}_C\) of partial sources can be used to
prepare a Hilbert space \(\mathcal{H}_C\) of states which decomposes into
\(\alpha\)-sectors \(\mathcal{H}_C^\alpha\) that are generically \emph{not}
one-dimensional. Conceptually, cuts are crucial for the existence of nontrivial
operators in quantum gravity because they are objects to which one can dress
such operators. Concretely, we will see that generically noncommuting operators
on \(\mathcal{H}_C^\alpha\) can be constructed by gluing certain two-sided
partial sources \(J_{C\otimes{}C^*}\in\mathcal{J}_{C\otimes{}C^*}\) to
states.\footnote{The cut of the partial source \(J_{C\otimes{}C^*}\) consists of
  two components: \(C\) and an appropriate dual \(C^*\). We will refrain from
  elaborating further here; see \cref{sec:generalpartialsources} for details.}

Even though the \(\alpha\)-sectors \(\mathcal{H}_C^\alpha\) need not be
one-dimensional, there are nonetheless useful bounds that one can place on the
size of \(\mathcal{H}_C^\alpha\). More precisely, we will show that the positive
definiteness of the gravitational inner product implies an upper bound on the
Hilbert space trace of certain positive operators over \(\mathcal{H}_C^\alpha\),
given by the value of a path integral evaluated with a corresponding source.
Together with an estimate of the size of eigenvalues, these traces can then be
used to roughly bound the number of states in appropriate windows of
eigenvalues. In the case of universes with spatial boundaries, considering the
Euclidean boundary time evolution operator \(e^{-\beta\, H}\) for example, one
recovers the expected bound on the canonical partition function in terms of a
path integral with a boundary manifold containing a Euclidean time circle
\cite{Marolf:2020xie}.

Slightly pushing the envelope on ref.~\cite{Abdalla:2025gzn}'s prescription, the
closest analogue of this example would be a bound, on the dimension of the
closed universe Hilbert space with one observer, given by a gravitational path
integral involving a circular observer worldline. Indeed, this is reminiscent of
the path integral studied by ref.~\cite{Maldacena:2024spf}, which emphasized the
importance of observers for extracting a state-counting interpretation in de
Sitter spacetime.\footnote{The second version of ref.~\cite{Maldacena:2024spf}
  reports that their path integral calculation still leaves an extraneous sign,
  which remains to be understood. \label{foot:dssign}} In AdS Jackiw-Teitelboim
(JT) gravity, we will however find that both sides of this bound seem to diverge
\cite{Abdalla:2025gzn,Blommaert:2025bgd}. The JT closed universe Hilbert space
with one observer is infinite dimensional in each \(\alpha\)-sector. Meanwhile,
the JT path integral with a circular observer worldline receives an infinite
contribution from the degenerate torus, where the observer worldline wraps the
degenerate cycle.

To get a notion of the density of states in typical \(\alpha\)-sectors,
ref.~\cite{Abdalla:2025gzn} further restricts to finite-dimensional subspaces
corresponding to microcanonical windows of the random matrix theory dual to JT
gravity. Using partial sources involving observer worldlines attached to
asymptotic boundaries, we will construct operators that project down to such
subspaces. When applied to these operators, our general bound on the Hilbert
space trace in each \(\alpha\)-sector becomes nontrivial. Indeed, we will find
that ref.~\cite{Abdalla:2025gzn}'s result for the (ensemble-averaged) size of
the one-observer closed universe Hilbert space in each subspace is roughly
consistent with the saturation of our bound. Through an analogous analysis with
analogous partial sources, we will also find rough agreement between the
saturation of our bound and ref.~\cite{Harlow:2025pvj}'s reported size of the
Hilbert space under their distinct prescription. As these examples demonstrate,
our general framework provides a unified language in which we can examine
various prescriptions for including observers, even if the details vary from
prescription to prescription.

The remainder of this paper is organized as follows. We start in
\cref{sec:gpibaby} by discussing the gravitational path integral \(\zeta\) as a
map from a set \(\mathcal{J}\) of sources to complex numbers. In
\cref{sec:exsources}, we first provide some examples of sources
\(J\in\mathcal{J}\) that are motivated from viewing \(\zeta\) as a literal path
integral over bulk configurations. This will hopefully ease the reader into the
more abstract construction in \cref{sec:generalprops} of general structures and
properties which we will take to be our definitions of a set \(\mathcal{J}\) of
sources and a gravitational path integral \(\zeta\). We will then use these
general structures in \cref{sec:babysector} to construct a Hilbert space
\(\mathcal{H}_\emptyset\) of baby universes and decompose
\(\mathcal{H}_\emptyset\) into one-dimensional \(\alpha\)-sectors
\(\mathcal{H}_\emptyset^\alpha\). There is a similar organization of
\cref{sec:partialsourcesectors}, which concerns cuts and partial sources. We
start in \cref{sec:expartialsources} with examples, before moving on in
\cref{sec:generalpartialsources} to abstract structures which define a set of
cuts and sets of partial sources more generally. In \cref{sec:cutsectors}, we
construct Hilbert spaces \(\mathcal{H}_C\) labelled by cuts, which decompose
into \(\alpha\)-sectors \(\mathcal{H}_C^\alpha\) that are generically not
one-dimensional. We consider operators that act on these Hilbert spaces in
\cref{sec:operatorsdimbounds}. After introducing operators constructed from
gluing partial sources in \cref{sec:operators}, we derive a bound on the Hilbert
space trace of certain positive operators in \cref{sec:bounds}. We conclude
with a discussion in \cref{sec:discuss}. This includes more explicit
explanations of how our general framework applies to universes with spatial
boundaries in \cref{sec:spatialbdy} and how it applies to the prescriptions of
refs.~\cite{Abdalla:2025gzn,Harlow:2025pvj} for including observers in closed
universes in \cref{sec:applications}. We end in \cref{sec:future} with a
discussion of future directions.

%% file: sections/section02_gpiBabyUniverses.tex
\section{The gravitational path integral and the baby universe Hilbert space}
\label{sec:gpibaby}

In this section, we will begin reviewing and extending the framework formulated
in ref.~\cite{Marolf:2020xie} for describing the quantum gravity Hilbert space
obtained from a gravitational path integral.

Let us begin by sketching what we mean by a gravitational path integral.
Abstractly, a gravitational path integral is a function
\begin{align}
  \zeta : \mathcal{J} \to \mathbb{C}
\end{align}
from a set \(\mathcal{J}\), of what we will call sources, to complex numbers
\(\mathbb{C}\). In \cref{sec:exsources}, we will first give some examples of
possible sources \(J\in\mathcal{J}\) which serve as inputs for the path
integral. It may be helpful to bear these examples in mind as we proceed with an
abstract discussion of \(\mathcal{J}\) in \cref{sec:generalprops}. There, we
will abstractly describe the defining properties that we demand of
\(\mathcal{J}\), in particular, giving \(\mathcal{J}\) the structure of a
commutative \(*\)-algebra. Along the way, we will also state axioms that we
require the path integral \(\zeta\) to satisfy. Finally, in
\cref{sec:babysector}, we will use \(\mathcal{J}\) to construct a Hilbert space
\(\mathcal{H}_\emptyset\) of freely fluctuating closed universes --- so-called
baby universes --- prepared by sources \(J\in\mathcal{J}\). Notably,
\(\mathcal{H}_\emptyset\) consists of one-dimensional superselection sectors ---
so-called \(\alpha\)-sectors.

We will later see in \cref{sec:partialsourcesectors} that Hilbert spaces
\(\mathcal{H}_C\) with nontrivial \(\alpha\)-sectors can instead be constructed
from objects that we call partial sources, which form sources when glued
together.

\subsection{Possible examples of sources \(J\in\mathcal{J}\)}
\label{sec:exsources}

The set \(\mathcal{J}\) of sources for the path integral will be
theory-dependent. Here, we will give only some illustrative examples of sources
\(J\in\mathcal{J}\) that might be natural to consider in a theory of quantum
gravity. These examples are intended to be neither inclusive nor exclusive ---
that is, for a given theory, \(\mathcal{J}\) need not include all of the sources
described below and \(\mathcal{J}\) may contain sources outside the scope of the
following examples.

The examples here are motivated by the intuition that \(\zeta\) may indeed be a
literal path integral\footnote{For concreteness, the weight \(e^{-S[\Phi]}\) of
  the path integral has been written to suggest Euclidean signature, but our
  general framework will be largely insensitive to this distinction.}
\begin{align}
  \zeta(J)
  &= \int_{\Phi\sim J}
    \mathcal{D}\Phi\,
    e^{-S[\Phi]}
    \label{eq:pathintegral}
\end{align}
over some bulk fields \(\Phi\) that, in particular, describe the geometry of the
bulk spacetime. The notation \(\Phi\sim J\) indicates that the integral is
restricted by certain conditions or given certain weights specified by the
source \(J\in\mathcal{J}\), as we will describe shortly. Let us emphasize that
\cref{eq:pathintegral} serves only as a motivation for the examples of sources
\(J\) described below and for some of the general properties of \(\mathcal{J}\)
and \(\zeta\) that we will formulate in \cref{sec:generalprops}. Ultimately, we
can consider any \(\mathcal{J}\) and \(\zeta\) that satisfy those more general
properties, regardless of whether they are related to an actual integral like
\labelcref{eq:pathintegral}.

\begin{figure}
  \centering
  \begin{subfigure}[t]{0.3\textwidth}
    \centering
    \includegraphics[scale=\figscale]{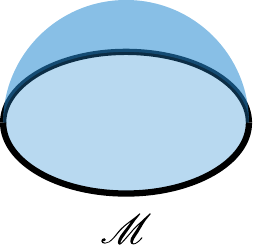}
    \caption{A source \(J=\mathscr{M}\) introducing a boundary manifold
      \(\mathscr{M}\) on which boundary conditions for bulk fields are
      specified.}
    \label{fig:sourcebdycond}
  \end{subfigure}
  \hfill
  \begin{subfigure}[t]{0.3\textwidth}
    \centering
    \includegraphics[scale=\figscale]{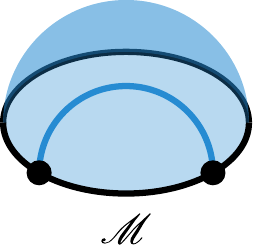}
    \caption{A source \(J=\mathscr{M}\) where \(\mathscr{M}\) is now a boundary
      manifold which creates matter excitations in the bulk, modelled as a
      worldline.}
    \label{fig:sourcebdydefect}
  \end{subfigure}
  \hfill
  \begin{subfigure}[t]{0.3\textwidth}
    \centering \includegraphics[scale=\figscale]{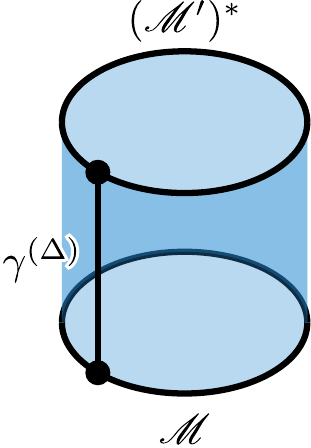}
    \caption{A source \(J=(\mathscr{M}')^*\cdot\gamma^{(\Delta)}\cdot\mathscr{M}\)
      specifying conditions on a boundary manifold with pieces \(\mathscr{M}\)
      and \((\mathscr{M}')^*\) joined by a geodesic \(\gamma^{(\Delta)}\). The length
      \(\ell\) of the geodesic is to be integrated over with an extra weight
      \(e^{-\Delta\ell}\).}
    \label{fig:sourcegeodesic}
  \end{subfigure}
  \caption{Possible examples of sources \(J\in\mathcal{J}\) (black) for
    the gravitational path integral. In each panel, a possible bulk
    configuration that might appear in the path integral \(\zeta(J)\) is
    illustrated in blue.}
  \label{fig:sources}
\end{figure}

If we do assume that \(\zeta\) takes the form of an integral
\labelcref{eq:pathintegral}, then one example of a source \(J\in\mathcal{J}\) is
boundary conditions on a boundary manifold of the bulk spacetime. We will use
\(J=\mathscr{M}\) in the following discussion to refer to both the boundary
manifold and the data on this manifold describing the boundary conditions
satisfied by the bulk fields. For example, \(\mathscr{M}\) it might specify the
boundary configurations that fields such as the metric must asymptotically
approach. In \cref{fig:sourcebdycond}, we illustrate a one-dimensional circular
boundary manifold \(\mathscr{M}\) on which boundary conditions are specified for
a two-dimensional bulk theory.
The boundary condition specified by such a source
can in general be nontrivial and can be engineered to create
excitations that propagate into the bulk. In particular, as illustrated in
\cref{fig:sourcebdydefect}, one can insert localized sources --- in the
conventional (A)dS/CFT sense of the word --- on the boundary manifold
\(\mathscr{M}\) which produces excitations of bulk matter
fields. In simple theories, such as two-dimensional topological models
\cite{Marolf:2020xie} or JT gravity, the matter excitations might be modelled
simply as worldlines in the bulk emanating from defects on
the boundary manifold.

The boundary manifold need not be connected. For example, a source
\begin{align}
  \mathscr{M}\sqcup\mathscr{M}'
  \in \mathcal{J}
  \label{eq:mcupm}
\end{align}
would specify boundary conditions on the disjoint union of two boundary
manifolds \(\mathscr{M}\) and \(\mathscr{M}'\). The path integral
\labelcref{eq:pathintegral} can then include bulk spacetimes with arbitrary
connections between the mutually disconnected pieces of the boundary manifold.
Thus, generically,
\begin{align}
  \zeta(\mathscr{M} \sqcup \mathscr{M}')
  &\ne \zeta(\mathscr{M})\, \zeta(\mathscr{M}')
    \;.
\end{align}
We also allow the data on the boundary manifold to have nonlocal correlations.
We can define, for example, a new source as the superposition of \(d\)-many
sources of the kind \labelcref{eq:mcupm},
\begin{align}
  \sum_{i=1}^d
  \mathscr{M}^{(i)} \sqcup \mathscr{M}^{\prime(i)}
  \in \mathcal{J}
  \;.
\end{align}
This formal sum is defined as the instruction to evaluate the path integral for
each term individually and add the results together:
\begin{align}
  \zeta\left(
  \sum_{i=1}^d
  \mathscr{M}^{(i)} \sqcup \mathscr{M}^{\prime(i)}
  \right)
  &=
  \sum_{i=1}^d
  \zeta\left(
  \mathscr{M}^{(i)} \sqcup \mathscr{M}^{\prime(i)}
    \right)
    \;.
\end{align}

So far, we have described examples where sources supply data on boundary
manifolds. More generally, a source can specify the existence and properties of
other objects in bulk configurations to be included in the path integral
\labelcref{eq:pathintegral}. An example of this is illustrated in
\cref{fig:sourcegeodesic}. Here, a source which we will write as
\(J=(\mathscr{M}')^*\cdot\gamma\cdot\mathscr{M}\) includes not only a boundary
manifold, given by pieces \(\mathscr{M}\) and \((\mathscr{M}')^*\), but also a
worldline \(\gamma\) in the bulk that is required to connect prescribed points
on the boundary pieces. (The \(*\) operation will be described in
\cref{sec:generalprops} and is used here only to match later notation; the
reader can safely regard \((\mathscr{M}')^*\) here as an arbitrary boundary
manifold.) The source might further specify that the worldline must be a
geodesic, fix a prescribed length \(\ell\) of the geodesic, or provide
instructions to integrate over \(\ell\) with some extra weight, say
\(e^{-\Delta\,\ell}\). The path integral \labelcref{eq:pathintegral} will then
be restricted to configurations possessing the prescribed geodesic and
satisfying the prescribed boundary conditions on the boundary manifold.

\subsection{The commutative \(*\)-algebra \(\mathcal{J}\) of sources and
  properties of the path integral \(\zeta\)}
\label{sec:generalprops}

We will now proceed with an abstract discussion of the properties we require of
\(\mathcal{J}\) and \(\zeta\). These properties will serve as the definition of
a set \(\mathcal{J}\) of sources and a function
\(\zeta:\mathcal{J}\to\mathbb{C}\) which we will call a gravitational path
integral, irrespective of whether \(\zeta\) truly comes from an integral
\labelcref{eq:pathintegral} over bulk field configurations. Since this
discussion will be quite abstract, the reader may find it helpful to keep
\cref{eq:pathintegral} and the examples of sources \(J\in\mathcal{J}\) discussed
in \cref{sec:exsources} in mind.

At the outset, \(\mathcal{J}\) is just a set of elements, each of which we call
a source. Below, we will describe in turn each additional structure that we
require \(\mathcal{J}\) to have. At each step, we will also describe relevant
properties required of the path integral with respect to this structure.

Firstly, we require \(\mathcal{J}\) to be a vector space, typically of infinite
dimension, over \(\mathbb{C}\). That is, given any two sources
\(J,J'\in\mathcal{J}\) and scalars \(c,c'\in\mathbb{C}\), we also include
the formal linear combination\footnote{This formal linear combination should not
  be confused with taking a point-wise linear combination of field profiles,
  \eg{} boundary conditions, specified by the sources \(J\) and \(J'\).}
\begin{align}
  c J + c' J' \in \mathcal{J}
  \;.
  \label{eq:sourcelincomb}
\end{align}
(Otherwise, we simply redefine \(\mathcal{J}\) to include all such linear
combinations.) By an abuse of notation, we will denote the additive identity in
\(\mathcal{J}\) with the same symbol \(0\) as in \(\mathbb{C}\). We will require
the path integral \(\zeta\) to act linearly, such that
\begin{align}
  \zeta(c J + c' J')
  &= c\, \zeta(J) + c'\, \zeta(J')
    \;.
\end{align}
Conceptually, this relation defines what it means to take a formal linear
combination of sources.

We further require \(\mathcal{J}\) to have a multiplication operation
\begin{align}
  \sqcup : \mathcal{J} \times \mathcal{J}
  &\to \mathcal{J}
  \;,
  &
  J_1, J_2
    &\mapsto J_1 \sqcup J_2
      \;,
      \label{eq:jmult}
\end{align}
which is monoidal\footnote{This means the multiplication is associative and has
  an identity element.}, distributive over addition, and commutative, thus
giving \(\mathcal{J}\) the structure of a commutative ring. We will denote the
\(\sqcup\)-multiplicative identity as \(\varnothing\in\mathcal{J}\). We will
further require \(\sqcup\) to be compatible with scalar multiplication by
complex numbers \(c\in\mathbb{C}\),
\begin{align}
  c\,(J\sqcup J')
  &=
  (c J)\sqcup J'
  = J\sqcup (c\,J')
    \;.
\end{align}
Because $\mathcal{J}$ is a vector space and a commutative ring with compatible
ring and scalar multiplication, it is a commutative algebra.

The symbols chosen above are purposely suggestive. In particular, suppose that
\(J\) and \(J'\) specify certain conditions on objects found in bulk
configurations (\eg{} boundary conditions on boundary manifolds). Then
\(J\sqcup J'\) is to be interpreted as a source specifying those conditions
on the disjoint union of the respective objects.

With respect to the path integral \(\zeta\), let us remark firstly that implicit
from the commutativity of \(\sqcup\)-multiplication is the requirement for
\(\zeta\) to be invariant,
\begin{align}
  \zeta(J\sqcup J')
  &= \zeta(J'\sqcup J)
    \;,
\end{align}
under the permutation of sources \(J,J'\in\mathcal{J}\). A second remark is
that we do not require \(\zeta\) to factorize between sources. That is, generically,
\begin{align}
  \zeta(J \sqcup J')
  &\ne \zeta(J)\, \zeta(J')
    \;.
    \label{eq:nofactor}
\end{align}
This is because, if \(\zeta(J\sqcup J')\) is a literal path integral
\labelcref{eq:pathintegral} that includes a sum over different spacetime
geometries, then it is natural to allow arbitrary topologies in the sum,
including those that connect the objects (\eg{} boundary manifolds) in \(J\) and
\(J'\). Due to the generic lack of factorization, the path integral $\zeta$ is
not necessarily an algebra homomorphism.

The final structure that we require $\mathcal{J}$ to have is an involution and
automorphism \(*\)
satisfying
\begin{align}
    (J^*)^*
  &= J
    \;,
    &
  (J + J')^*
  &= J^* + (J')^*
    \;,
  &
    (J \sqcup J')^*
  &= J^* \sqcup (J')^*
    \;,
  &
    \varnothing^*
  &= \varnothing
    \;.
\end{align}
This gives $\mathcal{J}$ the structure of a commutative \(*\)-ring.\footnote{For
  a general \(*\)-ring, one usually requires \(*\) to be an anti-automorphism.
  With a commutative multiplication \(\sqcup\) however, an automorphism is the
  same as an anti-automorphism.} We further require \(*\) to be compatible with
scalar multiplication,
\begin{align}
  (c\,J)^*
  &= c^* J^*
    \;,
\end{align}
where we have denoted complex conjugation in \(\mathbb{C}\) by \(*\) as well.
Thus, $\mathcal{J}$ is a commutative \(*\)-algebra.

The interpretation of \(*\) is a complex conjugation and orientation- or
time-reversal of sources.\footnote{More concretely, let us consider a \(J\) that
  is a boundary manifold, viewed as preparing an initial state of closed
  universes. Then the action of \(*\) on this state is the \(\mathcal{T}\) or
  \(\mathsf{T}\) transformation defined by ref.~\cite{Witten:2025ayw} in
  theories without or with time-reflection symmetry respectively.} We require
the path integral \(\zeta\) to commute with \(*\),
\begin{align}
  \zeta(J^*)
  &= \zeta(J)^*
    \;.
\end{align}
Additionally, we require \(\zeta\) to be reflection-positive, meaning
\begin{align}
  \zeta(J^*\sqcup J)
  &\ge 0
    \label{eq:reflpos}
\end{align}
is real and non-negative. We can therefore use \(\zeta\) to define a positive
semi-definite Hermitian form on \(\mathcal{J}\):
\begin{align}
  \braket{J'}{J}
  &= \zeta((J')^*\sqcup J)
    \;.
    \label{eq:innerprod}
\end{align}

In summary, the set \(\mathcal{J}\) of sources is a commutative \(*\)-algebra.
The path integral \(\zeta:\mathcal{J}\to\mathbb{C}\) is almost a
\(*\)-homomorphism, except it might fail to factorize, as expressed by the
generic inequality \labelcref{eq:nofactor}. Additionally, \(\zeta\) is required
to be reflection-positive \labelcref{eq:reflpos}, thus equipping \(\mathcal{J}\)
with a positive semi-definite Hermitian form \labelcref{eq:innerprod}. These
properties (together with a reflection-positivity condition generalizing
\cref{eq:reflpos} to be specified in \cref{sec:generalpartialsources}) can be
taken to be the definitions of a general set \(\mathcal{J}\) of sources and a
gravitational path integral \(\zeta\).

\subsection{The baby universe sector \(\mathcal{H}_\emptyset\) of the
  gravitational Hilbert space}
\label{sec:babysector}

\begin{figure}
  \centering
  \includegraphics[scale=\figscale]{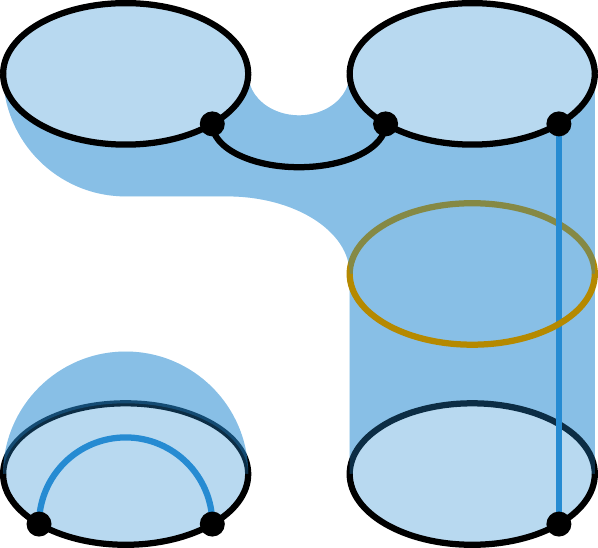}
  \caption{Slicing (orange) the gravitational path integral while avoiding
    cutting sources (black) gives a state in the baby universe Hilbert space.}
  \label{fig:babyuniverses}
\end{figure}

States of a quantum theory can be obtained from slicing the path integral. In
this subsection, we will construct a Hilbert space \(\mathcal{H}_\emptyset\) of
states obtained from slicing the gravitational path integral in a way that
avoids cutting sources, as illustrated in \cref{fig:babyuniverses}.\footnote{The
  subscript on \(\mathcal{H}_\emptyset\) has been introduced in anticipation of
  the more general sectors of the gravitational Hilbert space to be constructed
  in \cref{sec:cutsectors}. (The symbol \(\emptyset\) should not be confused
  with the multiplicative identity \(\varnothing\in\mathcal{J}\) introduced
  below \cref{eq:jmult}.)} In particular, the bulk spacetime slices do not have
spatial boundaries where boundary conditions are imposed by sources. The Hilbert
space \(\mathcal{H}_\emptyset\) therefore describes arbitrarily many closed
universes, whose topology and local fields vary across configurations in the
path integral, without being directly constrained by sources. We will loosely
refer to such universes as baby universes.

In \cref{sec:generalprops}, we defined the set \(\mathcal{J}\) of sources to be
a commutative \(*\)-algebra and therefore a vector space. We also required the
path integral \(\zeta:\mathcal{J}\to\mathbb{C}\) to be reflection-positive
\labelcref{eq:reflpos}, such that it can be used in \cref{eq:innerprod} to
define a positive semi-definite Hermitian form \(\braket{\bullet}{\bullet}\) on
\(\mathcal{J}\). We can therefore build a Hilbert space
\(\mathcal{H}_\emptyset\) from \(\mathcal{J}\) as follows. First, we quotient
\(\mathcal{J}\) by null vectors. In particular, we define the subspace
\(\mathcal{N}\subset\mathcal{J}\) of null vectors to be the kernel of
\(\braket{\bullet}{\bullet}\), meaning
\begin{align}
  \mathcal{N}
  &= \left\{
    J\in\mathcal{J}
    : \braket{J'}{J} = 0\quad \forall J'\in\mathcal{J}
    \right\}
    \;.
\end{align}
We then consider the quotient \(\mathcal{J}/\mathcal{N}\) which is a pre-Hilbert
space on which \(\braket{\bullet}{\bullet}\) is a (positive-definite) inner
product. Second, we take \(\mathcal{H}_\emptyset\) to be the completion of
\(\mathcal{J}/\mathcal{N}\) with respect to the norm obtained from
\(\braket{\bullet}{\bullet}\). We can construct this completion
\(\mathcal{H}_\emptyset\) as the set of equivalence classes of Cauchy sequences
in \(\mathcal{J}/\mathcal{N}\) where two sequences are equivalent if their
difference converges to \(0\in\mathcal{J}/\mathcal{N}\). Since
\(\mathcal{H}_\emptyset\) is a complete inner product space, it is a Hilbert
space and, by construction, \(\mathcal{J}/\mathcal{N}\) is a dense subspace of
\(\mathcal{H}_\emptyset\).

The states in \(\mathcal{H}_\emptyset\) have the following interpretation.
Consider a source \(J\in\mathcal{J}\), which picks out an element
\(\ket{J}\) of \(\mathcal{J}/\mathcal{N}\subset\mathcal{H}_\emptyset\). This is
the state of baby universes prepared by the gravitational path integral from
the source \(J\). For example, \(J\) might specify the boundary conditions at
the spacetime boundary at ``time-like \(\infty\)'' or the Euclidean past. In
particular, the special case where \(J=\varnothing\) gives the no-boundary or
Hartle-Hawking state \(\ket{\varnothing}\). The baby universe Hilbert space
\(\mathcal{H}_\emptyset\) contains all such states prepared by sources
\(J\in\mathcal{J}\), as well as limits of these states, with two states
identified as equal if their difference has vanishing norm under the inner
product \(\braket{\bullet}{\bullet}\) determined by the path integral \labelcref{eq:innerprod}.

For each source \(J\in\mathcal{J}\), we can also define an operator \(Z(J)\) on
\(\mathcal{J}/\mathcal{N}\). In particular, for any \(J'\in\mathcal{J}\),
we define
\begin{align}
  Z(J) \ket{J'}
  &= \ket{J \sqcup J'}
    \;.
\end{align}
This is a well-defined operator on equivalence classes under the quotient
\(\mathcal{J}/\mathcal{N}\) --- for any \(N \in\mathcal{N}\), we also have
\(J\sqcup N \in\mathcal{N}\). From the definition \labelcref{eq:innerprod} of
the inner product, we see that matrix elements of the adjoint is given by
\begin{align}
  \mel{J'}{Z(J)^\dagger}{J''}
  &= \mel{J'}{Z(J^*)}{J''}
    \;.
\end{align}
Moreover, because of the commutativity of \(\sqcup\)-multiplication, we find
that any two such operators commute,
\begin{align}
  Z(J)\, Z(J')
  &= Z(J')\, Z(J)
    \;.
    \label{eq:commutativity}
\end{align}
We have thus far described these operators on
\(\mathcal{J}/\mathcal{N}\).
Because these operators can have unbounded norm, it remains a nontrivial
question whether they have extensions to domains in \(\mathcal{H}_\emptyset\)
on which they satisfy \(Z(J)^\dagger=Z(J^*)\) and \cref{eq:commutativity}. The answer
can be model-dependent.

We will simply proceed as in refs.~\cite{Marolf:2020xie,Marolf:2024jze},
assuming that real\footnote{In the sense that
  \(f(\bullet)^\dagger=f(\bullet^\dagger)\).} bounded functions of
\begin{align}
  \Re Z(J)
  &= \frac{Z(J) + Z(J^*)}{2}
    \;,
  &
    \Im Z(J)
  &= \frac{Z(J) - Z(J^*)}{2 i}
\end{align}
define operators with unique self-adjoint extensions to
\(\mathcal{H}_\emptyset\) which commute with each other and between all
\(J\in\mathcal{J}\). We can therefore simultaneously diagonalize \(Z(J)\) for
all \(J\in\mathcal{J}\). Denoting the joint spectrum of all \(Z(J)\) by
\(\mathcal{A}\), the baby universe Hilbert space \(\mathcal{H}_\emptyset\)
decomposes into a direct sum or integral,\footnote{Taking the Cauchy completion
  is implicitly included in our definition for direct sums and integrals.}
\begin{align}
  \mathcal{H}_\emptyset
  &= \int_{\mathcal{A}}^\oplus\dd{\alpha} \mathcal{H}_\emptyset^\alpha
    \;,
    \label{eq:babydecomp}
\end{align}
into superselection sectors \(\mathcal{H}_\emptyset^\alpha\), or
``\(\alpha\)-sectors'' for brevity, in which the operators \(Z(J)\) take
definite eigenvalues \(Z_\alpha(J)\). In fact, each baby universe
\(\alpha\)-sector \(\mathcal{H}_\emptyset^\alpha\) is one dimensional. In
particular, it is spanned by a single state \(\ket{\alpha}\) --- a so-called
``\(\alpha\)-state'' \cite{Coleman:1988cy,Giddings:1988cx} --- uniquely determined
(up to scalar multiplication) by the values of the inner product
\begin{align}
  \braket{\alpha}{J}
  &= \mel{\alpha}{Z(J)}{\varnothing}
    = Z_\alpha(J) \braket{\alpha}{\varnothing}
    \label{eq:jalpha}
\end{align}
with states \(\ket{J}\) spanning the dense subspace
\(\mathcal{J}/\mathcal{N}\subset\mathcal{H}_\emptyset\). In particular, the
linear combination
\(\braket{\varnothing}{\alpha}'\ket{\alpha}-\braket{\varnothing}{\alpha}\ket{\alpha}'\)
of any two \(\ket{\alpha}, \ket{\alpha}'\in\mathcal{H}_\emptyset^\alpha\) has
vanishing inner product with all \(\ket{J}\) by \cref{eq:jalpha} and must
therefore be identified with \(0\).

Since \(\alpha\)-states form an orthonormal basis spanning
\(\mathcal{H}_\emptyset\), we can express the path integral \(\zeta\) as an
ensemble-average over \(\alpha\)-sectors,
\begin{align}
  \zeta(J)
  &= \int_{\mathcal{A}} \dd{\alpha}
    \abs{\braket{\alpha}{\varnothing}}^2\,
    \zeta_\alpha(J)
    \;,
    \label{eq:zetazetaalpha}
\end{align}
where
\begin{align}
  \zeta_\alpha(J)
  &\coloneqq \frac{\mel{\alpha}{Z(J)}{\alpha}}{\braket{\alpha}{\alpha}}
    = Z_\alpha(J)
    \;,
    \label{eq:alphapathintegral}
\end{align}
and we have chosen to normalize \(\ket{\alpha}\) such that
\(\braket{\alpha'}{\alpha}\) is the \(\delta\)-function on \(\mathcal{A}\) with
measure \(\dd{\alpha}\).\footnote{Note that \(\braket{\alpha}{\alpha}\) can be
  infinite. Nonetheless, for brevity, we will loosely say that \(\ket{\alpha}\)
  is a ``state'' and write ``\(\ket{\alpha}\in\mathcal{H}_\emptyset^\alpha\)'',
  as we have done in the previous paragraph. In \cref{sec:cutsectors}, the need
  will arise to consider inner products within nontrivial \(\alpha\)-sectors
  \(\mathcal{H}_C^\alpha\), in which case we will incorporate a rescaling
  \labelcref{eq:rescaledstate}.\label{foot:nonnormalizable}} As explained in
refs.~\cite{Marolf:2020xie,Marolf:2024jze}, each \(\zeta_\alpha(J)\) is itself a
path integral in the sense that it is equal to a limit of the original path
integral \(\zeta(J\sqcup J^\alpha_\nu)\) with the insertion of extra sources
\(J^\alpha_\nu\in\mathcal{J}\) whose corresponding operators converge
to\footnote{See \eg{} refs.~\cite{Blommaert:2021fob,Blommaert:2022ucs} for
  constructions of \(\zeta_\alpha(J)\) in JT gravity with branes providing the
  extra insertions \(J^\alpha_\nu\).\label{foot:alphabrane}}
\begin{align}
  \lim_\nu Z(J^\alpha_\nu)
  &= \frac{
    \ketbra{\alpha}{\alpha}
    }{
    \abs{\braket{\alpha}{\varnothing}}^2
    }
    \;.
\end{align}
In contrast to \cref{eq:nofactor}, these path integrals over individual
\(\alpha\)-sectors do factorize:
\begin{align}
  Z_\alpha(J \sqcup J')
  &= Z_\alpha(J)\, Z_\alpha(J')
    \;.
    \label{eq:factorization}
\end{align}

Let us comment briefly on, \eg{} holographic, contexts in which the sources
\(J\in\mathcal{J}\) specify objects, \eg{} boundary manifolds, which host
theories, \eg{} CFTs, dual to the bulk gravitational theory. The partition
function of a local dual theory must factorize as in \cref{eq:factorization}
where \(J\) and \(J'\) specify mutually disconnected objects. The gravitational
bulk dual of a single dual theory must correspond to an individual
\(\alpha\)-sector, with a factorizing path integral \(\zeta_\alpha\). The
generically\footnote{Of course, if the ensemble is trivial with a unique
  \(\alpha\)-sector, then \(\zeta\propto\zeta_\alpha\) again corresponds to a
  single dual theory. See \cref{foot:uniquealpha}.} non-factorizing
gravitational path integral \(\zeta\) then corresponds to an ensemble average of
the partition functions of dual theories.

%% file: sections/section03_partialSourceSectors.tex
\section{Partial sources and the Hilbert spaces of states they prepare}
\label{sec:partialsourcesectors}

In \cref{sec:gpibaby}, we described the set \(\mathcal{J}\) of sources which
serve as inputs to the path integral \(\zeta:\mathcal{J}\to\mathbb{C}\). Each
such source \(J\in\mathcal{J}\) is completely specified in the sense that it
contains the full input needed for the path integral to output a complex number
\(\zeta(J)\). A goal of this section is to describe objects which we will call
partial sources. A partial source \(J_C\) is characterized by the property that
it produces a complete source, written as \((J_C'|J_C)\in\mathcal{J}\), once it
is glued across what we will call a cut \(C\) with another, compatible partial
source \((J_C')^*\).

The outline of this section is much like that of \cref{sec:gpibaby}. In
\cref{sec:expartialsources}, we will first provide some possible examples of
partial sources. The reader may find it helpful to bear these in mind as we
proceed with a more abstract discussion in \cref{sec:generalpartialsources}.
There, we describe the general defining properties we demand of the set
\(\mathcal{C}\) of cuts \(C\) and the set \(\mathcal{J}_C\) of partial sources
\(J_C\) associated with each cut \(C\). Finally, in \cref{sec:cutsectors}, we
will use \(\mathcal{J}_C\) to construct a Hilbert space \(\mathcal{H}_C\) of
states prepared by partial sources \(J_C\in\mathcal{J}_C\). Notably, the
\(\alpha\)-sectors \(\mathcal{H}_C^\alpha\) of \(\mathcal{H}_C\) need not be one-dimensional.

\subsection{Possible examples of cuts \(C\in\mathcal{C}\) and partial sources
  \(J_C\in\mathcal{J}_C\)}
\label{sec:expartialsources}

As with \(\mathcal{J}\), the set \(\mathcal{C}\) of cuts \(C\) and the set
\(\mathcal{J}_C\) of partial sources \(J_C\) will be theory-dependent. We will
now give some illustrative examples of these objects, which may or may not be in
a given theory.

\begin{figure}
  \centering
  \begin{subfigure}[t]{0.3\textwidth}
    \centering
    \includegraphics[scale=\figscale]{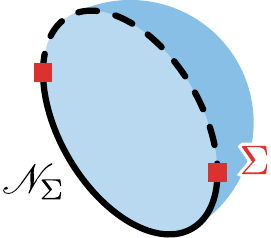}
    \caption{A partial boundary manifold \(J_\Sigma=\mathscr{N}_{\Sigma}\) with boundary \(C=\Sigma\).}
    \label{fig:psourcebdycond}
  \end{subfigure}
  \hfill
  \begin{subfigure}[t]{0.3\textwidth}
    \centering
    \includegraphics[scale=\figscale]{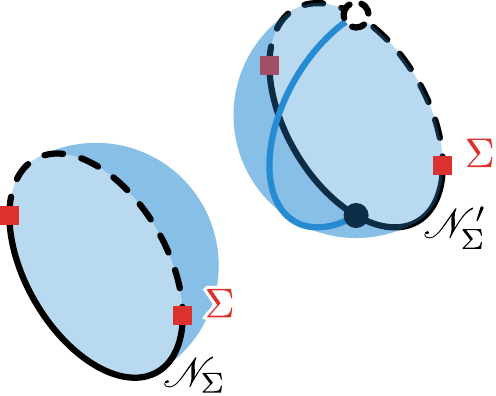}
    \caption{A partial boundary manifold
      \(J_C^{(\subref{fig:psourceotimes1})}=\mathscr{N}_\Sigma\otimes\mathscr{N}_\Sigma'\)
      with boundary \(C=\Sigma\otimes\Sigma\). (The superscript on
      \(J_C^{(\subref{fig:psourceotimes1})}\) is introduced for disambiguation
      with \cref{fig:psourceotimes2}.)}
    \label{fig:psourceotimes1}
  \end{subfigure}
  \hfill
  \begin{subfigure}[t]{0.3\textwidth}
    \centering
    \includegraphics[scale=\figscale]{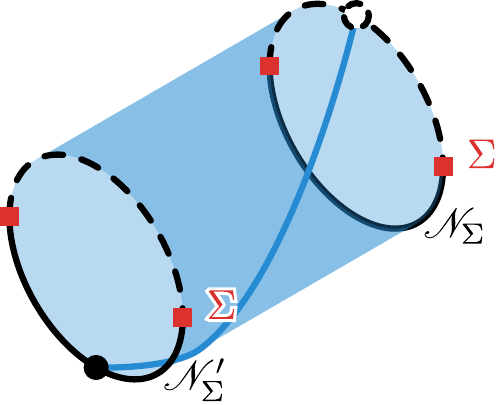}
    \caption{A partial boundary manifold
      \(J_C^{(\subref{fig:psourceotimes2})}=\mathscr{N}_\Sigma'\otimes\mathscr{N}_\Sigma\)
      with boundary \(C=\Sigma\otimes\Sigma\). The cut \(C\) here is the same
      as in \cref{fig:psourceotimes1}. However,
      \(\mathscr{N}_\Sigma\ne\mathscr{N}_\Sigma'\) so
      \(J_C^{(\subref{fig:psourceotimes1})}\ne{}J_C^{(\subref{fig:psourceotimes2})}\)
      and
      \((J_C'|J_C^{(\subref{fig:psourceotimes1})})\ne{}(J_C'|J_C^{(\subref{fig:psourceotimes2})})\)
      for the same \(J_C'\) as in \cref{fig:psourceotimes1}.}
    \label{fig:psourceotimes2}
  \end{subfigure}
  \par\bigskip
  \begin{subfigure}[t]{0.3\textwidth}
    \centering
    \includegraphics[scale=\figscale]{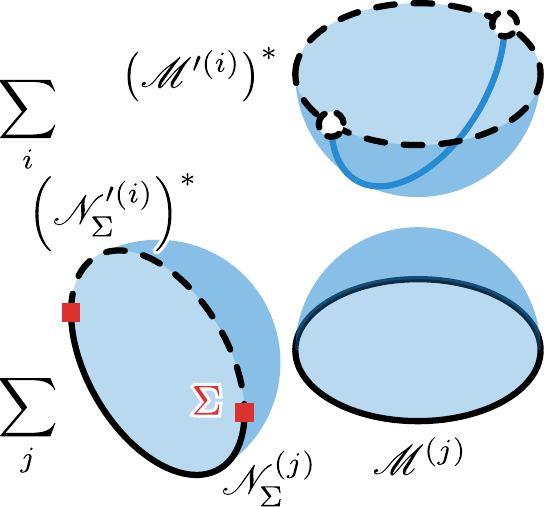}
    \caption{A formal linear combination \(J_\Sigma=\sum_{j=1}^d \mathscr{M}^{(j)}
      \sqcup \mathscr{N}_\Sigma^{(j)}\) of partial boundary manifolds where each
      \(\mathscr{N}_\Sigma^{(j)}\) has boundary \(C=\Sigma\).}
    \label{fig:psourcelincomb}
  \end{subfigure}
  \hfill
  \begin{subfigure}[t]{0.3\textwidth}
    \centering
    \includegraphics[scale=\figscale]{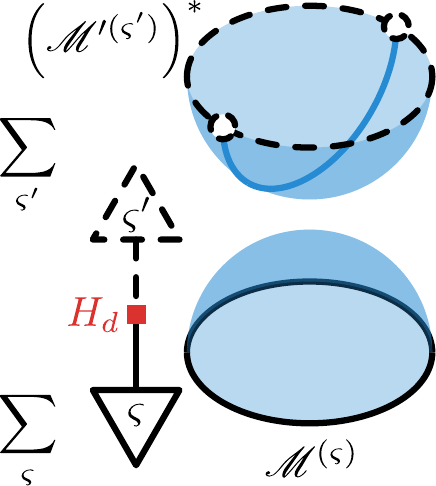}
    \caption{A formal linear combination \(J_d=\sum_{\varsigma=1}^d
      \mathscr{M}^{(\varsigma)} \ket{\varsigma}\) built from states
      \(\ket{\varsigma}\), spanning a nongravitational Hilbert space \(C=H_d\),
      and boundary manifolds \(\mathscr{M}^{(\varsigma)}\in\mathcal{J}\), playing
      the role of scalars.}
    \label{fig:psourcehilb}
  \end{subfigure}
  \hfill
  \begin{subfigure}[t]{0.3\textwidth}
    \centering
    \includegraphics[scale=\figscale]{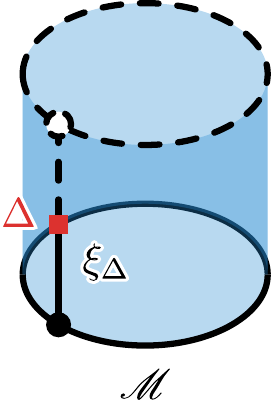}
    \caption{A partial source \(J_\Delta = \xi_{\Delta} \cdot \mathscr{M}\),
      obtained from cutting the geodesic in \cref{fig:sourcegeodesic}. The value
      of \(\Delta\) in this example is specified by the cut \(C\).}
    \label{fig:psourcegeodesic}
  \end{subfigure}
  \caption{Possible examples of cuts \(C\in\mathcal{C}\) (red) and partial
    sources \(J_C\in\mathcal{J}_C\) (solid black). In each panel, the complete
    source \((J_C'|J_C)\in\mathcal{J}\) obtained from gluing with another
    partial source \((J_C')^*\) (dashed black) is also shown. A possible bulk
    configuration that might appear in the path integral \(\zeta(J_C'|J_C)\) is
    illustrated in blue.}
  \label{fig:psources}
\end{figure}

The perhaps most straightforward example of a partial source is the
specification of boundary conditions for bulk fields on a part
\(J_C=\mathscr{N}_\Sigma\) of a boundary manifold. This is illustrated in
\cref{fig:psourcebdycond}. The partial boundary manifold \(\mathscr{N}_\Sigma\) is a
manifold itself with a boundary \(C=\Sigma\). (We will use \(\mathscr{N}_\Sigma\) and
\(\Sigma\) to refer to both the partial boundary manifold and its boundary, as
well as data specified there.)

If we have two such partial sources \(\mathscr{N}_\Sigma\) and \(\mathscr{N}_\Sigma'\)
associated with the same cut \(\Sigma\), then we can form a complete source
\((\mathscr{N}_\Sigma'|\mathscr{N}_\Sigma)\in\mathcal{J}\) as follows. Firstly, we apply a
complex conjugation and time- or orientation-reversal
transformation \(\mathscr{N}_\Sigma'\mapsto(\mathscr{N}_\Sigma')^*\).\footnote{In particular, if
  \(\Sigma\) has an orientation, then the image \(\Sigma^*\) has the opposite
  orientation.} We then glue each point \(p\in\Sigma=\partial\mathscr{N}\) to
its image \(p^*\) on the boundary \(\Sigma^*=\partial(\mathscr{N}_\Sigma')^*\) of
the partial boundary manifold \((\mathscr{N}_\Sigma')^*\), thus producing a complete
source \((\mathscr{N}_\Sigma'|\mathscr{N}_\Sigma)\in\mathcal{J}\) specifying boundary
conditions on a boundary manifold itself without boundary. Let us emphasize that
the points in \(\Sigma\) are marked by unique labels, such that each point
\(p\in\Sigma\) corresponds and is glued to a \emph{unique} point
\(p^*\in\Sigma^*\). The gluing is therefore unique, even if \(\Sigma\) possesses
isometries.

In general, a cut \(C\) provides specifications to ensure that two partial
sources \(J_C,J_C' \in\mathcal{J}_C\) associated with the same cut can be joined
to produce a source \((J_C|J_C')\in\mathcal{J}\). In the current example, if
\(\mathscr{N}_\Sigma,\mathscr{N}_\Sigma'\in\mathcal{J}_\Sigma\) specify local
background fields (conventionally called sources of the boundary theory in
(A)dS/CFT) which parametrize boundary conditions for bulk fields, then \(C=\Sigma\)
might specify the values these fields on the cut, to ensure continuity in
\((\mathscr{N}_\Sigma|\mathscr{N}_\Sigma')\in\mathcal{J}\). Higher
derivatives would also be needed to meet stronger smoothness requirements that
might be required for sources in \(\mathcal{J}\).

Let us comment briefly on the holographic interpretation of the above
construction, where we suppose the dual boundary theories are non-gravitational
QFTs. In each QFT, the overlap of two states, prepared by QFT path integrals on
the backgrounds \(\mathscr{N}_\Sigma\) and \(\mathscr{N}_\Sigma'\), is computed
by a partition function over the background
\((\mathscr{N}_\Sigma|\mathscr{N}_\Sigma')\). Ensuring that
\((\mathscr{N}_\Sigma)^*\) and \(\mathscr{N}_\Sigma'\) can be glued together is
equivalent to ensuring that the prepared states belong to the same Hilbert
space. Thus, \(C=\Sigma\) in this holographic context can be interpreted as
specifying the QFT Hilbert space for each boundary theory (in the ensemble),
which can depend on the value of background fields, \eg{} the metric, near the
cut as described in the previous paragraph.

As for boundary manifolds, a partial boundary manifold need not be connected.
Given a source \(\mathscr{M}\in\mathcal{J}\) specifying boundary conditions on a
boundary manifold \(\mathscr{M}\) and a partial source
\(\mathscr{N}_\Sigma\in\mathcal{J}_\Sigma\) specifying boundary conditions on a partial
boundary manifold \(\mathscr{N}_\Sigma\), we can also consider a partial source
\begin{align}
  \mathscr{M} \sqcup \mathscr{N}_\Sigma
  \in \mathcal{J}_\Sigma
  \;,
  \label{eq:mcupn}
\end{align}
specifying the boundary conditions on the disjoint union of \(\mathscr{M}\) and
\(\mathscr{N}_\Sigma\). Given another partial source \(\mathscr{N}_{\Sigma'}\in\mathcal{J}_{\Sigma'}\), we can also consider a partial source
\begin{align}
  \mathscr{N}_\Sigma \otimes \mathscr{N}_{\Sigma'}
  \in \mathcal{J}_{\Sigma\otimes\Sigma'}
  \;,
\end{align}
with a cut \(\Sigma\otimes\Sigma'\). Here, \(\otimes\) can be thought of as
disjoint union with the understanding that points in \(\Sigma\otimes\Sigma'\)
are marked by unique labels. Even if \(\Sigma=\Sigma'\) so that
\(\Sigma\otimes\Sigma'=\Sigma'\otimes\Sigma\), we still have
\(\mathscr{N}_\Sigma\otimes\mathscr{N}_\Sigma'\ne\mathscr{N}_\Sigma'\otimes\mathscr{N}_\Sigma\)
for \(\mathscr{N}_\Sigma\ne\mathscr{N}_\Sigma'\) in \(\mathcal{J}_\Sigma\) ---
see \cref{fig:psourceotimes1,fig:psourceotimes2}. The choice of symbol
\(\otimes\) is purposely evocative of a tensor product, as we will explain in
our abstract discussion of partial sources in \cref{sec:generalpartialsources}.
In particular, the gluing of partial sources described above is reminiscent of
tensor contraction, which we shall denote like
\begin{align}
  (\mathscr{N}_\Sigma|\mathscr{N}_\Sigma')
  &= (\mathscr{N}_{\tikzmarknode{C571}{\Sigma}})^*
    \otimes
    \mathscr{N}_{\tikzmarknode{C572}{\Sigma}}'
    \in \mathcal{J}
    \;.
    \begin{tikzpicture}[overlay, remember picture, every path/.style = {shorten <=1pt,shorten >=1pt}]
      \draw[rounded corners] (C571.south) -- ++ (0,-0.2) -| (C572);
    \end{tikzpicture}
\end{align}

As for boundary manifolds, we can also consider superpositions of partial
boundary manifolds. To prepare a convenient segue in our discussion, let us
consider superpositions of \(d\)-many partial boundary manifolds of the kind
\labelcref{eq:mcupn}:
\begin{align}
  J_\Sigma
  &=
    \sum_{j=1}^d \mathscr{M}^{(j)} \sqcup \mathscr{N}_\Sigma^{(j)}
  \in \mathcal{J}_\Sigma
  \;,
  &
    J_\Sigma'
    &=
    \sum_{i=1}^d \mathscr{M}^{\prime(i)} \sqcup \mathscr{N}_\Sigma^{\prime(i)}
    \in \mathcal{J}_\Sigma
         \;.
\end{align}
See \cref{fig:psourcelincomb}. Again, we can define \((\bullet|\bullet)\) to
mean taking the \(*\)-conjugation of the first argument and gluing to the second
argument, giving a Hermitian form
\begin{align}
  (J_\Sigma' | J_\Sigma)
  &= \sum_{i,j=1}^d
    (\mathscr{M}^{\prime(i)})^* \sqcup \mathscr{M}^{(j)} \sqcup
    (\mathscr{N}_\Sigma^{\prime(i)}|\mathscr{N}_\Sigma^{(j)})
    \in \mathcal{J}
    \;.
    \label{eq:jsigmasuminnerprod}
\end{align}

The above formulas suggest a natural generalization to other possible
types of partial sources that the theory might allow. In particular, let the cut
\(C=H_d\) be any \(d\)-dimensional Hilbert space of a non-gravitational system,
spanned by states \(\ket{\varsigma}\). (From here on, when it appears in a subscript, we
will abreviate \(H_d\) as simply \(d\).) We can then consider the set of formal
linear combinations
\begin{align}
  \mathcal{J}_d
  =\left\{
  \sum_{\varsigma=1}^d J^{(\varsigma)} \ket{\varsigma}
  : J^{(\varsigma)} \in \mathcal{J}
  \right\}
  \;.
  \label{eq:jd}
\end{align}
Some examples of these partial sources \(J_d\in\mathcal{J}_d\) are illustrated
in \cref{fig:psourcehilb}, where each \(J^{(\varsigma)}\) is a boundary manifold
\(\mathscr{M}^{(\varsigma)}\). With the obvious replacements in
\cref{eq:jsigmasuminnerprod}, for any two \(J_d, J_d'\in\mathcal{J}_d\), we can
define the Hermitian form
\begin{align}
  (J_d' | J_d)
  &= \sum_{\varsigma,\varsigma'=1}^d
    (J^{\prime(\varsigma')})^* \sqcup J^{(\varsigma)}
    \braket{\varsigma'}{\varsigma}
    \in \mathcal{J}
    \;.
    \label{eq:jdhermform}
\end{align}
This can again be viewed as taking a \(*\)-conjugation of the first argument and
gluing its cut to the cut of second argument, if we define
\begin{align}
  \ket{\varsigma}^*
  &= \bra{\varsigma}
\end{align}
and the gluing to mean\footnote{Strictly speaking, the RHS should be
  \(\braket{i}{j} \varnothing \in\mathcal{J}\).}
\begin{align}
  \bra{\varsigma'}\otimes\ket{\varsigma}
  &\mapsto
    \tikzmarknode{C891}{\bra{\varsigma'}}
    \otimes\tikzmarknode{C892}{\ket{\varsigma}}
    =
    \braket{\varsigma'}{\varsigma}
    \;.
    \label{eq:jdgluing}
    \begin{tikzpicture}[overlay, remember picture, every path/.style = {shorten <=1pt,shorten >=1pt}]
      \draw[rounded corners] (C891.south) -- ++ (0,-0.2) -| (C892);
    \end{tikzpicture}
\end{align}
For simplicity, when we continue our discussion of this example in
\cref{sec:obsclone}, we will choose the states \(\ket{\varsigma}\) to be orthonormal:
\begin{align}
  \braket{\varsigma'}{\varsigma}
  &= \delta_{\varsigma' \varsigma}
    \;.
    \label{eq:orthonormal}
\end{align}

More examples of partial sources might arise from cutting geometric objects
other than boundary manifolds specified by complete sources found in
\(\mathcal{J}\). For example, as illustrated in \cref{fig:psourcegeodesic}, if
\(\mathcal{J}\) contains sources of the form
\((\mathscr{M}')^*\cdot\gamma\cdot\mathscr{M}\) illustrated in
\cref{fig:sourcegeodesic}, we can cut a bulk worldline \(\gamma\) to obtain a
partial source, which we write as \(\xi\cdot\mathcal{M}\), with a partial
worldline \(\xi\). By definition, a partial worldline \(\xi\) with the same cut
\(C\) can be glued to its \(*\)-conjugate to give a full worldline \(\gamma\)
that would appear in a complete source, say
\begin{align}
  (\xi\cdot\mathscr{M}'|\xi\cdot\mathscr{M})
  =(\mathscr{M}')^*\cdot\gamma\cdot\mathscr{M}\in\mathcal{J}
  \;.
  \label{eq:wlhermitian}
\end{align}
In the example with a geodesic \(\gamma^{(\Delta)}\) whose length \(\ell\) is to be
integrated over with weight \(e^{-\Delta\,\ell}\), the cut \(C\) would in
particular specify the value of \(\Delta\) and perhaps the flavour \(\sigma\) of
the worldline if there exist multiple flavours. When \(C\) appears in a
subscript in this case, we will write it simply as \(\Delta\) or \(\sigma\) as
appropriate --- \eg{},
we might use \(\mathcal{J}_\Delta\) or \(\mathcal{J}_\sigma\) to denote the set of
partial sources with partial worldlines terminating at a cut corresponding to a
value of \(\Delta\) or a worldline flavour \(\sigma\).

\subsection{The monoid \(\mathcal{C}\) of
  cuts \(C\) and the \(\mathcal{J}\)-modules \(\mathcal{J}_C\) of partial sources}
\label{sec:generalpartialsources}

We will now proceed with an abstract discussion of the properties we require of
\(\mathcal{C}\) and \(\mathcal{J}_C\). These properties can be taken to be our
general definition for a set \(\mathcal{C}\) of cuts \(C\) and the set
\(\mathcal{J}_C\) of partial sources associated with a cut \(C\), irrespective
of how they are concretely realized in the examples discussed in
\cref{sec:expartialsources}. The reader may nonetheless find it helpful to bear
those examples in mind to conceptually motivate the abstract structures we will
discuss.

At the outset, \(\mathcal{C}\) is just a set of elements, each of which we call
a cut. For each cut \(C\in\mathcal{C}\), \(\mathcal{J}_C\) is itself a set of
elements, each of which we will call a partial source associated with the cut
\(C\). Below, we will describe in turn each additional
structure that we require \(\mathcal{J}_C\) and \(\mathcal{C}\) to have.

Firstly, we require each \(\mathcal{J}_C\) to be a \(\mathcal{J}\)-module. That is,
\(\mathcal{J}_C\) is equipped with addition \(+\) and scalar
multiplication \(\sqcup\) operations, where elements of the ring \(\mathcal{J}\)
play the role of scalars. For all \(J,J'\in\mathcal{J}\) and
\(J_C,J_C'\in\mathcal{J}_C\), these operations satisfy
\begin{align}
  J\sqcup(J_C + J_C')
  &= J\sqcup J_C + J\sqcup J_C'
    \;,
  \\
    (J+J')\sqcup J_C
  &= J\sqcup J_C + J'\sqcup J_C
    \;,
  \\
    (J\sqcup J') \sqcup J_C
  &= J \sqcup (J'\sqcup J_C)
    \;,
  \\
  \varnothing \sqcup J_C
  &= J_C
    \;.
\end{align}
(Because the ring \(\mathcal{J}\) is commutative, left \(\mathcal{J}\)-modules
are the same as right \(\mathcal{J}\)-modules.) Again, by an abuse of notation,
we will denote the additive identity in \(\mathcal{J}_C\) with the same symbol
\(0\) as in \(\mathbb{C}\).

Next, we require \(\mathcal{C}\) to have an associative (but not necessarily
commutative) multiplication operation, which we denote as
\begin{align}
  \otimes : \mathcal{C} \times \mathcal{C}
  &\to \mathcal{C}
    \;,
    &
      C, C'
    &\mapsto C\otimes C'
      \;.
      \label{eq:otimescuts}
\end{align}
This gives \(\mathcal{C}\) the structure of a semigroup. We will use the same
symbol \(\otimes\) to denote the tensor product on vector spaces or, more
generally, modules. In particular,
\begin{align}
  \otimes:
  \mathcal{J}_C \times \mathcal{J}_{C'}
  &\to
    \mathcal{J}_{C} \otimes \mathcal{J}_{C'}
  \;,
  &
    J_C, J_{C'}
    &\mapsto J_C \otimes J_{C'}
    \;,
\end{align}
is the tensor product on the \(\mathcal{J}\)-modules
\(\mathcal{J}_C,\mathcal{J}_{C'}\) of partial sources. The relation to the
multiplication \labelcref{eq:otimescuts} of cuts is that we require
\(\mathcal{J}_{C} \otimes \mathcal{J}_{C'}\) to be a submodule of
\(\mathcal{J}_{C\otimes C'}\). It may be the case that
\(\mathcal{J}_{C\otimes{}C'}\) contains more partial sources, \eg{} connecting
the two pieces \(C\) and \(C'\) of the cut \(C\otimes C'\), so
\(\mathcal{J}_{C}\otimes\mathcal{J}_{C'}\) might be a proper submodule.

We require \(\mathcal{C}\) to have a (left and right) \(\otimes\)-multiplicative
identity element \(\emptyset\in\mathcal{C}\). Thus, \(\mathcal{C}\) is a
monoid\footnote{A monoid is a semigroup with a left and right identity.}. For
this trivial cut \(\emptyset\),
we set
\begin{align}
  \mathcal{J}_\emptyset = \mathcal{J}
  \;,
  \label{eq:trivialpsources}
\end{align}
with \(\mathcal{J}\) considered as a module over itself.
Thus, complete sources are partial sources associated with the trivial cut
\(\emptyset\). For all \(J\in\mathcal{J}\), \(C\in\mathcal{C}\), and
\(J_C\in\mathcal{J}_C\), we in particular have
\begin{subequations}
\begin{align}
  \mathcal{J}_C \otimes \mathcal{J}_\emptyset
  &= \mathcal{J}_\emptyset \otimes \mathcal{J}_C
  = \mathcal{J}_C
  \label{eq:otimestrivial1}
  \;,
  \\
  J_C \otimes J
    &= J \otimes J_C
      = J\sqcup J_C
      \;.
  \label{eq:otimestrivial2}
\end{align}
\label{eq:otimestrivial}
\end{subequations}

The next structure we require \(\mathcal{C}\) to have is an involution and automorphism \(*\),
\ie{},
\begin{align}
  (C^*)^*
  &= C \;,
  &
    (C\otimes C')^*
  &=  C^* \otimes (C')^*
    \;,
  &
    \emptyset^*
    &= \emptyset
\end{align}
for all \(C,C'\in\mathcal{C}\). We will
correspondingly require \(\mathcal{J}_C\) and \(\mathcal{J}_{C^*}\) to be
related by an involution and homomorphism \(*\), \ie{}
\begin{align}
  (J_C^*)^*
  &= J_C
    \;,
    &
  (J\sqcup J_C + J_C')^*
  &= J^* \sqcup J_C^* +(J_C')^*
    \in \mathcal{J_{C^*}}
    \;,
    &
      (J_C\otimes J_{C'})^*
  &= J_C^* \otimes J_{C'}^*
    \;.
    \label{eq:jast}
\end{align}

To glue together cuts of partial sources, we require a linear gluing operation
which is analogous to tensor contraction,
\begin{align}
  \cup: \mathcal{J}_{C_1\otimes C \otimes C_2 \otimes C^* \otimes C_3}
  &\to \mathcal{J}_{C_1\otimes C_2 \otimes C_3}
    \;,
  &
    J_{C_1\otimes C \otimes C_2 \otimes C^* \otimes C_3}
    \mapsto J_{C_1\otimes \tikzmarknode{C01}{C} \otimes C_2 \otimes \tikzmarknode{C02}{C}^* \otimes C_3}
    \;.
  \label{eq:gluing}
    \begin{tikzpicture}[overlay, remember picture, every path/.style = {shorten <=1pt,shorten >=1pt}]
      \draw[rounded corners] (C01.south) -- ++ (0,-0.2) -| (C02);
    \end{tikzpicture}
\end{align}
Naturally, we require gluing operations to commute with each other, with the
\(*\)-involution of the partial source, and with taking the tensor product with
another partial source. Additionally, we require gluing to distribute over
products of cuts, meaning a gluing of \(C=C'\otimes{}C''\) to
\(C^*=(C')^*\otimes{}(C'')^*\) is equal to the gluing of \(C'\) to \((C')^*\)
and of \(C''\) to \((C'')^*\). Finally, we require any partial source whose cut
has all its components glued to other cuts to commute in tensor products with
other partial sources.

It will also be convenient for our later analysis in
\cref{sec:operatorsdimbounds} to introduce a swap operation, which is analogous
to permuting tensor indices,
\begin{align}
  \swap
  : \mathcal{J}_{C_1\otimes C \otimes C_2 \otimes C' \otimes C_3}
  &\to \mathcal{J}_{C_1\otimes C' \otimes C_2 \otimes C \otimes C_3}
    \;,
  &
    J_{C_1\otimes C \otimes C_2 \otimes C' \otimes C_3}
    \mapsto J_{C_1\otimes \tikzmarknode{C11}{C} \otimes C_2 \otimes \tikzmarknode{C12}{C}' \otimes C_3}
    \;.
  \begin{tikzpicture}[overlay, remember picture, every path/.style = {shorten <=1pt,shorten >=1pt}]
    \draw[rounded corners] (C12.south) -- ++ (0,-0.2) -| ([shift=({0,-0.4})]C11.south);
    \draw[rounded corners, white, line width=3] %
    (C11.south) -- ++ (0,-0.2) -| ([shift=({0,-0.4})]C12.south);%
    \draw[rounded corners] (C11.south) -- ++ (0,-0.2) -| ([shift=({0,-0.4})]C12.south);
  \end{tikzpicture}
\end{align}
Like the gluing operation, we require the swap to commute with the
\(*\)-involution of the partial source, commute with taking the tensor product
with another partial source, and distribute over products of cuts. The rules we
require for composing swaps with each other and with gluing are exactly those
intuited from the ``circuit diagram'' built by successively attaching pieces of ``wire'' \(\cup\)
and \(\swap\), \eg{}
\begin{align}
  \vcenter{\hbox{
  \begin{tikzpicture}[every path/.style = {rounded corners}]
    \node (C071) at (0,0) {}; %
    \node (C072) at (1,0) {}; %
    \node (C073) at (2,0) {}; %
    \draw[shorten <=1pt, shorten >=1pt, solarizedGreen] %
    (C072.center) -- ++ (0,-0.2) -| ([shift=({0,-0.4})]C071.center); %
    \draw[shorten <=1pt, white, line width=3] %
    (C071.center) -- ++ (0,-0.2) -| ([shift=({0,-0.4})]C072.center);%
    \draw[shorten <=1pt, solarizedBlue] (C071.center) -- ++ (0,-0.2) -| ([shift=({0,-0.4})]C072.center);
    \draw[shorten <=1pt, shorten >=1pt, solarizedRed] %
    (C073.center) -- ++ (0,-0.6) -| ([shift=({0,-0.8})]C072.center); %
    \draw[shorten >=1pt, white, line width=3] %
    ([shift=({0,-0.4})]C072.center) -- ++ (0,-0.2) -| ([shift=({0,-0.8})]C073.center); %
    \draw[shorten >=1pt, solarizedBlue] %
    ([shift=({0,-0.4})]C072.center) -- ++ (0,-0.2) -| ([shift=({0,-0.8})]C073.center); %
  \end{tikzpicture}
  }}
  &=
    \vcenter{\hbox{
    \begin{tikzpicture}[every path/.style = {rounded corners}]
      \node (C081) at (0,0) {}; %
      \node (C082) at (1,0) {}; %
      \node (C083) at (2,0) {}; %
      \draw[shorten <=1pt, solarizedRed] %
      (C083.center) -- ++ (0,-0.2) -| ([shift=({0,-0.4})]C081.center); %
      \draw[shorten <=1pt, white, line width=3] %
      (C081.center) -- ++ (0,-0.2) -| ([shift=({0,-0.4})]C083.center);%
      \draw[shorten <=1pt, solarizedBlue] (C081.center) -- ++ (0,-0.2) -| ([shift=({0,-0.4})]C083.center);
      \draw[shorten <=1pt, shorten >=1pt, white, line width=3] %
      (C082.center) -- ++ (0,-0.6) -| ([shift=({0,-0.8})]C081.center); %
      \draw[shorten <=1pt, shorten >=1pt, solarizedGreen] %
      (C082.center) -- ++ (0,-0.6) -| ([shift=({0,-0.8})]C081.center); %
      \draw[shorten >=1pt, white, line width=3] %
      ([shift=({0,-0.4})]C081.center) -- ++ (0,-0.2) -| ([shift=({0,-0.8})]C082.center); %
      \draw[shorten >=1pt, solarizedRed] %
      ([shift=({0,-0.4})]C081.center) -- ++ (0,-0.2) -| ([shift=({0,-0.8})]C082.center); %
    \end{tikzpicture}
    }}
    \;,
  &
    \vcenter{\hbox{
    \begin{tikzpicture}[every path/.style = {rounded corners}]
      \node (C091) at (0,0) {}; %
      \node (C092) at (1,0) {}; %
      \node (C093) at (2,0) {}; %
      \node (C094) at (3,0) {}; %
      \draw[shorten <=1pt, solarizedRed] %
      (C092.center) -- ++ (0,-0.2) -| ([shift=({0,-0.4})]C091.center); %
      \draw[shorten <=1pt, white, line width=3] %
      (C091.center) -- ++ (0,-0.2) -| ([shift=({0,-0.4})]C092.center);%
      \draw[shorten <=1pt, solarizedBlue] (C091.center) -- ++ (0,-0.2) -| ([shift=({0,-0.4})]C092.center); %
      \draw[shorten <=1pt, solarizedRed] %
      (C094.center) -- ++ (0,-0.2) -| ([shift=({0,-0.4})]C093.center); %
      \draw[shorten <=1pt, white, line width=3] %
      (C093.center) -- ++ (0,-0.2) -| ([shift=({0,-0.4})]C094.center);%
      \draw[shorten <=1pt, solarizedBlue] (C093.center) -- ++ (0,-0.2) -| ([shift=({0,-0.4})]C094.center); %
      \draw[rounded corners, solarizedRed] %
      ([shift=({0,-0.4})]C091.center) -- ++ (0,-0.2) -| ([shift=({0,-0.4})]C093.center); %
      \draw[rounded corners, white, line width=3] %
      ([shift=({0,-0.4})]C092.center) -- ++ (0,-0.4) -| ([shift=({0,-0.4})]C094.center); %
      \draw[rounded corners, solarizedBlue] %
      ([shift=({0,-0.4})]C092.center) -- ++ (0,-0.4) -| ([shift=({0,-0.4})]C094.center); %
    \end{tikzpicture}
    }}
  &=
    \vcenter{\hbox{
    \begin{tikzpicture}[every path/.style = {rounded corners}]
      \node (C091) at (0,0) {}; %
      \node (C092) at (1,0) {}; %
      \node (C093) at (2,0) {}; %
      \node (C094) at (3,0) {}; %
      \draw[rounded corners, solarizedBlue] %
      (C091.center) -- ++ (0,-0.2) -| (C093.center); %
      \draw[rounded corners, white, line width=3] %
      (C092.center) -- ++ (0,-0.4) -| (C094.center); %
      \draw[rounded corners, solarizedRed] %
      (C092.center) -- ++ (0,-0.4) -| (C094.center); %
    \end{tikzpicture}
    }}
    \;,
\end{align}
where we have used colours to emphasize that the wires on both sides of each
equality connect the same cuts.

Using the \(*\)-involution and the gluing operation, we can construct a
Hermitian form on each \(\mathcal{J}_C\),
\begin{align}
  (\bullet|\bullet)
  : \mathcal{J}_C \times \mathcal{J}_C
  &\to \mathcal{J}
    \;,
  &
    J_C', J_C
  &\mapsto
    (J_C'|J_C)
    =(J_{\tikzmarknode{C21}{C}}')^* \otimes J_{\tikzmarknode{C22}{C}}
    \;.
    \label{eq:hermitian}
    \begin{tikzpicture}[overlay, remember picture, every path/.style = {shorten <=1pt,shorten >=1pt}]
      \draw[rounded corners] (C21.south) -- ++ (0,-0.2) -| (C22);
    \end{tikzpicture}
\end{align}
It follows from the properties of the gluing operation that \((\bullet|\bullet)\) is
compatible with the tensor product \(\otimes\), meaning
\begin{align}
  (J_C'\otimes J_{C'}' | J_C\otimes J_{C'})
  &= (J_C'| J_C)\sqcup
    (J_{C'}'|J_{C'})
    \;.
    \label{eq:gluingtensorprod}
\end{align}
For sources \(J',J\in\mathcal{J}=\mathcal{J}_\emptyset\), gluing acts trivially
on the trivial cut \(\emptyset\) and \(\otimes\) reduces to \(\sqcup\) by
\cref{eq:otimestrivial}, so we simply have
\begin{align}
  (J'|J)
  = (J')^* \sqcup J
  \;.
  \label{eq:trivialgluing}
\end{align}

In addition to the properties of the path integral \(\zeta\) laid out in
\cref{sec:generalprops}, we further require \(\zeta\) to be reflection-positive
with respect to partial sources. That is, for all \(C\in\mathcal{C}\) and
\(J_C\in\mathcal{J}_C\),
\begin{align}
  \zeta(J_C|J_C)
  &\ge 0
    \;.
    \label{eq:preflpos}
\end{align}
This allows us to define a positive semi-definite Hermitian form on each
\(\mathcal{J}_C\ni J_C',J_C\) as a vector space over \(\mathbb{C}\):
\begin{align}
  \braket{J_C'}{J_C}
  = \zeta(J_C'|J_C)
  \;.
  \label{eq:pinnerprod}
\end{align}
(Since \(\mathcal{J}\) is a vector space over \(\mathbb{C}\), so is
\(\mathcal{J}_C\).) Given \cref{eq:trivialgluing}, note that
\cref{eq:reflpos,eq:innerprod} are special cases of
\cref{eq:preflpos,eq:pinnerprod}.

In summary, the set \(\mathcal{C}\) of cuts is a monoid and the set
\(\mathcal{J}_C\) of partial sources for each cut \(C\in\mathbb{C}\) is a
\(\mathcal{J}\)-module, such that the tensor product
\(\mathcal{J}_C\otimes\mathcal{J}_{C'}\) is required to be a submodule of
\(\mathcal{J}_{C\otimes C'}\). The identity element \(\emptyset\in\mathcal{C}\)
is the trivial cut, for which \(\mathcal{J}_\emptyset=\mathcal{J}\) corresponds
to the set of complete sources and the tensor product \(\otimes\) acts as ring
multiplication \(\sqcup\). We further specified a \(*\)-involution which is an
automorphism on \(\mathcal{C}\) and a homomorphism relating \(\mathcal{J}_C\) to
\(\mathcal{J}_{C^*}\). We also introduced gluing and swap operations for partial
sources which behave similarly to tensor contraction and permuting tensor
indices. Combining the \(*\)-involution and the gluing operation, we obtain a
\(\mathcal{J}\)-valued Hermitian form \((\bullet|\bullet)\) on each
\(\mathcal{J}_C\). Finally, composing this with the path integral, we obtain a
\(\mathbb{C}\)-valued Hermitian form \(\braket{\bullet}{\bullet}\) which we
require to be positive semi-definite.

\subsection{Hilbert space sectors \(\mathcal{H}_C\) labelled by cuts
  \(C\in\mathcal{C}\) of sources}
\label{sec:cutsectors}

\begin{figure}
  \centering
  \includegraphics[scale=\figscale]{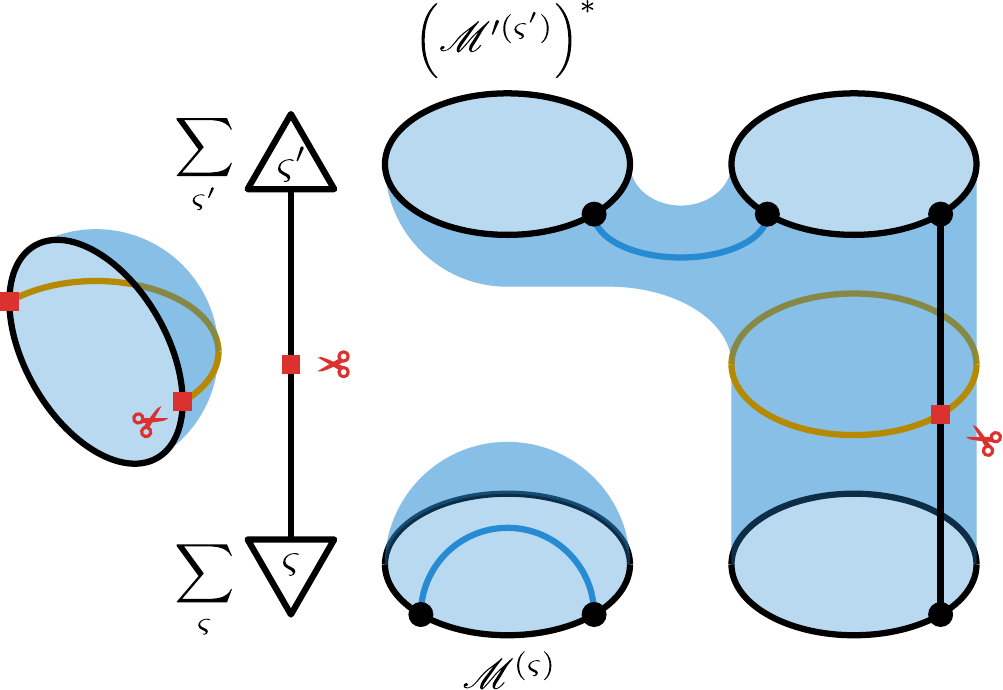}
  \caption{Slicing (orange and red) the gravitational path integral through cuts
    \(C\in\mathcal{C}\) (red) of sources (black) gives states in sectors
    \(\mathcal{H}_C\) of the gravitational Hilbert space.}
  \label{fig:cutsectors}
\end{figure}

Previously in \cref{sec:babysector}, we constructed the baby universe sector
\(\mathcal{H}_\emptyset\) of the gravitational Hilbert space, describing states
of closed universes on bulk slices where topology and local fields fluctuate
freely without being directly constrained by sources. In this subsection, we
will analogously construct sectors \(\mathcal{H}_C\) associated to various cuts
\(C\in\mathcal{C}\) through sources of the gravitational path integral.
The full gravitational Hilbert space will then be given by the sum of all such
sectors:
\begin{align}
  \mathcal{H}
  &= \bigoplus_{C\in\mathcal{C}} \mathcal{H}_C
\end{align}
The baby universe sector \(\mathcal{H}_\emptyset\) is a special case associated
with the trivial cut \(\emptyset\in\mathcal{C}\). More generally, as illustrated
in \cref{fig:cutsectors}, \(\mathcal{H}_C\) can describe states entangled with
external non-gravitational systems and states on bulk slices that intersect
sources at the cut \(C\). Here, the sources can specify boundary conditions at
spatial boundaries or dictate the presence of observer worldlines.

In \cref{sec:generalpartialsources}, we defined the set \(\mathcal{J}_C\) of
partial sources associated with each cut \(C\in\mathcal{C}\) to be a
\(\mathcal{J}\)-module and thus a vector space over \(\mathbb{C}\). We also
required the path integral \(\zeta:\mathcal{J}\to\mathbb{C}\) to be
reflection-positive \labelcref{eq:preflpos}, such that it can be used in
\cref{eq:pinnerprod} to define a \(\mathbb{C}\)-valued positive semi-definite
Hermitian form \(\braket{\bullet}{\bullet}\) on \(\mathcal{J}_C\). We can
therefore build a Hilbert space \(\mathcal{H}_C\) from \(\mathcal{J}_C\) in the
same manner as in \cref{sec:babysector}. First, we quotient \(\mathcal{J}_C\) by
the subspace
\begin{align}
  \mathcal{N}_C
  &= \left\{
    J_C\in\mathcal{J}_C
    : \braket{J_C'}{J_C} = 0\quad \forall J_C'\in\mathcal{J}_C
    \right\}
\end{align}
of null vectors, obtaining a pre-Hilbert space \(\mathcal{J}_C/\mathcal{N}_C\)
with inner product \(\braket{\bullet}{\bullet}\). Then, we take the Hilbert
space \(\mathcal{H}_C\) to be the completion of \(\mathcal{J}_C/\mathcal{N}_C\)
with respect to the norm obtained from \(\braket{\bullet}{\bullet}\).

We would now like to construct \(\alpha\)-sectors \(\mathcal{H}_C^\alpha\) of
\(\mathcal{H}_C\). One can do so like in \cref{sec:babysector} by defining
operators \(Z(J)\) on \(\mathcal{J}_C/\mathcal{N}_C\) as
\begin{align}
  Z(J) \ket{J_C}
  &= \ket{J \sqcup J_C}
    \;.
    \label{eq:zjonc}
\end{align}
Then, after making appropriate assumptions about extensions of bounded functions
of operators to \(\mathcal{H}_C\), one can again decompose \(\mathcal{H}_C\)
into \(\alpha\)-sectors \(\mathcal{H}_C^\alpha\) where \(Z(J)\) take definite
eigenvalues \(Z_\alpha(J)\). However, let us instead construct
\(\mathcal{H}_C^\alpha\) directly from the \(\alpha\)-sectors of the baby universe
Hilbert space \(\mathcal{H}_\emptyset^\alpha\), thus making it clear that the
spectrum \(\mathcal{A}\) of \(Z(J)\) is the same on \(\mathcal{H}_C\) as on the
baby universe Hilbert space \(\mathcal{H}_\emptyset\).

To do so, for all \(C,C'\in\mathcal{C}\) and \(J_C\in\mathcal{J}_C\), let us start
by defining an operator
\begin{align}
  \psi(J_C):
  \mathcal{J}_{C'}/\mathcal{N}_{C'}
  &\to \mathcal{J}_{C\otimes C'}/\mathcal{N}_{C\otimes C'}
  \;,
  &
  \ket{J_{C'}}
  &\mapsto \ket{J_C\otimes J_{C'}}
    \;.
    \label{eq:psijc}
\end{align}
A side remark is that \cref{eq:zjonc} is a special case of \cref{eq:psijc} with
\begin{align}
Z(J)=\psi(J)
\end{align}
for \(J\in\mathcal{J}=\mathcal{J}_\emptyset\). In the following, we will instead
focus on the case \(C'=\emptyset\) while \(C\in\mathcal{C}\) remains arbitrary
in \cref{eq:psijc}. In this case, we extend \(\psi(J_C)\) to a domain including
all \(\alpha\)-states\footnote{We expect that the domain of \(\psi(J_C)\) can be
  extended to also include integrals or sums of \(\ket{\alpha}\) over bounded
  regions in \(\mathcal{A}\). However, we do not expect to be able to extend
  \(\psi(J_C)\) to all of \(\mathcal{H}_\emptyset\). The danger is that
  inserting \((J_C|J_C)\in\mathcal{J}\) into the path integral produces a factor
  \(Z_\alpha(J_C|J_C)\) which can be unbounded in \(\mathcal{A}\) and thus spoil
  the normalizability for some putative image states.} in
\(\mathcal{H}_\emptyset\), such that the following continuity condition holds.
Since \(\alpha\)-states \(\ket{\alpha}\) span \(\mathcal{H}_\emptyset\), we can
decompose \(\ket{J}\) for any \(J\in\mathcal{J}\) as a sum or integral of
\(\ket{\alpha}\). The continuity condition we require is for \(\psi(J_C)\) to
commute with performing this sum or integral --- \ie{}, after acting on the
summand or integrand with \(\psi(J_C)\), the sum or integral converges to
\(\psi(J_C)\ket{J}=\ket{J\sqcup J_C}\) as expected from
\cref{eq:psijc}.\footnote{We have chosen \cref{eq:psijc} as our starting point
  because it illustrates the action of \(\psi(J_C)\) on states most intuitively.
  An alternative approach, which satisfies our continuity condition by
  construction, is the following. Let us forget the above definition of
  \(\mathcal{H}_C\) and \cref{eq:psijc} for \(C\ne\emptyset\) (but continue to
  assume the spectral decomposition of \(\mathcal{H}_\emptyset\) into
  \(\alpha\)-states \(\ket{\alpha}\) as in \cref{sec:babysector}). Instead, we
  define \(\mathcal{H}_C\) to be a Hilbert space \labelcref{eq:hcdecomp} with
  sectors \(\mathcal{H}_C^\alpha\) densely spanned by states
  \labelcref{eq:densespace} with inner product
  \labelcref{eq:innerdifferentalpha}. At this point \(\psi(J_C)\) are just maps
  taking \(\ket{\alpha}\) to abstractly defined states \(\psi(J_C)\ket{\alpha}\)
  with inner products \labelcref{eq:innerdifferentalpha}. We then define the
  action of \(\psi(J_C)\) on a state \(\ket{J}\in\mathcal{J}/\mathcal{N}\) by
  our continuity condition: decompose \(\ket{J}\) into \(\alpha\)-states
  \(\ket{\alpha}\) and define \(\psi(J_C)\ket{J}\) by the linear combination of
  \(\psi(J_C)\ket{\alpha}\). The resulting linear combinations
  \(\psi(J_C)\ket{J}\) are states with the same inner products as
  \cref{eq:psijc}, so this alternative construction is equivalent to the one
  described in the main text.}

We now define the Hilbert space \(\mathcal{H}_C^\alpha\) to be the Cauchy
completion of\footnote{As elsewhere in this paper, states written in the form of
  kets \(\ket{\bullet}\) and bras \(\ket{\bullet}\) are understood as
  equivalence classes under the addition of null states.}
\begin{align}
  \mathcal{D}_C^\alpha
  &= \{
    \ket{J_C, \alpha}_{\mathrm{r}}
    : J_C\in\mathcal{J}_C
    \}
    \;,
    &
    \ket{J_C, \alpha}
  &= \psi(J_C)\ket{\alpha}
    \label{eq:densespace}
\end{align}
(where the subscript on \(\ket{J_C, \alpha}_{\mathrm{r}}\) indicates a rescaling
to be explained in the next paragraph). To see that \(\mathcal{H}_C\) can be
fully decomposed into such sectors, recall that the set of
\(\ket{J_C}=\psi(J_C)\ket{\varnothing}\) for \(J_C\in\mathcal{J}_C\) is a dense
subspace. We can then apply our continuity condition to a decomposition of
\(\ket{\varnothing}\) in terms of \(\ket{\alpha}\). It follows that summing over
\(\mathcal{D}_C^\alpha\) for all \(\alpha\) and taking the Cauchy completion
recovers the full Hilbert space
\begin{align}
  \mathcal{H}_C
  &=
    \int_{\mathcal{A}}^\oplus \dd{\alpha}
    \mathcal{H}_C^\alpha
    \;.
    \label{eq:hcdecomp}
\end{align}

As expressed above, the \(\alpha\)-sectors are mutually orthogonal. Indeed, as
required by \cref{eq:pinnerprod} and our continuity condition,
\begin{align}
  \braket{J_C',\alpha'}{J_C,\alpha}
  &= Z_\alpha(J_C'|J_C) \braket{\alpha'}{\alpha}
    \;,
    \label{eq:innerdifferentalpha}
\end{align}
for states in the dense subspaces \(\mathcal{D}_C^{\alpha'}\) and
\(\mathcal{D}_C^\alpha\); the inner products of more general states are limits
of the above. Recall that \(\braket{\alpha'}{\alpha}\) is equal to the
\(\delta\)-function on \(\mathcal{A}\). It will be convenient therefore to
introduce the notation
\begin{align}
  \ket{\bullet,\alpha}_{\mathrm{r}}
  &= \norm{\alpha}^{-1} \ket{\bullet,\alpha}
    \label{eq:rescaledstate}
\end{align}
to denote states rescaled by the norm
\(\norm{\alpha}=\sqrt{\braket{\alpha}{\alpha}}\) so that this possibly infinite
factor does not appear in inner products within \(\alpha\)-sectors, \eg{}
\begin{align}
  \tensor[_{\mathrm{r}}]{
  \braket{J_C',\alpha}{J_C,\alpha}
  }{_{\mathrm{r}}}
  &= Z_\alpha(J_C'|J_C)
    \;.
    \label{eq:alphainner}
\end{align}

In \cref{sec:babysector}, baby universe \(\alpha\)-sectors
\(\mathcal{H}_\emptyset^\alpha\) were defined to be superselection sectors in
which the operators \(Z(J)\) take definite eigenvalues \(Z_\alpha(J)\). It is
clear from \cref{eq:zjonc,eq:psijc} that \(Z(J)\) and \(\psi(J_C)\) commute on
\(\mathcal{J}/\mathcal{N}\). Therefore, it is natural to define an extension of
\(Z(J)\) which acts on each \(\mathcal{H}_C^\alpha\) as multiplication by the
constant \(Z_\alpha(J)\). The consistency of this definition with
\cref{eq:zjonc} follows from our continuity condition. Thus, the
\(\alpha\)-sectors \(\mathcal{H}_C^\alpha\) are again superselection sectors
where the commuting operators \(Z(J)\) take the definite eigenvalues
\(Z_\alpha(J)\).

An important point of departure from \(\mathcal{H}_\emptyset^\alpha\), however,
is that \(\mathcal{H}_C^\alpha\) for \(C\ne\emptyset\) is generically not
one-dimensional. In particular, one can consider the closest analogue of
\cref{eq:jalpha},
\begin{align}
  \braket{\psi_C,\alpha}{J\sqcup J_C}
  &= \mel{\psi_C,\alpha}{Z(J)}{J_C}
    = Z_\alpha(J) \braket{\psi_C,\alpha}{J_C}
    \;,
    \label{eq:jpsialpha}
\end{align}
for any \(J\in\mathcal{J}\), \(J_C\in\mathcal{J}_C\), and
\(\ket{\psi_C,\alpha}_{\mathrm{r}}\in\mathcal{H}_C^\alpha\). For \(C\ne\emptyset\), the
\(\mathcal{J}\)-module \(\mathcal{J}_C\) need not be cyclic\footnote{A cyclic
  module is one which can be generated by a single element.}, and
\cref{eq:jpsialpha} can be consistent with the existence of
many linearly independent states
\(\ket{\psi_C,\alpha}_{\mathrm{r}}\in\mathcal{H}_C^\alpha\).

Let us end this section with two further remarks which will be of use in
\cref{sec:bounds}. Firstly, we note that there is a natural isometric embedding
\(\mathcal{H}_C^\alpha\otimes\mathcal{H}_{C'}^\alpha\to\mathcal{H}_{C\otimes{}C'}^\alpha\).
This is just the continuous extension of the isometric embedding
\begin{align}
  \mathcal{D}_C^\alpha\otimes \mathcal{D}_{C'}^\alpha
  &\to \mathcal{D}_{C\otimes C'}^\alpha
    \;,
    &
  \ket{J_C,\alpha}_{\mathrm{r}} \otimes \ket{J_C,\alpha}_{\mathrm{r}}
  &\mapsto
    \ket{J_C \otimes J_{C'}, \alpha}_{\mathrm{r}}
    \label{eq:embedding1}
\end{align}
More generally, we will write the embedding of arbitrary states similarly as
\begin{align}
  \mathcal{H}_C^\alpha\otimes\mathcal{H}_{C'}^\alpha
  &\to\mathcal{H}_{C\otimes C'}^\alpha
    \;,
  &
  \ket{\psi_C,\alpha}_{\mathrm{r}} \otimes \ket{\psi_{C'},\alpha}_{\mathrm{r}}
  &\mapsto
    \ket{\psi_C \otimes \psi_{C'}, \alpha}_{\mathrm{r}}
    \;.
    \label{eq:embedding2}
\end{align}
This embedding need not be onto, however. The Hilbert space
\(\mathcal{H}_{C\otimes{}C'}^\alpha\) can thus fail to factorize into a tensor
product of Hilbert spaces associated respectively with \(C\) and
\(C'\).\footnote{This kind of Hilbert space factorization is sometimes referred
  to as Harlow \cite{Harlow:2015lma} factorization or ``factorisation with an
  S''. We have explained how the Hilbert space can be decomposed into
  \(\alpha\)-sectors, in each of which the path integral factorizes
  \labelcref{eq:factorization} --- sometimes referred to as ``factorization with
  a Z''. Similarly, we expect that \(\mathcal{H}_{C\otimes{}C'}^\alpha\) can be
  further decomposed as a direct sum of sectors which exhibit Harlow
  factorization
  \cite{Colafranceschi:2023moh,Marolf:2024adj}. \label{foot:harlowfactor}}

Secondly, we can construct an anti-isometry relating
\(\mathcal{H}_C^\alpha\) and \(\mathcal{H}_{C^*}^\alpha\). Continuously extending
\begin{align}
  \mathcal{D}_C^\alpha
  &\to \mathcal{D}_{C^*}^\alpha
    \;,
    &
  \ket{J_C,\alpha}_{\mathrm{r}}
  &\mapsto
    \ket{J_C^*, \alpha}_{\mathrm{r}}
    \;,
    \label{eq:staralpha1}
\end{align}
we will more generally write
\begin{align}
  \mathcal{H}_C^\alpha
  &\to\mathcal{H}_{C^*}^\alpha
    \;,
  &
  \ket{\psi_C,\alpha}_{\mathrm{r}}
  &\mapsto
    \ket{(\psi_C)^*, \alpha}_{\mathrm{r}}
    \;.
    \label{eq:staralpha2}
\end{align}

%% file: sections/section04_operatorsDimBounds.tex
\section{A noncommutative algebra of operators and bounds on
  \(\alpha\)-sectors}
\label{sec:operatorsdimbounds}

In \cref{sec:babysector}, we showed that \(\alpha\)-sectors
\(\mathcal{H}_\emptyset^\alpha\) in the baby universe Hilbert space
\(\mathcal{H}_\emptyset\) must be one-dimensional. The fact that each
\(\mathcal{H}_\emptyset^\alpha\) is one-dimensional is consistent with the lack
of nontrivial operators on \(\mathcal{H}_\emptyset^\alpha\). Recall that
\(\mathcal{H}_\emptyset\) is comprised of states of baby universes, which are
closed universes on slices where sources for the gravitational path integral do
not prescribe the existence of objects like spatial boundaries, observer
worldlines, or nongravitational external systems. The lack of nontrivial
operators can therefore be understood as a consequence of the lack of objects on
these slices to which one can dress such operators.

In \cref{sec:cutsectors}, we explained that the \(\alpha\)-sectors
\(\mathcal{H}_C^\alpha\), in Hilbert spaces \(\mathcal{H}_C\) of states prepared
by partial sources with nontrivial cuts \(C\ne\emptyset\), need not be
one-dimensional. These states describe slices where sources for the
gravitational path integral prescribe the existence of objects located at \(C\),
on which one can construct operators which remain generically nontrivial in each
superselection \(\mathcal{H}_C^\alpha\).

In \cref{sec:operators}, we will construct some of these generically noncommuting
operators on \(\mathcal{H}_C^\alpha\) which act by gluing partial sources
\(J_{C\otimes{}C^*}\in\mathcal{J}_{C\otimes{}C^*}\) to states --- this
construction will parallel that of ref.~\cite{Colafranceschi:2023moh}. In
\cref{sec:bounds}, we will then derive a bound on the Hilbert space trace of
certain such operators in each \(\mathcal{H}_C^\alpha\) using the Cauchy-Schwarz
inequality. One can view the trace of positive operators as a proxy for measuring
the size of \(\mathcal{H}_C^\alpha\) and accordingly interpret our result as an
upper bound on this size. This analysis will be a hybrid of similar analyses
found in refs.~\cite{Marolf:2020xie,Colafranceschi:2023moh}.

\subsection{Noncommuting operators}
\label{sec:operators}

Let us begin by defining some gluing operators on the dense domain
\(\mathcal{D}_C^\alpha\) of \(\mathcal{H}_C^\alpha\). For each
\(J_{C\otimes{}C^*}\in\mathcal{J}_{C\otimes{}C^*}\), we define the operator
\(O(J_{C\otimes{}C^*})\) on this domain using the gluing operation introduced in
\cref{eq:gluing},
\begin{align}
  O(J_{C\otimes{}C^*})
  : \mathcal{D}_C^\alpha
  &\to \mathcal{D}_C^\alpha
  \;,
  &
    \ket{J_C,\alpha}_{\mathrm{r}}
    \mapsto \Big|
    J_{C\otimes\tikzmarknode{C01}{C}^*}\otimes J_{\tikzmarknode{C02}{C}}
    ,\alpha
    \Big\rangle_{\mathrm{r}}
    \;.
    \begin{tikzpicture}[overlay, remember picture, every path/.style = {shorten <=1pt,shorten >=1pt}]
    \draw[rounded corners] (C01.south) -- ++ (0,-0.2) -| (C02);
  \end{tikzpicture}
    \label{eq:ooperator}
\end{align}
Before extending \(O(J_{C\otimes{}C^*})\) to \(\mathcal{H}_C^\alpha\), let us
make some preliminary observations about these operators defined on
\(\mathcal{D}_C^\alpha\).

The first is that the product of two such operators \labelcref{eq:ooperator} is
another,
\begin{align}
  O(J_{C\otimes{}C^*})
  O(J_{C\otimes{}C^*}')
  &= O\Big(
    J_{C\otimes{}\tikzmarknode{C731}{C}^*}
    \otimes
    J_{\tikzmarknode{C732}{C}\otimes{}C^*}'
    \Big)
    \;.
    \label{eq:oopprod}
    \begin{tikzpicture}[overlay, remember picture, every path/.style = {shorten <=1pt,shorten >=1pt}]
      \draw[rounded corners] (C731.south) -- ++ (0,-0.2) -| (C732);
    \end{tikzpicture}
\end{align}
We see from this expression that, with the exception of \(C=\emptyset\) (in
which case \(O(J)=Z(J)\) for \(J\in\mathcal{J}\)), the operators \(O\) are not
generically expected to commute,
\begin{align}
  O(J_{C\otimes{}C^*})\,
  O(J_{C\otimes{}C^*}')
  &\ne
    O(J_{C\otimes{}C^*}')\,
    O(J_{C\otimes{}C^*})
    \;,
\end{align}
for distinct
\(J_{C\otimes{}C^*},J_{C\otimes{}C^*}'\in\mathcal{J}_{C\otimes{}C^*}\). If this
is the case on \(\mathcal{D}_C^\alpha\), then any extensions of the operators
will also fail to commute.

Secondly, for states in \(\mathcal{D}_C^\alpha\), it is straightforward to show
from \cref{eq:pinnerprod,eq:hermitian} that matrix elements of the adjoint
\(O(J_{C\otimes{}C^*})^\dagger\) are equal to
\begin{align}
  \tensor[_{\mathrm{r}}]{
  \mel{J'_C,\alpha}{O(J_{C\otimes{}C^*})^\dagger}{J_C,\alpha}
  }{_{\mathrm{r}}}
  &= \tensor[_{\mathrm{r}}]{
    \mel{J'_C,\alpha}{O\Big((J_{\tikzmarknode{C71}{C}\otimes{}\tikzmarknode{C72}{C}^*})^*\Big)}{J_C,\alpha}
    }{_{\mathrm{r}}}
    \;.
    \label{eq:meloadjoint}
    \begin{tikzpicture}[overlay, remember picture, every path/.style = {shorten <=1pt,shorten >=1pt}]
      \draw[rounded corners=3] (C72.south) -- ++ (0,-0.15) -| ([shift=({0,-0.3})]C71.south);
      \draw[rounded corners=3, white, line width=3] %
      (C71.south) -- ++ (0,-0.15) -| ([shift=({0,-0.3})]C72.south);%
      \draw[rounded corners=3] (C71.south) -- ++ (0,-0.15) -| ([shift=({0,-0.3})]C72.south);
    \end{tikzpicture}
\end{align}
Therefore, if the reality condition
\begin{align}
  J_{C\otimes{}C^*}
  &= (J_{\tikzmarknode{C81}{C}\otimes{}\tikzmarknode{C82}{C}^*})^*
    \label{eq:realpsource}
    \begin{tikzpicture}[overlay, remember picture, every path/.style = {shorten <=1pt,shorten >=1pt}]
      \draw[rounded corners=3] (C82.south) -- ++ (0,-0.15) -| ([shift=({0,-0.3})]C81.south);
      \draw[rounded corners=3, white, line width=3] %
      (C81.south) -- ++ (0,-0.15) -| ([shift=({0,-0.3})]C82.south);%
      \draw[rounded corners=3] (C81.south) -- ++ (0,-0.15) -| ([shift=({0,-0.3})]C82.south);
    \end{tikzpicture}
\end{align}
is satisfied, then the matrix elements of \(O(J_{C\otimes{}C^*})\) in
\(\mathcal{D}_C^\alpha\) form a Hermitian matrix. Moreover, if
\(J_{C\otimes{}C^*}\) takes the form
\begin{align}
  J_{C\otimes{}C^*}
  &= J_{C\otimes\tikzmarknode{C91}{C}'} \otimes (J_{C\otimes\tikzmarknode{C92}{C}'})^*
    \label{eq:pospsource}
    \begin{tikzpicture}[overlay, remember picture, every path/.style = {shorten <=1pt,shorten >=1pt}]
      \draw[rounded corners] (C91.south) -- ++ (0,-0.2) -| (C92);
    \end{tikzpicture}
\end{align}
for some \(J_{C\otimes{}C'}\in\mathcal{J}_{C\otimes{}C'}\), then it is
straightforward to show, using the positive definiteness
\labelcref{eq:pinnerprod} of the inner product, that expectation values of
\(O(J_{C\otimes{}C^*})\) in \(\mathcal{D}_C^\alpha\) are non-negative,
\begin{align}
  \tensor[_{\mathrm{r}}]{
  \langle
  J_C,\alpha
  |
  O\Big(
  J_{C\otimes\tikzmarknode{C001}{C}'} \otimes (J_{C\otimes\tikzmarknode{C002}{C}'})^*
  \Big)
  |
  J_C,\alpha
  \rangle
  }{_{\mathrm{r}}}
  &\ge 0 \;.
    \label{eq:evo}
  \begin{tikzpicture}[overlay, remember picture, every path/.style = {shorten <=1pt,shorten >=1pt}]
    \draw[rounded corners] (C001.south) -- ++ (0,-0.2) -| (C002);
  \end{tikzpicture}
\end{align}

We would now like to continuously extend \(O(J_{C\otimes{}C^*})\), for any
\(J_{C\otimes{}C^*}\in\mathcal{J}_{C\otimes{}C^*}\), to all of
\(\mathcal{H}_C^\alpha\), where we will argue by continuity that the above
statements continue to hold. To see that this is possible, we will bound the
operator norm of \(O(J_{C\otimes{}C^*})\). Consider therefore the squared norm
of an image state, which we can write using \cref{eq:alphainner} as
\begin{align}
  \tensor[_{\mathrm{r}}]{
  \Big\langle
  J_{C\otimes\tikzmarknode{C11}{C}^*}\otimes J_{\tikzmarknode{C12}{C}},\alpha
  \Big|
  J_{C\otimes\tikzmarknode{C21}{C}^*}\otimes J_{\tikzmarknode{C22}{C}},\alpha
  \Big\rangle
  }{_{\mathrm{r}}}
  &=
    Z_\alpha\Big(
    (J_{\tikzmarknode{C51}{C}\otimes\tikzmarknode{C31}{C}^*})^*
    \otimes (J_{\tikzmarknode{C32}{C}})^*
    \otimes J_{\tikzmarknode{C52}{C}\otimes\tikzmarknode{C41}{C}^*}
    \otimes J_{\tikzmarknode{C42}{C}}
    \Big)
  \\[10pt]
  &=
    \tensor[_{\mathrm{r}}]{
    \Big\langle
    J_{C\otimes \tikzmarknode{C61}{C}^*}
    \otimes
    (J_{\tikzmarknode{C62}{C}})^*
    ,\alpha
    \Big|
    J_{C\otimes C^*}
    \otimes
    (J_C)^*,
    \alpha
    \Big\rangle
    }{_{\mathrm{r}}}
    \;.
    \label{eq:ostatenorm}
    \begin{tikzpicture}[overlay, remember picture, every path/.style = {shorten <=1pt,shorten >=1pt}]
      \draw[rounded corners] (C11.south) -- ++ (0,-0.2) -| (C12);
      \draw[rounded corners] (C21.south) -- ++ (0,-0.2) -| (C22);
      \draw[rounded corners] (C31.south) -- ++ (0,-0.2) -| (C32);
      \draw[rounded corners] (C41.south) -- ++ (0,-0.2) -| (C42);
      \draw[rounded corners] (C51.south) -- ++ (0,-0.3) -| (C52);
      \draw[rounded corners=3] (C62.south) -- ++ (0,-0.15) -| ([shift=({0,-0.3})]C61.south);
      \draw[rounded corners=3, white, line width=3] %
      (C61.south) -- ++ (0,-0.15) -| ([shift=({0,-0.3})]C62.south);%
      \draw[rounded corners=3] (C61.south) -- ++ (0,-0.15) -| ([shift=({0,-0.3})]C62.south);
    \end{tikzpicture}
\end{align}
We can now apply the Cauchy-Schwarz inequality,
\begin{align}
  \abs{\braket{\psi'}{\psi}}
  \le \norm{\psi'}\,\norm{\psi\vphantom{'}}
  \;,
  \label{eq:csineq}
\end{align}
which follows simply from the positive definiteness of the inner
product,
\begin{align}
  0
  &\le \min_{z\in\mathbb{C}}
    \norm{z\,\psi + \psi'}
    = \norm{
    -\frac{
    \abs{\braket{\psi'}{\psi}}^2
    }{
    \norm{\psi}^2\braket{\psi'}{\psi}
    }\,
    \psi + \psi'
    }
    =
    - \frac{\abs{\braket{\psi'}{\psi}}^2}{\norm{\psi}^2}
    + \norm{\psi'}^2
    \;.
    \label{eq:csineqderiv}
\end{align}
Upon applying the Cauchy-Schwarz inequality to the RHS of \cref{eq:ostatenorm},
a short calculation shows that the operator norm \(\norm{O(J_{C\otimes{}C^*})}\)
is bounded by
\begin{align}
  \norm{O(J_{C\otimes{}C^*})}
  \le
  \sqrt{
  \tensor[_{\mathrm{r}}]{
  \braket{J_{C\otimes C^*},\alpha}{J_{C\otimes C^*},\alpha}
  }{_{\mathrm{r}}}
  }
  =
  \sqrt{
  Z_\alpha(J_{C\otimes C^*}|J_{C\otimes C^*})
  }
  \;.
\end{align}
For fixed \(\alpha\), we expect \(Z_\alpha(J_{C\otimes C^*}|J_{C\otimes C^*})\)
to be finite.\footnote{Indeed, when \((J_{C\otimes C^*}|J_{C\otimes C^*})\) is a
  smooth boundary manifold, \cite{Colafranceschi:2023moh} takes this finiteness
  to be an axiom of the factorized path integral (which we call)
  \(\zeta_\alpha=Z_\alpha\). On the other hand, we do not necessarily expect the
  eigenvalues \(Z_\alpha\) to be bounded on all of \(\mathcal{A}\).} If this is
case, we can continuously extend the densely-defined bounded operator
\(O(J_{C\otimes{}C^*})\) to all of \(\mathcal{H}_C^\alpha\).

By continuity, \cref{eq:oopprod} applies to these extended operators on
\(\mathcal{H}_C^\alpha\). Continuity of \cref{eq:meloadjoint} implies that the
adjoint is indeed given by
\begin{align}
  O(J_{C\otimes{}C^*})^\dagger
  &= O\Big(
    (J_{\tikzmarknode{C011}{C}\otimes{}\tikzmarknode{C012}{C}^*})^*
    \Big)
    \;.
    \label{eq:oadjoint}
    \begin{tikzpicture}[overlay, remember picture, every path/.style = {shorten <=1pt,shorten >=1pt}]
      \draw[rounded corners=3] (C012.south) -- ++ (0,-0.15) -| ([shift=({0,-0.3})]C011.south);
      \draw[rounded corners=3, white, line width=3] %
      (C011.south) -- ++ (0,-0.15) -| ([shift=({0,-0.3})]C012.south);%
      \draw[rounded corners=3] (C011.south) -- ++ (0,-0.15) -| ([shift=({0,-0.3})]C012.south);
    \end{tikzpicture}
\end{align}
In particular, partial sources that satisfy \cref{eq:realpsource} give rise to
self-adjoint \(O(J_{C\otimes{}C^*})\), while continuity of \cref{eq:evo} shows
that those of the form \labelcref{eq:pospsource} give positive semi-definite
\(O(J_{C\otimes{}C^*})\).

\subsection{Bounds on \(\alpha\)-sectors \(\mathcal{H}_C^\alpha\)}
\label{sec:bounds}

The dimension of a finite-dimensional Hilbert space is simply given by the
Hilbert space trace of the identity operator. To get a finite notion of size in
an infinite-dimensional Hilbert space, however, it is often useful to look at
the trace of operators such as \(e^{-\beta H}\), where \(H\) is the Hamiltonian
of the system. We will return to this example in
\cref{sec:spatialbdy}.

In the remainder of this section, we will more generally consider operators
\(O(J_{C\otimes{}C^*})\) of the kind introduced in \cref{sec:operators}, in the
case where \(J_{C\otimes{}C^*}\) takes the form
\begin{align}
  J_{C\otimes{}C^*}
  &= J_{C\otimes\tikzmarknode{C171}{C}^*}^{\sfrac{1}{2}}
    \otimes
    J_{\tikzmarknode{C172}{C}\otimes{}C^*}^{\sfrac{1}{2}}
    \label{eq:jhalf}
    \begin{tikzpicture}[overlay, remember picture, every path/.style = {shorten <=1pt,shorten >=1pt}]
      \draw[rounded corners] (C171.south) -- ++ (0,-0.2) -| (C172);
    \end{tikzpicture}
\end{align}
for some
\begin{align}
  J_{C\otimes{}C^*}^{1/2}
  &= \Big(
    J_{\tikzmarknode{C191}{C}\otimes{}\tikzmarknode{C192}{C}^*}^{1/2}
    \Big)^*
    \in \mathcal{J}_{C\otimes{}C^*}
    \;.
    \label{eq:jhalfreal}
    \begin{tikzpicture}[overlay, remember picture, every path/.style = {shorten <=1pt,shorten >=1pt}]
      \draw[rounded corners=3] (C192.south) -- ++ (0,-0.15) -| ([shift=({0,-0.3})]C191.south);
      \draw[rounded corners=3, white, line width=3] %
      (C191.south) -- ++ (0,-0.15) -| ([shift=({0,-0.3})]C192.south);%
      \draw[rounded corners=3] (C191.south) -- ++ (0,-0.15) -| ([shift=({0,-0.3})]C192.south);
    \end{tikzpicture}
\end{align}
Then, \(O\left(J_{C\otimes{}C^*}^{1/2}\right)\) is self-adjoint and
\begin{align}
  O(J_{C\otimes{}C^*})
  &= O\left(
    J_{C\otimes{}C^*}^{1/2}
    \right)^2
\end{align}
is positive semi-definite. We will now show that the Hilbert space trace of
\(O(J_{C\otimes{}C^*})\) on \(\mathcal{H}_C^\alpha\) is bounded by a path
integral,
\begin{align}
  \sum_\gamma
  \lambda_\gamma^\alpha
  &\le
    Z_\alpha\Big(
    J_{\tikzmarknode{C091}{C}\otimes\tikzmarknode{C092}{C}^*}
    \Big)
    \;,
    \label{eq:tracebound}
    \begin{tikzpicture}[overlay, remember picture, every path/.style = {shorten <=1pt,shorten >=1pt}]
      \draw[rounded corners] (C091.south) -- ++ (0,-0.2) -| (C092);
    \end{tikzpicture}
\end{align}
where \(\lambda_\gamma^\alpha\) are eigenvalues of \(O(J_{C\otimes{}C^*})\) on
\(\mathcal{H}_C^\alpha\) and degeneracies are included in the sum.

Let \(\ket{\psi_C^\gamma,\alpha}_{\mathrm{r}}\in\mathcal{H}_C^\alpha\) be an
orthonormal basis of eigenstates for which\footnote{By definition,
  \(\left(\lambda_\gamma^\alpha \right)^{1/2}\) denotes the eigenvalue of
  \(O\left(J_{C\otimes{}C^*}^{1/2}\right)\), which might be negative.}
\begin{align}
  O\left(J_{C\otimes{}C^*}^{1/2}\right) \ket{\psi_C^\gamma,\alpha}_{\mathrm{r}}
  &= \left( \lambda_\gamma^\alpha \right)^{1/2} \ket{\psi_C^\gamma,\alpha}_{\mathrm{r}}
    \;,
    &
    O\left(J_{C\otimes{}C^*}\right) \ket{\psi_C^\gamma,\alpha}_{\mathrm{r}}
  &= \lambda_\gamma^\alpha \ket{\psi_C^\gamma,\alpha}_{\mathrm{r}}
    \;.
\end{align}
We would like to apply the
Cauchy-Schwarz inequality \labelcref{eq:csineq} to the states
\(\ket{J_{C\otimes{}C^*}^{1/2},\alpha}_{\mathrm{r}}\) and
\begin{align}
  \ket{\mathrm{TFD}_{C\otimes C^*},\alpha}_{\mathrm{r}}
  &= \sum_\gamma
    \left( \lambda_\gamma^\alpha \right)^{1/2} \,
    \ket{\psi_C^\gamma \otimes (\psi_C^\gamma)^*,\alpha}_{\mathrm{r}}
    \;.
\end{align}
To that end, we evaluate the inner products
\begin{align}
  \tensor[_{\mathrm{r}}]{
  \braket{J_{C\otimes{}C^*}^{1/2},\alpha}{J_{C\otimes{}C^*}^{1/2},\alpha}
  }{_{\mathrm{r}}}
  &= Z_\alpha
    \left(
    J_{C\otimes{}C^*}^{1/2}
    \middle|
    J_{C\otimes{}C^*}^{1/2}
    \right)
    = Z_\alpha\Big(
    J_{\tikzmarknode{C911}{C}\otimes\tikzmarknode{C912}{C}^*}
    \Big)
    \;,
  \\[10pt]
  \tensor[_{\mathrm{r}}]{
  \braket{\mathrm{TFD}_{C\otimes C^*},\alpha}{J_{C\otimes{}C^*}^{1/2},\alpha}
  }{_{\mathrm{r}}}
  &= \sum_\gamma
    \left( \lambda_\gamma^\alpha \right)^{1/2} \,
    \tensor[_{\mathrm{r}}]{
    \braket{\psi_C^\gamma \otimes (\psi_C^\gamma)^*,\alpha}{J_{C\otimes{}C^*}^{1/2},\alpha}
    }{_{\mathrm{r}}}
  \\
  &= \sum_\gamma
    \left( \lambda_\gamma^\alpha \right)^{1/2} \,
    \tensor[_{\mathrm{r}}]{
    \mel{
    \psi_C^\gamma,\alpha
    }{
    O\left(J_{C\otimes{}C^*}^{1/2}\right)
    }{
    \psi_C^\gamma,\alpha
    }
    }{_{\mathrm{r}}}
    \label{eq:usescontinuity}
  \\
  &= \sum_\gamma \lambda_\gamma^\alpha
  \\
  &= \tensor[_{\mathrm{r}}]{
  \braket{\mathrm{TFD}_{C\otimes C^*},\alpha}{\mathrm{TFD}_{C\otimes C^*},\alpha}
    }{_{\mathrm{r}}}
    \;,
    \begin{tikzpicture}[overlay, remember picture, every path/.style = {shorten <=1pt,shorten >=1pt}]
      \draw[rounded corners] (C911.south) -- ++ (0,-0.2) -| (C912);
    \end{tikzpicture}
\end{align}
where \cref{eq:usescontinuity} follows from the continuity of
\(O\left(J_{C\otimes{}C^*}^{1/2}\right)\) and replacing
\(\ket{\psi_C^\gamma,\alpha}_{\mathrm{r}}\) by limits of
\(\ket{J_C,\alpha}_{\mathrm{r}}\in\mathcal{D}_C^\alpha\), which we can more
easily manipulate using \cref{eq:hermitian,eq:alphainner,eq:ooperator}. The
Cauchy-Schwarz inequality \labelcref{eq:csineq} then gives the desired bound
\labelcref{eq:tracebound} on the Hilbert space trace of \(O(J_{C\otimes{}C^*})\)
on \(\mathcal{H}_C^\alpha\).

The constructive derivation \labelcref{eq:csineqderiv} of the Cauchy-Schwarz
inequality also tells us that \cref{eq:tracebound} is saturated if and only if
\begin{align}
  \ket{J_{C\otimes{}C^*}^{1/2},\alpha}_{\mathrm{r}}
  = \ket{\mathrm{TFD}_{C\otimes C^*},\alpha}_{\mathrm{r}}
  \;.
  \label{eq:equivstates}
\end{align}
Recall, as described below \cref{eq:embedding2}, that the Hilbert space
\(\mathcal{H}_{C\otimes{}C^*}^\alpha\) need not factorize
\cite{Harlow:2015lma,Colafranceschi:2023moh} between the cut components \(C\)
and \(C^*\), and
\(\mathcal{H}_C^\alpha\otimes\mathcal{H}_{C^*}^\alpha\ni\ket{\mathrm{TFD}_{C\otimes{}C^*},\alpha}_{\mathrm{r}}\)
might be a proper subspace of \(\mathcal{H}_{C\otimes{}C^*}^\alpha\ni
\ket{J_{C\otimes{}C^*}^{1/2},\alpha}_{\mathrm{r}}\). Therefore,
\cref{eq:equivstates} need not hold and the bound \labelcref{eq:tracebound} need
not be saturated.

Let us remark finally that while path integrals \(\zeta_\alpha=Z_\alpha\)
restricted to \(\alpha\)-sectors are practically inaccessible (except in simple
toy theories), we can take the ensemble average of \cref{eq:tracebound} to
obtain a useful bound involving the original, generically non-factorizing path
integral \(\zeta\):
\begin{align}
  \int_{\mathcal{A}} \dd{\alpha}
  \abs{\braket{\alpha}{\varnothing}}^2
  \,
  \sum_\gamma
  \lambda_\gamma^\alpha
  &\le \zeta\Big(
    J_{\tikzmarknode{C721}{C}\otimes\tikzmarknode{C722}{C}^*}
    \Big)
    \;.
    \label{eq:traceboundavg}
    \begin{tikzpicture}[overlay, remember picture, every path/.style = {shorten <=1pt,shorten >=1pt}]
      \draw[rounded corners] (C721.south) -- ++ (0,-0.2) -| (C722);
    \end{tikzpicture}
\end{align}
This provides an upper bound on the \(\alpha\)-dependent trace
\(\sum_\gamma\lambda_\gamma^\alpha\) averaged over the Hartle-Hawking ensemble.

%% file: sections/section05_discussion.tex
\section{Discussion}
\label{sec:discuss}

In this paper, we have developed a framework which describes the gravitational
path integral and corresponding Hilbert spaces in terms of abstract sources and
partial sources.

In particular, we described in \cref{sec:generalprops} the gravitational path
integral as a function from the commutative \(*\)-algebra of sources
\(\mathcal{J}\) to \(\mathbb{C}\). In \cref{sec:babysector}, we then constructed
the Hilbert space \(\mathcal{H}_\emptyset\) of baby universe states prepared by
the insertion of sources \(J\in\mathcal{J}\) in the path integral. We saw in
particular that \(\mathcal{H}_\emptyset\) decomposes into superselection sectors
\(\mathcal{H}_\emptyset^\alpha\) which are one-dimensional. Viewing each \(\alpha\)-sector
\(\mathcal{H}_\emptyset^\alpha\) as describing a self-contained UV-complete theory of
gravity, this leads to the striking conclusion that there is a unique state of
freely fluctuating closed universes in each such theory.

In \cref{sec:generalpartialsources}, we generalized our framework to include
\(\mathcal{J}\)-modules \(\mathcal{J}_C\) of partial sources
\(J_C\in\mathcal{J}_C\) which, when appropriately glued across the cut
\(C\in\mathcal{C}\), produce complete sources \(J\in\mathcal{J}\). We then used
these partial sources in \cref{sec:cutsectors} to prepare states in Hilbert
space sectors \(\mathcal{H}_C\) labelled by cuts \(C\). Importantly,
\(\mathcal{H}_C\) decomposes into \(\alpha\)-sectors \(\mathcal{H}_C^\alpha\)
which can have dimension greater than one. In particular, we constructed in
\cref{sec:operators} a noncommutative algebra of operators
\(O(J_{C\otimes{}C^*}):\mathcal{H}_C^\alpha\to\mathcal{H}_C^\alpha\).

In \cref{sec:spatialbdy}, we will first review how spacetimes with spatial
boundaries fit into our framework of partial sources and cuts
\cite{Marolf:2020xie}. Then in \cref{sec:applications}, we will demonstrate how
our abstract formalism also encapsulates recent proposals for modifying rules of
the gravitational path integral, such that closed universes no longer have
one-dimensional Hilbert spaces when observers are included
\cite{Abdalla:2025gzn,Harlow:2025pvj}. We will moreover apply the bound we
derived in \cref{sec:bounds} on the Hilbert space trace of
\(O(J_{C\otimes{}C^*})\) in \(\mathcal{H}_C^\alpha\), finding rough agreement
with bounds on the dimension of \(\mathcal{H}_C^\alpha\) reported by
refs.~\cite{Abdalla:2025gzn,Harlow:2025pvj} in simple theories. We end in
\cref{sec:future} by pointing to directions for future work.

\subsection{Application to spatial boundaries}
\label{sec:spatialbdy}

As described in \cref{sec:exsources}, some examples of sources
\(J\in\mathcal{J}\) are boundary manifolds \(J=\mathscr{M}\) on which boundary
conditions are prescribed for a bulk path integral. As described in
\cref{sec:expartialsources}, examples of partial sources \(J_C\in\mathcal{J}_C\)
are partial boundary manifolds \(J_C=\mathscr{N}_\Sigma\). The cut \(C=\Sigma\)
in this case is the boundary \(\Sigma=\partial\mathscr{N}_\Sigma\) of the
partial boundary manifold. From the bulk perspective, \(\Sigma\) is also the
boundary of a bulk codimension-one slice; taking this bulk slice to be a Cauchy
slice of a Lorentzian section, \(\Sigma\) is then a spatial boundary of this
Cauchy slice. As described in \cref{sec:cutsectors}, we can construct the
Hilbert space \(\mathcal{H}_\Sigma\) of bulk spacetimes with spatial boundary
\(\Sigma\). In particular, its \(\alpha\)-sectors \(\mathcal{H}_\Sigma^\alpha\)
need not be one-dimensional, as was the case for baby universe, \ie{} freely
fluctuating closed universes.

Let us now consider in this context an example of the kind of operator
\(O(J_{C\otimes C^*})\) described in \cref{sec:operatorsdimbounds}. In
particular, let
\(J_{C\otimes{}C^*}=\mathrm{Cyl}_{\Sigma\otimes{}\Sigma^*}^\beta\) be a partial
boundary manifold which is a Euclidean cylinder, \(\Sigma\times(0,\beta)\),
prescribing boundary conditions which are invariant under translation along the
Euclidean time interval \((0,\beta)\). Then this cylinder can be decomposed into
two halves, as in \labelcref{eq:jhalf},
\begin{align}
  \mathrm{Cyl}_{\Sigma\otimes{}\Sigma^*}^\beta
  &=\mathrm{Cyl}_{\Sigma\otimes\tikzmarknode{C311}{\Sigma}^*}^{\beta/2}
    \otimes
  \mathrm{Cyl}_{\tikzmarknode{C312}{\Sigma}\otimes{}\Sigma^*}^{\beta/2}
    \;.
    \label{eq:hhalf}
    \begin{tikzpicture}[overlay, remember picture, every path/.style = {shorten <=1pt,shorten >=1pt}]
      \draw[rounded corners] (C311.south) -- ++ (0,-0.2) -| (C312);
    \end{tikzpicture}
\end{align}
Let us furthermore require the boundary conditions on the cylinder to be real,
in the sense of \cref{eq:jhalfreal},
\begin{align}
  \mathrm{Cyl}_{\tikzmarknode{C312}{\Sigma}\otimes{}\Sigma^*}^{\beta/2}
  &= \Big(
    \mathrm{Cyl}_{\tikzmarknode{C471}{C}\otimes{}\tikzmarknode{C472}{C}^*}^{\beta/2}
    \Big)^*
    \label{eq:hselfadjoint}
    \begin{tikzpicture}[overlay, remember picture, every path/.style = {shorten <=1pt,shorten >=1pt}]
      \draw[rounded corners=3] (C472.south) -- ++ (0,-0.15) -| ([shift=({0,-0.3})]C471.south);
      \draw[rounded corners=3, white, line width=3] %
      (C471.south) -- ++ (0,-0.15) -| ([shift=({0,-0.3})]C472.south);%
      \draw[rounded corners=3] (C471.south) -- ++ (0,-0.15) -| ([shift=({0,-0.3})]C472.south);
    \end{tikzpicture}
\end{align}
The operator \(O\left(\mathrm{Cyl}_{\Sigma\otimes{}\Sigma^*}^\beta\right)\) is then of the
particular kind studied in \cref{sec:bounds}, to which our bound
\labelcref{eq:tracebound} on the trace applies.

As expressed in \cref{eq:ooperator}, this operator
\(O\left(\mathrm{Cyl}_{\Sigma\otimes{}\Sigma^*}^\beta\right)\) is defined to act
by gluing the Euclidean cylinder
\(\mathrm{Cyl}_{\Sigma\otimes{}\Sigma^*}^\beta\) to partial boundary manifolds
\(\mathscr{N}_\Sigma\) preparing states in \(\mathcal{H}_C^\alpha\). Thus, in
more familiar language, we might describe this operator as a Euclidean time
evolution operator,
\begin{align}
  O\left(\mathrm{Cyl}_{\Sigma\otimes{}\Sigma^*}^\beta\right)
  &= e^{-\beta\, H}
    \;,
\end{align}
generated by some Hamiltonian \(H\). In particular, \cref{eq:hhalf} requires
\(e^{-\beta\, H}=e^{-\beta\, H/2}\,e^{-\beta\, H/2}\) while
\cref{eq:hselfadjoint} requires \(e^{-\beta H}\) to be self-adjoint --- see
\cref{eq:oadjoint}. In this language, the bound \labelcref{eq:tracebound} reads
\begin{align}
  \sum_\gamma e^{-\beta\, E_\gamma^\alpha}
  \le Z_\alpha(\beta)
  \;,
  \label{eq:partfuncbound}
\end{align}
where \(E_\gamma^\alpha\) are energies of eigenstates in
\(\mathcal{H}_\Sigma^\alpha\) and
\begin{align}
  Z_\alpha(\beta)
  &= Z_\alpha\Big(
    \mathrm{Cyl}_{\tikzmarknode{C571}{\Sigma}\otimes\tikzmarknode{C572}{\Sigma}^*}^\beta
    \Big)
    \begin{tikzpicture}[overlay, remember picture, every path/.style = {shorten <=1pt,shorten >=1pt}]
      \draw[rounded corners] (C571.south) -- ++ (0,-0.2) -| (C572);
    \end{tikzpicture}
\end{align}
is a gravitational path integral, restricted to an \(\alpha\)-sector, and subject to
boundary conditions on a boundary manifold with a Euclidean time circle of
length \(\beta\).

It is perhaps worth emphasizing again that \cref{eq:partfuncbound} need not be
saturated. The LHS is the thermal partition function on the one-sided Hilbert
space \(\mathcal{H}_\Sigma^\alpha\), given by the Hilbert space trace of
\(e^{-\beta\,H}\). Meanwhile, the RHS (perhaps modulo the restriction to an
\(\alpha\)-sector) is how one might ordinarily calculate a ``thermal partition
function'' using a gravitational path integral \cite{Gibbons:1976ue}, with a
trace defined by gluing boundaries of the partial boundary manifold
\cite{Colafranceschi:2023moh}. As described below \cref{eq:equivstates}, a
mismatch between the two sides of the inequality can occur if the two-sided
Hilbert space \(\mathcal{H}_{\Sigma\otimes{}\Sigma^*}^\alpha\) does not
factorize as a product
\(\mathcal{H}_\Sigma^\alpha\otimes\mathcal{H}_{\Sigma^*}^\alpha\) of one-sided
Hilbert spaces.\footnote{As mentioned in \cref{foot:harlowfactor}, we expect
  that \(\mathcal{H}_{\Sigma\otimes{}\Sigma^*}^\alpha\) can be expressed as a
  direct sum of sectors which exhibit Harlow factorization between \(\Sigma\)
  and \(\Sigma^*\). In each such sector, there can still be a mismatch by a
  proportionality constant between the Hilbert space trace and the path integral
  trace, which can be attributed to the existence of ``hidden sectors''
  \cite{Colafranceschi:2023moh}.}

\subsection{Application to proposed prescriptions for including observers}
\label{sec:applications}

Let us now see how our general framework applies to two recent proposals for
modifying rules of the gravitational path integral in the presence of observers
\cite{Abdalla:2025gzn,Harlow:2025pvj}, particularly in the context of closed
universes.

\subsubsection{Prescribed observer worldlines}
\label{sec:obsworldline}

One proposal, by Abdalla, Antonini, Iliesiu, and Levine \cite{Abdalla:2025gzn},
is to consider path integrals with sources
\(J=(\mathscr{M}^{(i)})^*\cdot\gamma^{(\Delta)}\cdot\mathscr{M}^{(j)}\) of the
kind illustrated in \cref{fig:sourcegeodesic}. In particular,
\((\mathscr{M}^{(i)})^*\cdot\gamma^{(\Delta)}\cdot\mathscr{M}^{(j)}\) is
comprised of two asymptotic boundary manifolds \((\mathscr{M}^{(i)})^*\) and
\(\mathscr{M}^{(j)}\) connected by a bulk geodesic \(\gamma^{(\Delta)}\), whose
Euclidean length \(\ell\) is to be integrated over with an extra weight
\(e^{-\Delta\,\ell}\). The interpretation is that \((\mathscr{M}^{(i)})^*\) and
\(\mathscr{M}^{(j)}\) prepare bra- and ket-states for a closed universe, each
including (possibly among other things) the insertion of an operator
\(O^\Delta\) of dimension \(\Delta\). This \(O^\Delta\) insertion is viewed as
preparing an excitation in the bulk which we declare to be an observer and is
modelled as a geodesic worldline \(\gamma^{(\Delta)}\) with action
\(e^{-\Delta\,\ell}\).\footnote{In general, a QFT propagator on a curved spacetime
  can be expressed as a sum over all worldlines, geodesic or not. However, on
  two-dimensional hyperbolic manifolds, asymptotic boundary-to-boundary
  propagators can be expressed as a sum of \(e^{-\Delta\,\ell}\) over geodesics
  of length \(\ell\) --- see \cite{Lin:2023wac}. In the present model, the
  observer worldline is modelled as such a geodesic. \label{foot:geodesic}}
Compared to other matter excitations, the worldline of the observer is
distinguished by the fact that the source
\((\mathscr{M}^{(i)})^*\cdot\gamma^{(\Delta)}\cdot\mathscr{M}^{(j)}\) requires
the observer worldline to connect \((\mathscr{M}^{(i)})^*\) and
\(\mathscr{M}^{(j)}\); \ie{} the observer is prescribed to survive and exist
continuously on all intermediate slices of bulk configurations in the path
integral between \(\mathscr{M}^{(j)}\) and \((\mathscr{M}^{(i)})^*\). By
evaluating the JT path integral with products
\((\mathscr{M}^{(i_1)})^*\cdot\gamma^{(\Delta)}\cdot\mathscr{M}^{(j_1)}
\sqcup\cdots\sqcup
(\mathscr{M}^{(i_n)})^*\cdot\gamma^{(\Delta)}\cdot\mathscr{M}^{(j_n)}\) of such
sources, ref.~\cite{Abdalla:2025gzn} was able to conclude that the Hilbert space
of closed universes is no longer one-dimensional in the presence of prescribed
observer worldlines.

More precisely, in our language, the gravitational states in question are those
prepared by partial sources
\(J_\Delta=\xi_\Delta\cdot\mathscr{M}\in\mathcal{J}_\Delta\) of the kind
illustrated in \cref{fig:psourcegeodesic} and are elements of
\(\mathcal{J}_\Delta\) described at the end of \cref{sec:expartialsources}. In
particular, \(\xi_\Delta\cdot\mathscr{M}\) is comprised of the boundary
manifold \(\mathscr{M}\) together with half \(\xi_\Delta\) of an observer
worldline terminating at a cut \(C\) labelled by the value of \(\Delta\). (We
will continue using our shorthand of denoting \(C\) in this case just by
\(\Delta\) when it appears in a subscript; in such contexts, \(\Delta^*\) shall refer to
\(C^*\).) As in \cref{eq:wlhermitian}, taking the complex conjugation and
orientation- or time-reversal of \(\xi_\Delta\cdot\mathscr{M}^{(i)}\) and gluing to
\(\xi_\Delta\cdot\mathscr{M}^{(j)}\) by definition gives a complete source
\begin{align}
  \left(
  \xi_\Delta\cdot\mathscr{M}^{(i)}
  \middle|
  \xi_\Delta\cdot\mathscr{M}^{(j)}
  \right)
  = (
  \xi_{\tikzmarknode{C731}{\Delta}}\cdot\mathscr{M}^{(i)}
  )^* \otimes
  \xi_{\tikzmarknode{C732}{\Delta}}\cdot\mathscr{M}^{(j)}
  =(\mathscr{M}^{(i)})^*\cdot\gamma^{(\Delta)}\cdot\mathscr{M}^{(j)}
    \begin{tikzpicture}[overlay, remember picture, every path/.style = {shorten <=1pt,shorten >=1pt}]
      \draw[rounded corners] (C731.south) -- ++ (0,-0.2) -| (C732);
    \end{tikzpicture}
  \label{eq:xigluing}
\end{align}
of the kind described in the previous paragraph.

Inserting products of such sources into the path integral, we see that
\begin{align}
  \begin{split}
    \MoveEqLeft\zeta\left(
  (\mathscr{M}^{(i_1)})^*\cdot\gamma^{(\Delta)}\cdot\mathscr{M}^{(j_1)}
  \sqcup\cdots\sqcup
  (\mathscr{M}^{(i_n)})^*\cdot\gamma^{(\Delta)}\cdot\mathscr{M}^{(j_n)}
    \right)
    \\
    &= \int_{\mathcal{A}} \dd{\alpha}
    \abs{\braket{\alpha}{\varnothing}}^2
    \tensor[_{\mathrm{r}}]{\braket{
    \xi_\Delta\cdot\mathscr{M}^{(i_1)},\alpha
    }{
    \xi_\Delta\cdot\mathscr{M}^{(j_1)},\alpha
    }}{_{\mathrm{r}}}
    \cdots
    \tensor[_{\mathrm{r}}]{\braket{
    \xi_\Delta\cdot\mathscr{M}^{(i_n)},\alpha
    }{
    \xi_\Delta\cdot\mathscr{M}^{(j_n)},\alpha
    }}{_{\mathrm{r}}}
  \end{split}
  \label{eq:grammatrixmoments}
\end{align}
gives moments of the \(\alpha\)-sector \(\mathcal{H}_\Delta^\alpha\) inner
products
\(\tensor[_{\mathrm{r}}]{\braket{\xi_\Delta\cdot\mathscr{M}^{(i)},\alpha}{\xi_\Delta\cdot\mathscr{M}^{(j)},\alpha}}{_{\mathrm{r}}}\)
in the Hartle-Hawking ensemble. The analysis of ref.~\cite{Abdalla:2025gzn} can
be summarized as calculating the LHS and using the result to deduce the rank of
the Gram matrix
\(\tensor[_{\mathrm{r}}]{\braket{\xi_\Delta\cdot\mathscr{M}^{(i)},\alpha}{\xi_\Delta\cdot\mathscr{M}^{(j)},\alpha}}{_{\mathrm{r}}}\)
for typical \(\alpha\in\mathcal{A}\). From our general analysis in
\cref{sec:cutsectors}, the result that \(\mathcal{H}_\Delta^\alpha\) can have
dimension greater than one is no more surprising than the same statement about
the Hilbert space \(\mathcal{H}_\Sigma^\alpha\) of universes with spatial
boundary \(\Sigma\).

As in the case of spacetimes with spatial boundaries in \cref{sec:spatialbdy},
the bounds \labelcref{eq:tracebound,eq:traceboundavg} can be used to estimate
the size of \(\mathcal{H}_\Sigma^\alpha\) in terms of path integrals. To that
end, we would like to consider a partial source
\(J_{\Delta\otimes{}\Delta^*}^{1/2}\) satisfying the reality condition
\cref{eq:jhalfreal}.

By analogy to \(\mathrm{Cyl}_{\Sigma\otimes{}\Sigma^*}^{\beta/2}\) in
\cref{sec:spatialbdy}, it is natural to consider taking
\(J_{\Delta\otimes\Delta^*}^{1/2}\) to be an interval, denoted
\(I_{\Delta\otimes\Delta^*}\), of an observer worldline running between two
cuts. At the moment, \(I_{\Delta\otimes\Delta^*}\) lacks a precise definition;
rather, it will be defined by the gluing properties to be described below. In
contrast to the canonical description in \cref{sec:spatialbdy} parametrized by a
temperature \(1/\beta\), the microcanonical observer described here has fixed
energy (corresponding to a fixed \(\Delta\)) while the length \(\ell\) of the
worldline should eventually be integrated over with weight \(e^{-\Delta\,\ell}\).
Therefore, it is natural to demand that gluing \(I_{\Delta\otimes\Delta^*}\) to
any partial source
\(J_{\Delta\otimes\cdots}\in\mathcal{J}_{\Delta\otimes\cdots}\) should act
identically \cite{Blommaert:2025bgd},
\begin{align}
  I_{\Delta\otimes\tikzmarknode{C371}{\Delta}^*}
  \otimes
  J_{\tikzmarknode{C372}{\Delta}\otimes\cdots}
  &=J_{\Delta\otimes\cdots}
    \;,
  &
    O(I_{\Delta\otimes\Delta^*})
  &=
    \mathds{1}
    \;.
    \begin{tikzpicture}[overlay, remember picture, every path/.style = {shorten <=1pt,shorten >=1pt}]
      \draw[rounded corners] (C371.south) -- ++ (0,-0.2) -| (C372);
    \end{tikzpicture}
\end{align}
In particular,
\begin{align}
  I_{\Delta\otimes\tikzmarknode{C841}{\Delta}^*}
  \otimes
  I_{\tikzmarknode{C842}{\Delta}\otimes\Delta^*}
  &=
    I_{\Delta\otimes\Delta^*}
    \;.
    \begin{tikzpicture}[overlay, remember picture, every path/.style = {shorten <=1pt,shorten >=1pt}]
      \draw[rounded corners] (C841.south) -- ++ (0,-0.2) -| (C842);
    \end{tikzpicture}
\end{align}

The bound \labelcref{eq:tracebound} and its ensemble average
\labelcref{eq:traceboundavg} in this case read
\begin{align}
  \dim(\mathcal{H}_\Delta^\alpha)
  &\le Z_\alpha\Big(
    I_{\tikzmarknode{C091}{\Delta}\otimes\tikzmarknode{C092}{\Delta}^*}
    \Big)
    \;,
    \label{eq:boundcircobs}
  \\
  \int_{\mathcal{A}} \dd{\alpha}
  \abs{\braket{\alpha}{\varnothing}}^2
  \,
  \dim(\mathcal{H}_\Delta^\alpha)
  &\le \zeta\Big(
    I_{\tikzmarknode{C381}{\Delta}\otimes\tikzmarknode{C382}{\Delta}^*}
    \Big)
    \;.
    \label{eq:avgboundcircobs}
    \begin{tikzpicture}[overlay, remember picture, every path/.style = {shorten <=1pt,shorten >=1pt}]
      \draw[rounded corners] (C091.south) -- ++ (0,-0.2) -| (C092);
      \draw[rounded corners] (C381.south) -- ++ (0,-0.2) -| (C382);
    \end{tikzpicture}
\end{align}
The path integrals on the RHSs involve a source
\(\vphantom{\Big(}I_{\tikzmarknode{C531}{\Delta}\otimes\tikzmarknode{C532}{\Delta}^*}\),
which is yet unspecified. However, to the extent that
\(I_{\Delta\otimes\Delta^*}\) has the interpretation as an interval of the
observer worldline,
\(\vphantom{\Big(}I_{\tikzmarknode{C571}{\Delta}\otimes\tikzmarknode{C572}{\Delta}^*}\)
should describe a circular worldline for the observer. It is natural
to integrate over different circular worldlines, in particular with different
lengths, including some action for the worldline.\footnote{Because the worldline
  no longer connects asymptotic boundary points --- see \cref{foot:geodesic} ---
  it might be natural to integrate over all circular worldlines (geodesic or
  not) perhaps with an action \(e^{-m\,\ell}\) where \(m^2=\Delta(\Delta-1)\) is
  the mass squared of the bulk field dual to \(O^\Delta\).}
\begin{tikzpicture}[overlay, remember picture, every path/.style = {shorten <=1pt,shorten >=1pt}]
  \draw[rounded corners=3] (C531.south) -- ++ (0,-0.15) -| (C532);
  \draw[rounded corners=3] (C571.south) -- ++ (0,-0.15) -| (C572);
\end{tikzpicture}

Indeed, the path integral studied by ref.~\cite{Maldacena:2024spf} for observers
in de Sitter (dS) spacetimes is similar to the kind appearing on the RHS of
\cref{eq:avgboundcircobs}. Ref.~\cite{Maldacena:2024spf} emphasized that, to
recover an interpretation for the path integral as a count of states in dS while
avoiding extraneous phase factors \cite{Polchinski:1988ua}, it is crucial to
include the observer and moreover integrate over the \emph{Lorentzian} length of
the observer worldline.\footnote{See, however, \cref{foot:dssign}.} This means
one should take the integration contour of \(\ell\) to be the imaginary axis.
Conceptually, this integral has the effect of introducing a \(\delta\)-function
imposing the energy constraint associated with time translation
diffeomorphisms.\footnote{See \cite{Blommaert:2025bgd,Nomura:2025whc} for
  further comments along this line. See also \cite{Casali:2021ewu} for a similar
  discussion about the integration contour for lapse, in relation to baby
  universes and worldline field theories.} We will leave it for future work to
precisely define the path integrals, in particular the integration contours, on
the RHSs of \cref{eq:boundcircobs,eq:avgboundcircobs}.

In AdS JT gravity, both sides of the bounds
\labelcref{eq:boundcircobs,eq:avgboundcircobs} in fact diverge
\cite{Abdalla:2025gzn,Blommaert:2025bgd}. Assuming minimal gravitational
coupling to matter and the observer, performing the dilaton path integral
reduces
\(\zeta\Big(I_{\tikzmarknode{C371}{\Delta}\otimes\tikzmarknode{C372}{\Delta}^*}\Big)\)
to an integral over hyperbolic geometries and circular observer worldlines.
There exist (nondegenerate) hyperbolic manifolds with no boundaries only at
genus two and higher, and geometries of genus \(g\) are in general suppressed by
\(\sim e^{S_0(2-2\,g)}\) in the JT path integral. So, one might initially worry
that \cref{eq:avgboundcircobs} predicts a nonsensically small typical dimension
\(\sim{}e^{-2S_0}\) for \(\mathcal{H}_\Delta^\alpha\). In actuality, at \(g=1\),
there is an infinite contribution to the path integral from the degenerate
torus, on which the observer worldline wraps a degenerate cycle. The length of
the remaining nondegenerate circle on the torus is then a free modulus, whose
integration leads to a positive divergence on the RHS of the bound
\labelcref{eq:avgboundcircobs}.
\begin{tikzpicture}[overlay, remember picture, every path/.style = {shorten <=1pt,shorten >=1pt}]
  \draw[rounded corners=3] (C371.south) -- ++ (0,-0.15) -| (C372);
\end{tikzpicture}

On the LHSs of \cref{eq:boundcircobs,eq:avgboundcircobs}, the divergence of
\(\dim(\mathcal{H}_\Delta^\alpha)\) corresponds to the infinite dimension of the
Hilbert space in the JT (double scaled) random matrix theory, a double copy of
which encodes \(\mathcal{H}_\Delta^\alpha\) \cite{Abdalla:2025gzn}. To get a
useful sense of the density of states in the one-observer closed universe
Hilbert space, ref.~\cite{Abdalla:2025gzn} restricts to finite-dimensional
subspaces of \(\mathcal{H}_\Delta^\alpha\) corresponding to microcanonical
windows of the matrix theory. The dimensions of such subspaces are found to scale
like \(\sim{}e^{2S_0}\). We will now show that this result of
ref.~\cite{Abdalla:2025gzn} roughly corresponds to the saturation of a
nontrivial instance of our bound \labelcref{eq:traceboundavg} for appropriately
chosen operators \(O\left(J_{\Delta\otimes{}\Delta^*}^{1/2}\right)\) that
effectively project down to these subspaces.

\begin{figure}
  \centering
  \begin{subfigure}[t]{0.475\textwidth}
    \centering
    \includegraphics[scale=\figscale]{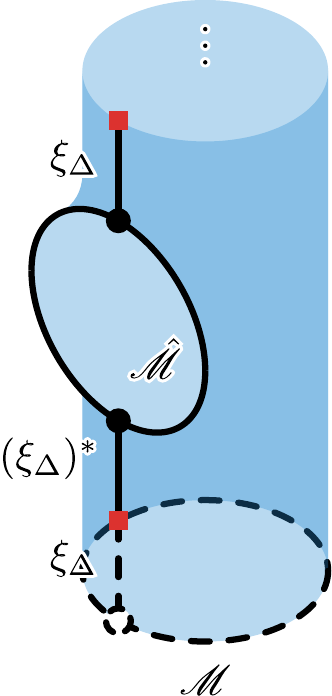}
    \caption{\(
      \vphantom{\biggl(}
      J_{\Delta\otimes{}\tikzmarknode{C131}{\Delta}^*}^{1/2} \otimes J_{\tikzmarknode{C132}{\Delta}}
      \), where \(
      J_\Delta=\xi_\Delta\cdot\mathscr{M}
      \)}
    \label{fig:ximgammam}
  \end{subfigure}
  \hfill
  \begin{subfigure}[t]{0.475\textwidth}
    \centering
    \includegraphics[scale=\figscale]{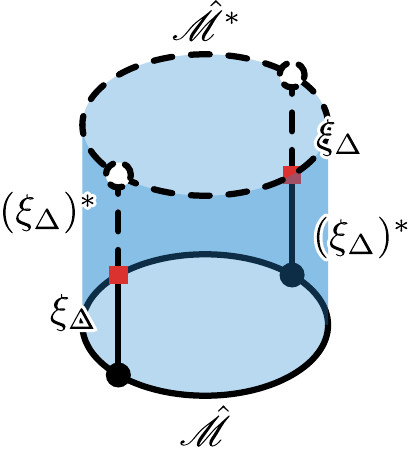}
    \caption{\(
      \vphantom{\biggl(}
      J_{\tikzmarknode{C721}{\Delta}\otimes\tikzmarknode{C722}{\Delta}^*}
      =
      J_{\tikzmarknode{C091}{\Delta}\otimes{}\tikzmarknode{C181}{\Delta}^*}^{1/2} \otimes
      J_{\tikzmarknode{C182}{\Delta}\otimes{}\tikzmarknode{C092}{\Delta}^*}^{1/2}
      \)}
    \label{fig:gammamgammam}
  \end{subfigure}
  \caption{A partial source
    \(J_{\Delta\otimes{}\Delta^*}^{1/2}=\xi_\Delta\cdot\hat{\mathscr{M}}\cdot(\xi_\Delta)^*\)
    (solid black), shown in each panel glued to another partial source (dashed
    black). A (part of a) possible bulk configuration that might appear in the
    path integral involving each of these glued partial sources is illustrated
    in blue.}
  \label{fig:ximxi}
  \begin{tikzpicture}[overlay, remember picture, every path/.style = {shorten <=1pt,shorten >=1pt}]
    \draw[rounded corners=3] (C131.south) -- ++ (0,-0.15) -| (C132);
    \draw[rounded corners=3] (C091.south) -- ++ (0,-0.20) -| (C092);
    \draw[rounded corners=3] (C721.south) -- ++ (0,-0.15) -| (C722);
    \draw[rounded corners=3] (C181.south) -- ++ (0,-0.15) -| (C182);
  \end{tikzpicture}
\end{figure}
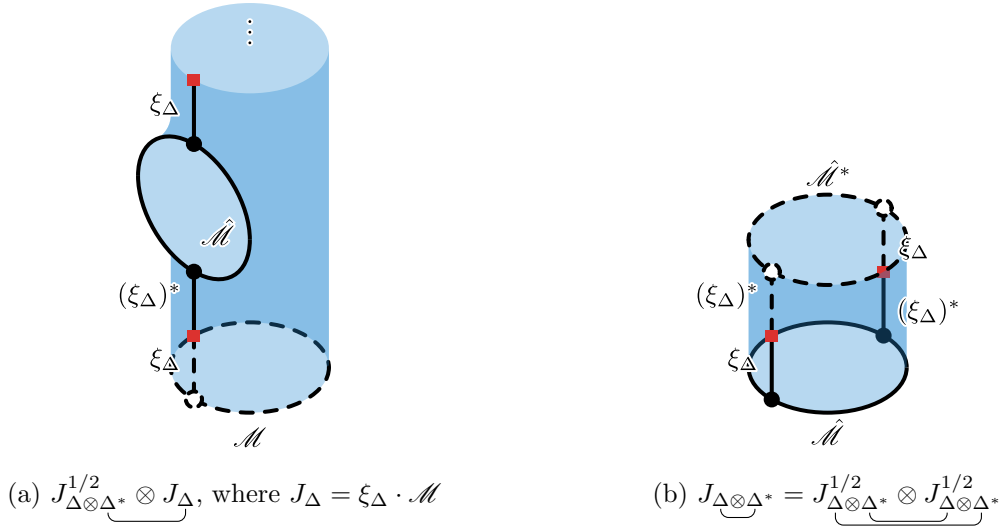

Let us denote the new partial sources we wish to consider by
\begin{align}
  J_{\Delta\otimes{}\Delta^*}^{1/2}
  &=\xi_\Delta\cdot\hat{\mathscr{M}}\cdot(\xi_\Delta)^*
    \in\mathcal{J}_{\Delta\otimes\Delta^*}
    \;,
    \label{eq:jdeltadeltahalf}
  \\
  J_{\Delta\otimes\Delta^*}
  &= J_{\Delta\otimes{}\tikzmarknode{C381}{\Delta}^*}^{1/2} \otimes
    J_{\tikzmarknode{C382}{\Delta}\otimes{}\Delta^*}^{1/2}
    = \xi_\Delta\cdot\hat{\mathscr{M}}
    \cdot\gamma^{(\Delta)}\cdot\hat{\mathscr{M}}\cdot(\xi_\Delta)^*
    \in\mathcal{J}_{\Delta\otimes\Delta^*}
    \;.
    \label{eq:jdeltadeltageo}
    \begin{tikzpicture}[overlay, remember picture, every path/.style = {shorten <=1pt,shorten >=1pt}]
      \draw[rounded corners] (C381.south) -- ++ (0,-0.2) -| (C382);
    \end{tikzpicture}
\end{align}
As illustrated in \cref{fig:ximxi},
\(\xi_\Delta\cdot\hat{\mathscr{M}}\cdot(\xi_\Delta)^*\) consists of two half
worldlines \(\xi_\Delta\) and \((\xi_\Delta)^*\) for the observer, attached to
an asymptotic boundary manifold \(\hat{\mathscr{M}}\). We will require
\(\hat{\mathscr{M}}\) to be real in the sense that
\(J_{\Delta\otimes{}\Delta^*}^{1/2}\) satisfies \cref{eq:jhalfreal}. The rules
for gluing the half worldlines in this partial source to other half worldlines
are what one expects from generalizing \cref{eq:xigluing} in the obvious way.
For example, \cref{fig:ximgammam} illustrates the gluing relevant
for\footnote{Note that, by including partial sources like the kind in the RHS of
  \cref{eq:ximxiop}, we are including in \(\mathcal{J}_\Delta\) partial sources
  which are not exactly of the kind \(\xi_\Delta\cdot\mathscr{M}^{(j)}\) in the
  Gram matrices appearing in \cref{eq:grammatrixmoments} and studied by
  ref.~\cite{Abdalla:2025gzn}. In principle, the Hilbert space
  \(\tilde{\mathcal{H}}_\Delta^\alpha\) constructed from the submodule
  \(\tilde{\mathcal{J}}_\Delta\), generated only by partial sources of the kind
  \(\xi_\Delta\cdot\mathscr{M}^{(j)}\), might be a proper subspace of the
  Hilbert space \(\mathcal{H}_\Delta^\alpha\), generated by the enlarged set
  \(\mathcal{J}_\Delta\) of partial sources. However, we do not expect the
  mismatch, if any, to be significant at the level of precision of this
  discussion.}
\begin{align}
  O\left(
  \xi_\Delta\cdot\hat{\mathscr{M}}\cdot(\xi_\Delta)^*
  \right)
  \ket{
  \xi_\Delta\cdot\mathscr{M}
  , \alpha
  }
  &=
    \ket{
  \xi_\Delta\cdot\hat{\mathscr{M}}\cdot\gamma^{(\Delta)}\cdot\mathscr{M}
    , \alpha
    }
    \label{eq:ximxiop}
\end{align}
while \cref{fig:gammamgammam} illustrates the complete source
\begin{align}
  J_{\tikzmarknode{C241}{\Delta}\otimes\tikzmarknode{C242}{\Delta}^*}
  &= J_{\tikzmarknode{C831}{\Delta}\otimes{}\tikzmarknode{C281}{\Delta}^*}^{1/2} \otimes
    J_{\tikzmarknode{C282}{\Delta}\otimes{}\tikzmarknode{C832}{\Delta}^*}^{1/2}
    =
    \vphantom{\bigg(}
    \xi_{\tikzmarknode{C291}{\Delta}}\cdot\hat{\mathscr{M}}\cdot\gamma^{(\Delta)}
    \cdot\hat{\mathscr{M}}\cdot(\xi_{\tikzmarknode{C292}{\Delta}})^*
    \begin{tikzpicture}[overlay, remember picture, every path/.style = {shorten <=1pt,shorten >=1pt}]
      \draw[rounded corners] (C241.south) -- ++ (0,-0.2) -| (C242);
      \draw[rounded corners] (C831.south) -- ++ (0,-0.3) -| (C832);
      \draw[rounded corners] (C281.south) -- ++ (0,-0.2) -| (C282);
      \draw[rounded corners] (C291.south) -- ++ (0,-0.3) -| (C292);
    \end{tikzpicture}
\end{align}
for the path integrals appearing in \cref{eq:tracebound,eq:traceboundavg}.

If the boundary manifold \(\hat{\mathscr{M}}\) is a circle with some fixed
lengths \(\beta_1,\beta_2>0\) for its two pieces (demarked by the black dots in
\cref{fig:ximxi}), then
\(O\left(\xi_\Delta\cdot\hat{\mathscr{M}}\cdot(\xi_\Delta)^*\right)\) will not
project states of \(\mathcal{H}_\Delta^\alpha\) to a subspace corresponding to
a microcanonical energy window of the JT matrix theory (but larger energies will
be suppressed). If \(\hat{\mathscr{M}}\) further includes inverse Laplace
transformations from \(\beta_1\) and \(\beta_2\) to a variable \(E\), then
\(O\left(\xi_\Delta\cdot\hat{\mathscr{M}}\cdot(\xi_\Delta)^*\right)\) will have
support only on the subspace of \(\mathcal{H}_\Delta^\alpha\) corresponding to
the energy \(E\) eigenspace of the matrix theory. One can further take
superpositions over small intervals of \(E\) to obtain microcanonical windows of
the matrix theory, as considered by ref.~\cite{Abdalla:2025gzn}.

To use \cref{eq:traceboundavg} to roughly bound the typical number of states in
such a subspace of \(\mathcal{H}_\Delta^\alpha\), we make the following three
observations. Firstly, the topological suppression of bulk spacetimes of genus
\(g\) with \(n\) boundaries goes like \(\sim e^{S_0(2-2g-n)}\) in JT gravity.
Secondly, gluing \(J_{\Delta\otimes\Delta^*}^{1/2}\) given in
\cref{eq:jdeltadeltahalf} to a partial source \(J_\Delta\in\mathcal{J}_\Delta\),
viewed as preparing a state, effectively introduces an extra asymptotic boundary
to the component of the bulk connected to \(J_\Delta\). This is illustrated in
\cref{fig:ximgammam}. Therefore, we expect typical eigenvalues of
\(O\left(J_{\Delta\otimes\Delta^*}\right)\), in the sense of dominating the LHS
of \cref{eq:traceboundavg}, to be of size \(\sim (e^{-S_0})^2\). Thirdly, the
path integral
\(\zeta\Big(J_{\tikzmarknode{C341}{C}\otimes\tikzmarknode{C342}{C}^*}\Big)\)
\begin{tikzpicture}[overlay, remember picture, every path/.style = {shorten <=1pt,shorten >=1pt}]
  \draw[rounded corners=3] (C341.south) -- ++ (0,-0.15) -| (C342);
\end{tikzpicture}
has a leading order contribution, in the topological expansion, whose
connected\footnote{To obtain a normalized average over \(\alpha\)-sectors, we
  should divide \cref{eq:traceboundavg} by
  \(\braket{\varnothing}{\varnothing}=\zeta(\varnothing)\). On the RHS, this
  divides out the multiplicative contribution from components of the bulk
  disconnected from the source.\label{foot:dividehh}} bulk component is the
cylinder illustrated in \cref{fig:gammamgammam}. The contribution of this
cylinder is \(\sim(e^{S_0})^0\). Altogether, it follows that the
ensemble-average number of eigenstates of
\(O\left(J_{\Delta\otimes\Delta^*}\right)\), in the window where eigenvalues are
of the typical size \(\sim{}e^{-2 S_0}\) in \cref{eq:traceboundavg}, is bounded
by \(\sim{}e^{2S_0}\). This is consistent with the finding of
ref.~\cite{Abdalla:2025gzn} that the dimension of the closed universe Hilbert
space with one observer is \(\sim e^{2S_0}\), once a restriction is made to a
microcanonical window of the matrix theory.

In the above discussion, for simplicity, we have described observers with one
internal state. In fact, refs.~\cite{Maldacena:2024spf,Abdalla:2025gzn} consider
observers with some number \(d\) of orthogonal internal states \(\sigma\), with
different energies corresponding to different values of \(\Delta\), such that a
``clock'' can be constructed from these states. It is straightforward to
generalize to this case by taking a direct sum of the above discussion. That is,
the sector \(\mathcal{H}_\sigma^\alpha\) of the gravitational Hilbert space
associated with each internal state \(\sigma\) of an observer is just one
instance of the \(\mathcal{H}_\Delta^\alpha\) described above. One may say that
the Hilbert space \(\mathcal{H}_\sigma^\alpha\) is associated with a cut that is
labelled by the observer internal state \(\sigma\), which in particular
determines the value of \(\Delta=\Delta_\sigma\). The direct sum
\(\mathcal{H}_{\mathrm{obs}}=\bigoplus_{\sigma=1}^d \mathcal{H}_\sigma^\alpha\)
then gives the gravitational Hilbert space of closed universes with one
observer.\footnote{Subspaces of this Hilbert space, corresponding to
  microcanonical windows of the matrix theory, are then each
  \(\sim{}d\,e^{2S_0}\)-dimensional.} (Alternatively, one could have also
arrived at the Hilbert space \(\mathcal{H}_{\mathrm{obs}}\) by considering a cut
which allows arbitrary observer internal states to begin with and whose gluing
rules involve an inner product of the observer's internal Hilbert space --- our
approach in \cref{sec:obsclone} will more closely resemble this construction.)
The Hilbert space \(\mathcal{H}_{\mathrm{obs}}\) will support operators that can
act nontrivially on the observer internal state, \eg{} allowing interactions
with their environment to influence the observer's internal state and
allowing dressing to the observer's clock.

\subsubsection{External observer clones}
\label{sec:obsclone}

Next, let us see how our framework applies to another prescription, by Harlow,
Usatyuk, and Zhao \cite{Harlow:2025pvj}, for evaluating the gravitational path
integral in the presence of observers. Their proposal is to treat observers just
like any other bulk matter excitation, except that the source preparing the
state of the observer is ``entangled'' with a ``clone'' of the observer,
modelled as an external non-gravitational system.

\begin{figure}
  \centering
  \begin{subfigure}[t]{0.475\textwidth}
    \centering
    \includegraphics[scale=\figscale]{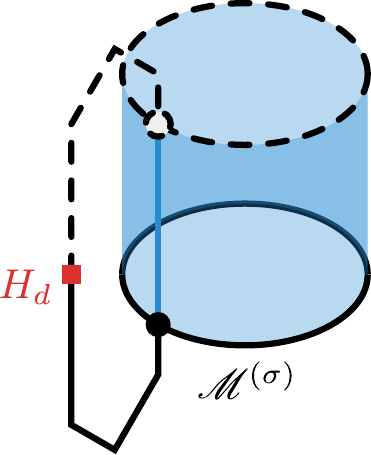}
    \caption{A partial source
      \(\tilde{J}_d=\sum_{\sigma=1}^d\mathscr{M}^{(\sigma)}\ket{\sigma}\in\tilde{\mathcal{J}}_d\)
      where \(\mathscr{M}^{(\sigma)}\) is required to excite an
      observer in state \(\sigma\).}
    \label{fig:jtilde}
  \end{subfigure}
  \hfill
  \begin{subfigure}[t]{0.475\textwidth}
    \centering
    \includegraphics[scale=\figscale]{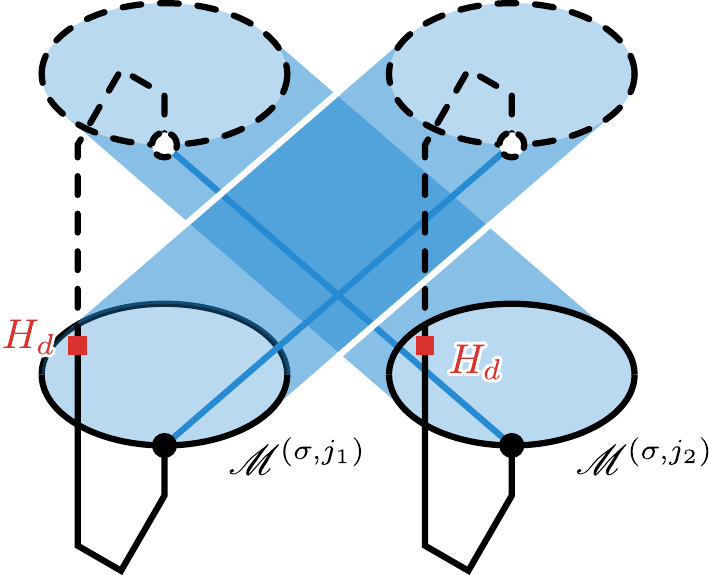}
    \caption{\(\tilde{J}_d^{(j_1)}\otimes\tilde{J}_d^{(j_2)}\) for
      \(\tilde{J}_d^{(j_1)}=\sum_{\sigma=1}^d\mathscr{M}^{(\sigma,j_1)}\ket{\sigma}\in\tilde{\mathcal{J}}_d\)
      and similarly \(\tilde{J}_d^{(j_2)}\in\tilde{\mathcal{J}}_d\).}
    \label{fig:jtildejtilde}
  \end{subfigure}
  \caption{Partial sources \(\tilde{J}_d\in\tilde{\mathcal{J}}_d\) and
    \(\tilde{J}_d^{(j_1)}\otimes\tilde{J}_d^{(j_2)}\in\tilde{\mathcal{J}}_d\otimes\tilde{\mathcal{J}}_d\)
    (solid black). We also illustrate \((\tilde{J}_d'|\tilde{J}_d),
    (\tilde{J}_d^{(i_1)}\otimes\tilde{J}_d^{(i_2)}|\tilde{J}_d^{(j_1)}\otimes\tilde{J}_d^{(j_2)})\in\mathcal{J}\)
    for some
    \(\tilde{J}_d',\tilde{J}_d^{(i_1)},\tilde{J}_d^{(i_2)}\in\tilde{\mathcal{J}}_d\)
    (dashed black).
    A possible bulk configuration that might appear in the path integral
    \(\zeta\) evaluated with each complete source is illustrated in blue.}
  \label{fig:cloneobs}
\end{figure}

Here, the set \(\mathcal{J}\) of complete sources for the path integral is
generated just by boundary manifolds \(\mathscr{M}\) preparing states of closed
universes. To introduce the observer and its clone, ref.~\cite{Harlow:2025pvj}
considers, in our language, a set \(\tilde{\mathcal{J}}_d\) of partial sources
which is a submodule of the \(\mathcal{J}_d\) introduced in \cref{eq:jd}. In
particular,
\begin{align}
  \tilde{\mathcal{J}}_d
  =\spn\left\{
  \sum_{\sigma=1}^d \mathscr{M}^{(\sigma)} \ket{\sigma}
  :
  \text{boundary manifold \(\mathscr{M}^{(\sigma)} \in \mathcal{J}\) excites observer in state \(\sigma\)}
  \right\}
  \label{eq:tildejd}
\end{align}
where \(\mathscr{M}^{(\sigma)}\) is a boundary manifold preparing an arbitrary
state of closed universes\footnote{Ref.~\cite{Harlow:2025pvj} focuses on cases
  where \(\mathscr{M}^{(\sigma)}\) is a connected boundary manifold preparing
  one closed universe, containing one observer in state \(\sigma\) and one
  other, arbitrary excitation of worldline matter. Firstly, allowing
  \(\mathscr{M}^{(\sigma)}\) to be disconnected does not alter the
  \(\alpha\)-sectors \(\tilde{\mathcal{H}}_d^\alpha\) constructed from
  \(\tilde{\mathcal{J}}_d\), because the gravitational path integral in
  \(\alpha\)-sectors factorizes between disconnected sources. Secondly, at least
  in the two-dimensional models which were the focus of
  ref.~\cite{Harlow:2025pvj}'s calculations, we do not expect the restriction to
  having one other worldline matter excitation to give any significant reduction
  to \(\tilde{\mathcal{H}}_d^\alpha\) (as long as there are enough possible
  states for the matter excitation).}, including excitations of at least some
matter constituting one observer in an internal state labelled by \(\sigma\). In
the two-dimensional topological \cite{Marolf:2020xie} and JT models for which
ref.~\cite{Harlow:2025pvj} performed explicit calculations, the observer is
simply modelled as a matter worldline of flavour \(\sigma\). The superposition
in \cref{eq:tildejd} indicates the sense in which the internal state of this
bulk observer is entangled with the state \(\ket{\sigma}\) of an external clone.
Because \(\tilde{\mathcal{J}}_d\) is a submodule of \(\mathcal{J}_d\), it
naturally inherits the gluing operation \labelcref{eq:jdgluing} and Hermitian
form \labelcref{eq:jdhermform}.\footnote{From here on, we will consider a basis
  of states for the external quantum mechanical system which is orthonormal
  \labelcref{eq:orthonormal}. A comment on notation: we will use
  \(\ket{\sigma},\ket{\sigma'},\ldots\) and
  \(\ket{\varsigma},\ket{\varsigma'},\ldots\) to denote elements of the same
  orthonormal basis of states in the external quantum mechanical Hilbert space
  \(H_d\); the use of \(\ket{\sigma},\ket{\sigma'},\ldots\) will be reserved for
  situations where this external system is a clone, \ie{} maximally
  ``entangled'' with the internal state, of a bulk observer.} In
\cref{fig:jtilde}, we illustrate a partial source
\(\tilde{J}_d\in\tilde{\mathcal{J}}_d\) and its gluing to another partial source
\((\tilde{J}_d')^*\in\tilde{\mathcal{J}}_{d^*}\).

From \(\tilde{\mathcal{J}}_d\), we can define \(\alpha\)-sector Hilbert spaces
\(\tilde{\mathcal{H}}_d^\alpha\) as described in \cref{sec:cutsectors}.
Analogous to \cref{eq:grammatrixmoments}, a path integral with insertions of
\(\tilde{J}_d^{(j_1)},\ldots,\tilde{J}_d^{(j_n)}\in\tilde{\mathcal{J}}_d\) glued
to the \(*\)-conjugates of
\(\tilde{J}_d^{(i_1)},\ldots,\tilde{J}_d^{(i_n)}\in\tilde{\mathcal{J}}_d\),
\begin{align}
  \begin{split}
    \MoveEqLeft\zeta\left(
    (\tilde{J}_d^{(i_1)}|\tilde{J}_d^{(j_1)})
  \sqcup\cdots\sqcup
    (\tilde{J}_d^{(i_n)}|\tilde{J}_d^{(j_n)})
    \right)
    \\
    &= \int_{\mathcal{A}} \dd{\alpha}
    \abs{\braket{\alpha}{\varnothing}}^2
    \tensor[_{\mathrm{r}}]{\braket{
    \tilde{J}_d^{(i_1)},\alpha
    }{
    \tilde{J}_d^{(j_1)},\alpha
    }}{_{\mathrm{r}}}
    \cdots
    \tensor[_{\mathrm{r}}]{\braket{
    \tilde{J}_d^{(i_n)},\alpha
    }{
    \tilde{J}_d^{(j_n)},\alpha
    }}{_{\mathrm{r}}}
      \;,
  \end{split}
  \label{eq:clonegrammom}
\end{align}
gives moments of the \(\alpha\)-sector \(\tilde{\mathcal{H}}_d^\alpha\) inner
products
\(\tensor[_{\mathrm{r}}]{\braket{\tilde{J}_d^{(i)},\alpha}{\tilde{J}_d^{(j)},\alpha}}{_{\mathrm{r}}}\)
in the Hartle-Hawking ensemble. Note that, unlike the prescription of
ref.~\cite{Abdalla:2025gzn} described in \cref{sec:obsworldline}, the bulk
observer worldline is not prescribed to connect certain bra- pieces of the
boundary manifold to certain ket- pieces --- see \cref{fig:jtildejtilde} for an
example of a bulk configuration that can contribute to the above path integral
for \(n=2\).\footnote{If an anti-observer state \(\sigma^*\) is also an observer
  state, then there is another bulk configuration in which the bra- and ket-
  boundary manifolds are disconnected through the bulk. The observers in
  this bulk topology annihilate and do not even survive for all time between the
  boundary manifolds preparing bra- and ket-states.}

Evaluating path integrals like \cref{eq:clonegrammom} in a two-dimensional model
of topological gravity \cite{Marolf:2020xie} and in AdS JT gravity, the analysis
of ref.~\cite{Harlow:2025pvj} can be understood as approximating the typical
rank of the Gram matrix
\(\tensor[_{\mathrm{r}}]{\braket{\tilde{J}_d^{(i)},\alpha}{\tilde{J}_d^{(j)},\alpha}}{_{\mathrm{r}}}\).
Their result, that \(\tilde{\mathcal{H}}_d^\alpha\) can have dimension greater
than one, is again unsurprising from our general analysis in
\cref{sec:cutsectors}. Again, we can use our general bound
\labelcref{eq:tracebound} and its ensemble average \labelcref{eq:traceboundavg}
to roughly constrain the size of \(\tilde{\mathcal{H}}_d^\alpha\), finding
agreement below with the quantitative results of ref.~\cite{Harlow:2025pvj}.

In preparation, let us first argue that
\begin{align}
  \dim \tilde{\mathcal{H}}_d^\alpha
  \le \dim \mathcal{H}_d^\alpha
  = d
  \;,
  \label{eq:cloneineq1}
\end{align}
where \(\mathcal{H}_d^\alpha\) is the \(\alpha\)-sector Hilbert space
constructed from the set \(\mathcal{J}_d\) of partial sources defined in
\cref{eq:jd}. The coefficients \(J^{(\varsigma)}\) in \cref{eq:jd} are arbitrary
elements of the set \(\mathcal{J}\) of sources, so \(\tilde{\mathcal{J}}_d\)
defined in \cref{eq:tildejd} is clearly a submodule of \(\mathcal{J}_d\). It
then follows that \(\tilde{\mathcal{H}}_d^\alpha\) is a subspace of
\(\mathcal{H}_d^\alpha\), hence the inequality in \cref{eq:cloneineq1}. Next, we
observe that \(\mathcal{H}_d^\alpha\) is isometric to the \(d\)-dimensional
Hilbert space \(H_d\) of the external quantum mechanical system:
\begin{align}
  \mathcal{H}_d^\alpha
  &\leftrightarrow
    H_d
    \;,
  &
  \ket{\sum_{\varsigma=1}^d
  J^{(\varsigma)} \ket{\varsigma},
  \alpha}_{\mathrm{r}}
  &\leftrightarrow
    \sum_{\varsigma=1}^d
    Z_\alpha(J^{(\varsigma)}) \ket{\varsigma}
    \;.
    \label{eq:hdalphatohd}
\end{align}

For a closer analogy to our discussion in \cref{sec:obsworldline}, however, let
us mention a more round-about derivation of the last equality in
\cref{eq:cloneineq1}. Let us consider the two-sided partial source\footnote{We
  have not given a precise definition for \(\mathcal{J}_{d\otimes d^*}\) as it
  is not crucial to our discussion. However, a natural definition would be
  \begin{align}
    \mathcal{J}_{d\otimes d^*}
    &=
      \left\{
  \sum_{\varsigma,\varsigma'=1}^d J^{(\varsigma,\varsigma')} \ketbra{\varsigma}{\varsigma'}
  : J^{(\varsigma,\varsigma')} \in \mathcal{J}
      \right\}
      = \mathcal{J}_d \otimes \mathcal{J}_{d^*}
      \;.
  \end{align}
  \label{foot:jdd}}
\begin{align}
  I_{d\otimes d^*}
  &= \sum_{\varsigma=1}^d \ketbra{\varsigma}{\varsigma} \in \mathcal{J}_{d\otimes d^*}
    \;.
    \label{eq:cloneobsid}
\end{align}
This is the identity operator on \(H_d\) and the corresponding
\(O(I_{d\otimes{}d^*})\), defined by \cref{eq:ooperator}, is the identity
operator on \(\mathcal{H}_d^\alpha\). The condition
\labelcref{eq:equivstates} for the saturation of the bound
\labelcref{eq:tracebound} is manifestly satisfied by the expression
\labelcref{eq:cloneobsid} for \(I_{d\otimes d^*}=I_{d\otimes d^*}^{1/2}\). The
saturated bound then gives
\begin{align}
  \dim \mathcal{H}_d^\alpha
    &= Z_\alpha\Big(
    I_{\tikzmarknode{C281}{d}\otimes\tikzmarknode{C282}{d}^*}
      \Big)
      = d\, Z_\alpha(\varnothing)
      = d
    \;,
    \begin{tikzpicture}[overlay, remember picture, every path/.style = {shorten <=1pt,shorten >=1pt}]
      \draw[rounded corners] (C281.south) -- ++ (0,-0.2) -| (C282);
    \end{tikzpicture}
\end{align}
as desired.

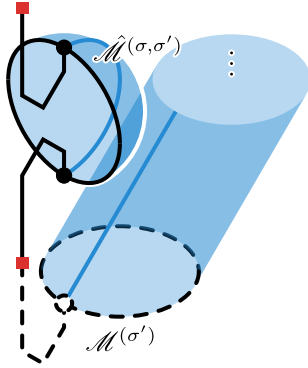
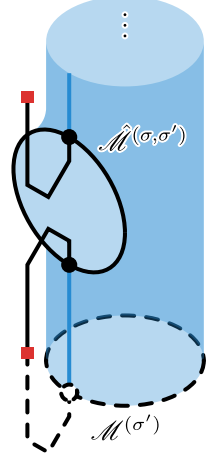
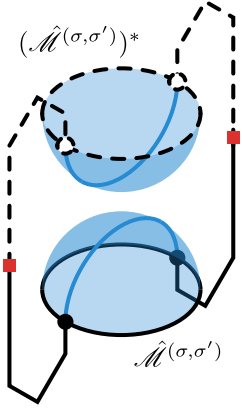
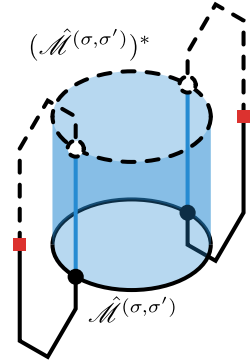
\begin{figure}
  \centering
  \begin{subfigure}[t]{0.475\textwidth}
    \centering
    \includegraphics[scale=\figscale]{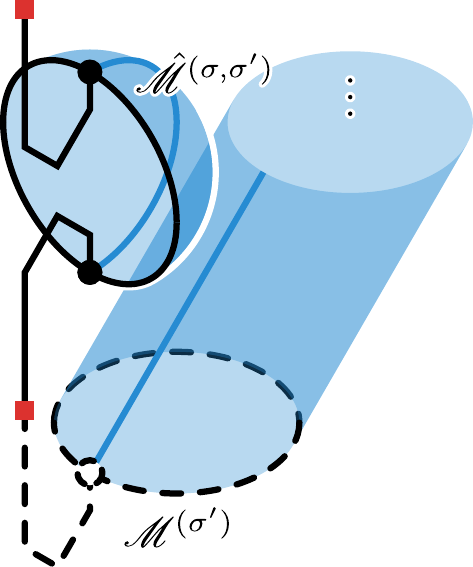}
    \caption{Part of a bulk topology in a path integral involving
      \(\vphantom{\biggl(}\tilde{J}_{d\otimes{}\tikzmarknode{C431}{d}^*}^{1/2}
      \otimes\tilde{J}_{\tikzmarknode{C432}{d}} =\sum_{\sigma,\sigma'=1}^d
      \hat{\mathscr{M}}^{(\sigma,\sigma')}\sqcup \mathscr{M}^{(\sigma')}
      \ket{\sigma}\) where
      \(\tilde{J}_d=\sum_{\sigma'=1}^d \mathscr{M}^{(\sigma')}\ket{\sigma'}\).
      When \(e^{2S_0}\gg d\), bulk topologies of this kind dominate the
      path integral.}
    \label{fig:jddhalfjddiscon}
  \end{subfigure}
  \hfill
  \begin{subfigure}[t]{0.475\textwidth}
    \centering
    \includegraphics[scale=\figscale]{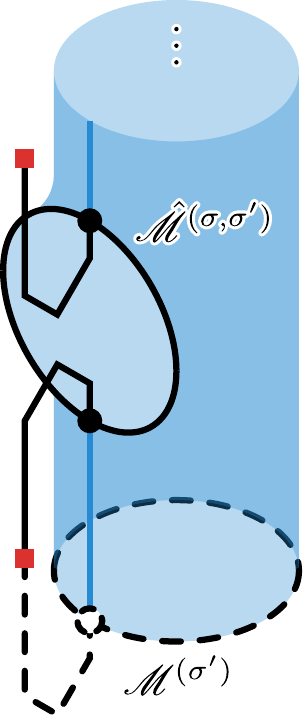}
    \caption{Part of another bulk topology in the path integral involving
      \(\vphantom{\biggl(}\tilde{J}_{d\otimes{}\tikzmarknode{C391}{d}^*}^{1/2}
      \otimes\tilde{J}_{\tikzmarknode{C392}{d}}\). When \(d\gg e^{2S_0}\gg 1\), bulk
      topologies connecting the boundary manifolds
      \(\hat{\mathscr{M}}^{(\sigma,\sigma')}\) and
      \(\hat{\mathscr{M}}^{(\sigma')}\) like this dominate the path integral.}
    \label{fig:jddhalfjdcon}
  \end{subfigure}
  \par\bigskip
  \begin{subfigure}[t]{0.475\textwidth}
    \centering
    \includegraphics[scale=\figscale]{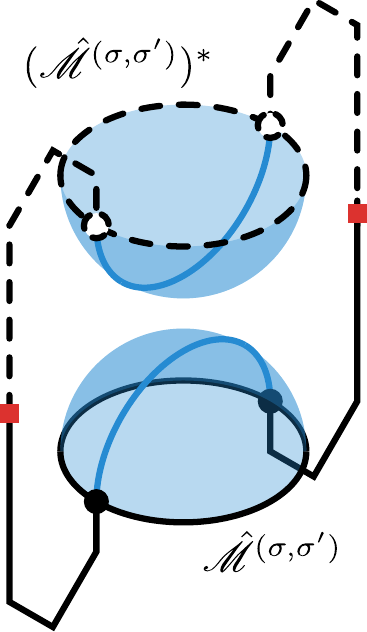}
    \caption{A bulk topology in the path integral with the source \(
      \vphantom{\bigg(}
      \tilde{J}_{\tikzmarknode{C291}{d}\otimes\tikzmarknode{C292}{d}^*}
      =\tilde{J}_{\tikzmarknode{C561}{d}\otimes{}\tikzmarknode{C901}{d}^*}^{1/2} \otimes
      \tilde{J}_{\tikzmarknode{C902}{d}\otimes{}\tikzmarknode{C562}{d}^*}^{1/2}
      = \sum_{\sigma,\sigma'=1}^d
      \hat{\mathscr{M}}^{(\sigma,\sigma')}
      \sqcup \hat{\mathscr{M}}^{(\sigma',\sigma)}
      \). When \(e^{2S_0}\gg d\), this bulk topology dominates the
      path integral.}
    \label{fig:jdddiscon}
  \end{subfigure}
  \hfill
  \begin{subfigure}[t]{0.475\textwidth}
    \centering
    \includegraphics[scale=\figscale]{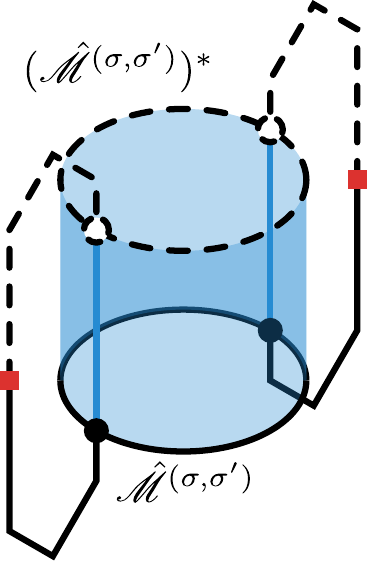}
    \caption{Another bulk topology for the path integral with source
      \(\vphantom{\biggl(}
      \tilde{J}_{\tikzmarknode{C711}{d}\otimes\tikzmarknode{C712}{d}^*}\). When
      \(d\gg e^{2S_0} \gg 1\), this bulk topology dominates the path integral.}
    \label{fig:jddcon}
  \end{subfigure}
  \caption{A partial source \(\tilde{J}_{d\otimes d^*}^{1/2} =
    \sum_{\sigma,\sigma'=1}^d \hat{\mathscr{M}}^{(\sigma,\sigma')}
    \ketbra{\sigma}{\sigma'}\) (solid black), shown in each panel glued to
    another partial source (dashed black). A (part of a) possible bulk topology
    that might appear in the path integral involving each of these glued partial
    sources is illustrated in blue.}
  \label{fig:jddhalfglue}
  \begin{tikzpicture}[overlay, remember picture, every path/.style = {shorten <=1pt,shorten >=1pt}]
    \draw[rounded corners=3] (C431.south) -- ++ (0,-0.15) -| (C432);
    \draw[rounded corners=3] (C391.south) -- ++ (0,-0.15) -| (C392);
    \draw[rounded corners=3] (C291.south) -- ++ (0,-0.15) -| (C292);
    \draw[rounded corners=3] (C561.south) -- ++ (0,-0.20) -| (C562);
    \draw[rounded corners=3] (C901.south) -- ++ (0,-0.15) -| (C902);
    \draw[rounded corners=3] (C711.south) -- ++ (0,-0.15) -| (C712);
  \end{tikzpicture}
\end{figure}

We saw in \cref{sec:obsworldline} that, by considering another two-sided partial
source \labelcref{eq:jdeltadeltageo}, the ensemble-averaged bound
\labelcref{eq:traceboundavg} gave a useful rough bound on the typical size of
\(\mathcal{H}_\Delta^\alpha\) (or rather, each subspace corresponding to a
microcanonical window of the JT matrix theory). Let us now demonstrate that a
completely analogous construction leads to a rough bound on
\(\tilde{\mathcal{H}}_d^\alpha\) (or an appropriate subspace in JT) which is
sometimes stronger than \cref{eq:cloneineq1} and roughly saturated by the
results of ref.~\cite{Harlow:2015lma}. The analogous partial sources in the
present context would be\footnote{Analogous to \cref{foot:jdd}, we can define
  \begin{align}
    \begin{split}
    \tilde{\mathcal{J}}_{d\otimes d^*}
      = \spn\Bigg\{
      \sum_{\sigma,\sigma'=1}^d \mathscr{M}^{(\sigma,\sigma')} \ketbra{\sigma}{\sigma'}
      &:\text{boundary manifold \(\mathscr{M}^{(\sigma,\sigma')} \in \mathcal{J}\) excites}\\
      &\phantom{{}:{}}\text{observer and anti-observer in states \(\sigma\) and \((\sigma')^*\)}
      \Bigg\}
      \;.
    \end{split}
  \end{align}
  Note that \(\tilde{\mathcal{J}}_d \otimes \tilde{\mathcal{J}}_{d^*}\) is a
  proper submodule of \(\tilde{\mathcal{J}}_{d\otimes d^*}\).}
\begin{align}
  \tilde{J}_{d\otimes{}d^*}^{1/2}
  &= \sum_{\sigma,\sigma'=1}^d
    \hat{\mathscr{M}}^{(\sigma,\sigma')} \ketbra{\sigma}{\sigma'}
    \in\tilde{\mathcal{J}}_{d\otimes{} d^*}
    \;,
    \label{eq:jddhalf}
  \\
  \tilde{J}_{d\otimes d^*}
  &= \tilde{J}_{d\otimes{}\tikzmarknode{C381}{d}^*}^{1/2} \otimes
    \tilde{J}_{\tikzmarknode{C382}{d}\otimes{}d^*}^{1/2}
    = \sum_{\sigma,\sigma',\sigma''=1}^d
    \hat{\mathscr{M}}^{(\sigma,\sigma')}\sqcup \hat{\mathscr{M}}^{(\sigma',\sigma'')}
    \ketbra{\sigma}{\sigma''}
    \in\tilde{\mathcal{J}}_{d\otimes{} d^*}
    \label{eq:jdd}
    \begin{tikzpicture}[overlay, remember picture, every path/.style = {shorten <=1pt,shorten >=1pt}]
      \draw[rounded corners] (C381.south) -- ++ (0,-0.2) -| (C382);
    \end{tikzpicture}
\end{align}
where \(\hat{\mathscr{M}}^{(\sigma,\sigma')}\in\mathcal{J}\) is a boundary
manifold which excites at least an observer and an anti-observer in states
\(\sigma\) and \((\sigma')^*\) respectively --- see \cref{fig:jddhalfglue}.
(Exciting an anti-observer in state \((\sigma')^*\) is equivalent to absorbing
an observer in state \(\sigma'\).) To ensure that
\(\tilde{J}_{d\otimes{}d^*}^{1/2}\) satisfies the reality condition
\labelcref{eq:jhalfreal}, we will further require
\(\hat{\mathscr{M}}^{(\sigma,\sigma')}\) to satisfy
\begin{align}
  \left(\hat{\mathscr{M}}^{(\sigma,\sigma')}\right)^*
  &= \hat{\mathscr{M}}^{(\sigma',\sigma)}
    \;.
\end{align}

The reason why the trace of \(O(\tilde{J}_{d\otimes d^*})\) is better suited for
estimating the size of \(\tilde{\mathcal{H}}_d^\alpha\) is because, unlike the
identity operator \(O(I_{d\otimes{}d^*})\) on \(\mathcal{H}_d^\alpha\), the
operator \(O(\tilde{J}_{d\otimes d^*})\) effectively projects
\(\mathcal{H}_d^\alpha\to\tilde{\mathcal{H}}_d^\alpha\). To use
\cref{eq:traceboundavg} to roughly bound the number of states in
\(\tilde{\mathcal{H}}_d^\alpha\) (or an appropriate subspace in JT gravity), we
make the following three observations. Firstly, the topological suppression of
bulk spacetimes of genus \(g\) with \(n\) boundaries goes like
\(\sim{}e^{S_0\,(2-2g-n) + S_\partial\,n}\) where \(S_\partial=0\) in JT gravity
and \(S_\partial=S_0\) in the two-dimensional topological model of gravity
\cite{Marolf:2020xie}.

Secondly, let us consider the gluing of \(\tilde{J}_{d\otimes{}d^*}^{1/2}\)
given in \cref{eq:jddhalf} to a partial source
\(\tilde{J}_d\in\tilde{\mathcal{J}}_d\) preparing a state, \eg{} as illustrated
in \cref{fig:jddhalfjddiscon,fig:jddhalfjdcon}. If \(e^{S_0}\gg d\), then due to
the strong topological suppression, we expect the leading order effect of
attaching \(\tilde{J}_{d\otimes{}d^*}^{1/2}\) to be the introduction of a
disconnected disk with boundary
\(\hat{\mathscr{M}}^{(\sigma,\sigma')}\) in the path integral, as illustrated in
\cref{fig:jddhalfjddiscon}. Therefore, we expect typical eigenvalues of
\(O(\tilde{J}_{d\otimes{}d^*})\), again in the sense of dominating the LHS of
\cref{eq:traceboundavg}, to be of size
\begin{align}
  &\sim \left(e^{S_0+S_\partial}\right)^2
  &
  &\text{if \(e^{2S_0}\gg d\)}
    \;.
    \label{eq:jddsize1}
\end{align}
On the other hand, if \(d\gg{}e^{S_0}\gg{}1\), then topologies which admit
observer/clone loops become more favourable, with each loop contributing a
factor of \(\dim{H_d}=d\) --- see \cref{fig:jddhalfjdcon}. Therefore, we expect
typical eigenvalues of \(O(\tilde{J}_{d\otimes{}d^*})\) to be of size
\begin{align}
  &\sim \left(d\,e^{-S_0+S_\partial}\right)^2
  &
  &\text{if \(d \gg e^{2S_0}\gg 1\)}
    \;.
    \label{eq:jddsize2}
\end{align}

Thirdly, let us consider the path integral
\(\zeta\Big(\tilde{J}_{\tikzmarknode{C861}{d}\otimes\tikzmarknode{C862}{d}^*}\Big)\)
appearing on the RHS of \cref{eq:traceboundavg}.
\begin{tikzpicture}[overlay, remember picture, every path/.style = {shorten <=1pt,shorten >=1pt}]
  \draw[rounded corners=3] (C861.south) -- ++ (0,-0.15) -| (C862);
\end{tikzpicture}
By similar reasoning as in the
previous paragraph, after dividing out bulk components disconnected from the
source\footnote{See \cref{foot:dividehh}.}, the dominant contribution to this
path integral comes from two disks
\begin{align}
  &\sim d\, (e^{S_0+S_\partial})^2
  &
  &\text{if \(e^{2S_0}\gg d\)},
    \label{eq:zetajdd1}
\end{align}
as illustrated in \cref{fig:jdddiscon}; or a cylinder
\begin{align}
  &\sim{}d^2\,e^{2S_\partial}
  &
  &\text{if \(d\gg e^{2S_0}\gg 1\)}
    \;,
    \label{eq:zetajdd2}
\end{align}
as illustrated in \cref{fig:jddcon}.

Finally, let us suppose that the ensemble-averaged size of
\(\tilde{\mathcal{H}}_d^\alpha\) (or an appropriate subspace in JT) is roughly
approximated by the number of eigenstates of \(O(\tilde{J}_{d\otimes{}d^*})\),
in the window where eigenvalues are of the typical size \labelcref{eq:jddsize1}
or \labelcref{eq:jddsize2}. Then, using \cref{eq:zetajdd1,eq:zetajdd2}, we see
that \cref{eq:traceboundavg} bounds this number of states from above by
\begin{align}
  \sim \min(d, e^{2S_0})
  \;.
  \label{eq:huzresult}
\end{align}
In fact, ref.~\cite{Harlow:2025pvj} reports the ensemble-averaged dimension of
\(\tilde{\mathcal{H}}_d^\alpha\) (or an appropriate subspace in JT\footnote{In
  JT gravity, \cref{eq:huzresult} should really be regarded as the approximate
  dimensions of subspaces of \(\tilde{\mathcal{H}}_d^\alpha\), for example,
  corresponding to microcanonical windows of the matrix theory
  \cite{Abdalla:2025gzn} as reviewed in \cref{sec:obsworldline}. The typical
  dimension of \(\tilde{\mathcal{H}}_d^\alpha\), without restriction to such a
  subspace, will be \(d\) in JT gravity even if \(d\gg{}e^{2S_0}\).}) to be
\(\sim \min(d, e^{2S_0})\).

\subsection{Future directions}
\label{sec:future}

Let us conclude with some remarks on future directions for exploration.

In \cref{sec:operators}, we introduced operators \(O(J_{C\otimes{}C^*})\) which
glue partial sources \(J_{C\otimes{}C^*}\) to states in
\(\mathcal{H}_C^\alpha\). It is straightforward to extend this definition to
two-sided Hilbert spaces \(\mathcal{H}_{C\otimes C'}^\alpha\). In close analogy
to ref.~\cite{Colafranceschi:2023moh}, one might consider taking the completion
of these operators in the weak operator topology to obtain a von Neumann
algebra. Ref.~\cite{Colafranceschi:2023moh} focused on the case where
\(J_{C\otimes{}C^*}\) is a partial boundary manifold and \(C=\Sigma\) is a cut
of a boundary manifold. By studying the structure of the resulting von Neumann
algebra \(\mathcal{A}_C\), it was found that much can be learned about the
structure of the Hilbert space on which the operators act. For example,
two-sided Hilbert spaces decompose \cite{Colafranceschi:2023moh,Marolf:2024adj}
\begin{align}
  \mathcal{H}_{C\otimes C'}^\alpha
  &= \bigoplus_\mu
    \mathcal{H}_C^{\alpha,\mu} \otimes \mathcal{H}_{C'}^{\alpha,\mu}
\end{align}
into factorizing eigenspaces of central elements of \(\mathcal{A}_C\). It would
be interesting to carry out an analogous analysis in the generalized framework
of this paper, including abstract partial sources for the path integral. In
particular, one might try to understand whether two-observer Hilbert spaces
decompose in a similar manner as two-boundary Hilbert spaces do above. We might
expect results analogous to refs.~\cite{Colafranceschi:2023moh,Marolf:2024adj}
since, at least superficially, almost all of their axioms (for each
\(\alpha\)-sector path integral \(\zeta_\alpha=Z_\alpha\)) seem to have obvious
analogues in the generalized framework considered in this paper, with the
possible exception of their continuity axiom.

In \cref{sec:obsworldline}, we raised the possibility of interpreting path
integrals with circular observer worldlines as partition functions counting
states in the Hilbert space of closed universes with an observer. Again, there
are several open questions here. Can we appropriately define such a path
integral in a realistic theory so that it gives a positive (as opposed to
complex \cite{Polchinski:1988ua} or negative \cite{Maldacena:2024spf}) number?
Would starting in Lorentzian signature help in this regard
\cite{Casali:2021ewu,Marolf:2022ybi}? We leave these questions for future work.

Finally, it would be interesting to explore the extent to which the gravitational
path integral can be characterized as a random non-isometric code
\cite{Akers:2022qdl,Harlow:2025pvj,Akers:2025ahe}.

To make some concrete comments in this direction, let us consider the
prescription of ref.~\cite{Harlow:2025pvj} described in \cref{sec:obsclone} for
introducing external observer clones entangled with observers in closed
universes. As we have described in \cref{sec:obsclone}, in our language, partial
sources \(\tilde{J}_d\in\tilde{\mathcal{J}}_d\) prepare states
\(\ket{\tilde{J}_d,\alpha}_{\mathrm{r}}\) in the \(\alpha\)-sector Hilbert space
\(\tilde{\mathcal{H}}_d^\alpha\) where an external observer clone is entangled
with a bulk observer. Let us consider a family of such partial sources
\begin{align}
  \tilde{J}_d^{(i)}
  &= \sum_{\sigma=1}^d \mathscr{M}^{(\sigma,i)} \ket{\sigma}
    \;,
\end{align}
where \(\mathscr{M}^{(\sigma,i)}\) is a boundary manifold preparing a bulk
observer in the state labelled by \(\sigma\) and other bulk degrees of freedom
in a semiclassical state labelled by \(i\). The naive inner product
\(\braket{i}{j}\) of perturbative semiclassical fields suggests a Hilbert space
of arbitrarily large dimension --- this naive description is referred to as the
``effective'' description. In contrast, as described in \cref{sec:obsclone},
each \(\alpha\)-sector Hilbert space \(\tilde{\mathcal{H}}_d^\alpha\) of the
full gravitational theory, with inner product
\(\tensor[_{\mathrm{r}}]{\braket{\tilde{J}_d^{(i)},\alpha}{\tilde{J}_d^{(j)},\alpha}}{_{\mathrm{r}}}\),
has bounded dimension. In particular, as described below \cref{eq:cloneineq1},
\(\tilde{\mathcal{H}}_d^\alpha\) is isometric to a subspace of the
\(d\)-dimensional Hilbert space \(H_d\) of the external observer clone.

Ref.~\cite{Harlow:2025pvj} describes their prescription as implementing a
non-isometric code mapping states \(\ket{i}\) in the effective description to a
``fundamental'' description by states in the Hilbert space \(H_d\) of the
external observer clone. With the aid of \cref{eq:hdalphatohd}, we see that this
non-isometric map \(V_\alpha\) (denoted \(\hat{V}\) in
ref.~\cite{Harlow:2025pvj}) can be written as
\begin{align}
  V_\alpha \ket{i}
  &= \sum_{\sigma=1}^d Z_\alpha(\mathscr{M}^{(\sigma,i)}) \ket{\sigma}
    \in H_d
    \;,
\end{align}
such that
\begin{align}
  \mel{i}{V_\alpha^\dagger V_\alpha}{j}
  &= \tensor[_{\mathrm{r}}]{\braket{\tilde{J}_d^{(i)},\alpha}{\tilde{J}_d^{(j)},\alpha}}{_{\mathrm{r}}}
    \;.
\end{align}
In particular, the matrix elements of \(V_\alpha\) are given by the
\(\alpha\)-sector path integrals
\begin{align}
  \mel{\sigma}{V_\alpha}{i}
  &= Z_\alpha(\mathscr{M}^{(\sigma,i)})
    \;.
\end{align}
From \cref{eq:clonegrammom}, we further see that the path integral \(\zeta\),
without the restriction to an \(\alpha\)-sector, corresponds to an ensemble
average over the non-isometric code \(V_\alpha\),
\begin{align}
    \zeta\left(
    (\tilde{J}_d^{(i_1)}|\tilde{J}_d^{(j_1)})
  \sqcup\cdots\sqcup
    (\tilde{J}_d^{(i_n)}|\tilde{J}_d^{(j_n)})
    \right)
    &= \int_{\mathcal{A}} \dd{\alpha}
    \abs{\braket{\alpha}{\varnothing}}^2
      \mel{i_1}{V_\alpha^\dagger V_\alpha}{j_1}
    \cdots
      \mel{i_n}{V_\alpha^\dagger V_\alpha}{j_n}
      \;.
\end{align}

In this sense, the path integral \(\zeta\) corresponds to a random non-isometric
code. Even in very simple toy models where the non-isometric code \(V_\alpha\)
is built from a handful of tensors and the ensemble average is taken as the Harr
measure integral over the tensors, the ensemble average produces contractions
between tensor legs which resemble bulk wormhole connections between source
components \cite{Akers:2022qdl,Harlow:2025pvj}. Ref.~\cite{Akers:2025ahe}
further incorporates locality into the encoding by considering non-isometric
codes built from tensor networks, finding that such toy models can also
reproduce some aspects of the prescription by ref.~\cite{Abdalla:2025gzn},
described in \cref{sec:obsworldline}, for including bulk observers with
prescribed worldlines. It would be interesting to explore the extent to which we
can push this apparent correspondence between gravitational path integrals
and random non-isometric codes.